\documentclass[aps,prd,twocolumn,reprint,amsmath,amssymb,preprintnumbers,superscriptaddress,nobibnotes,nofootinbib,floatfix]{revtex4-1}

\usepackage{graphicx}
\usepackage{dcolumn}
\usepackage{bm}
\usepackage{color}
\usepackage{amssymb}
\usepackage{amsmath}
\usepackage[colorlinks=true,allcolors=Purple,pdfusetitle]{hyperref}
\usepackage{rotating}
\usepackage{multirow}
\usepackage{graphicx}
\usepackage[dvipsnames]{xcolor}
\usepackage{svg}
\usepackage{esdiff}
\definecolor{Green}{RGB}{0, 128, 0}
\newcommand{\orcid}[1]{\href{https://orcid.org/#1}{\includegraphics[width=10pt]{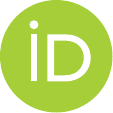}}}

\begin{document}
\preprint{N3AS-23-031}
\preprint{INT-PUB-23-042}
\preprint{CETUP-2023-010}

\title{Probing self-interacting sterile neutrino dark matter  \\ with the diffuse supernova neutrino background}

\author{A.~Baha Balantekin \orcid{0000-0002-2999-0111}}
\email{baha@physics.wisc.edu}
\affiliation{
Department of Physics, University of Wisconsin--Madison, Madison, Wisconsin 53706, USA}

\author{George M. Fuller \orcid{0000-0002-9716-9552}}
\email{gfuller@physics.ucsd.edu}
\affiliation{
Department of Physics, University of California, San Diego, La Jolla, CA 92093-0319, USA}

\author{Anupam Ray \orcid{0000-0001-8223-8239}}
\email{anupam.ray@berkeley.edu}
\affiliation{
Department of Physics, University of California Berkeley, Berkeley, California 94720, USA}
\affiliation{
School of Physics and Astronomy, University of Minnesota, Minneapolis, MN 55455, USA}

\author{Anna M. Suliga \orcid{0000-0002-8354-012X}}
\email{asuliga@berkeley.edu}
\affiliation{
Department of Physics, University of California Berkeley, Berkeley, California 94720, USA}
\affiliation{
Department of Physics, University of Wisconsin--Madison,
Madison, Wisconsin 53706, USA}
\affiliation{
Department of Physics, University of California, San Diego, La Jolla, CA 92093-0319, USA}

\date{October 11, 2023}


\begin{abstract}
The neutrinos in the diffuse supernova neutrino background (DSNB) travel over cosmological distances and this provides them with an excellent opportunity to interact with dark relics. We show that a cosmologically-significant relic population of keV-mass sterile neutrinos with strong self-interactions could imprint their presence in the DSNB. The signatures of the self-interactions would be ``dips" in the otherwise smooth DSNB spectrum. Upcoming large-scale neutrino detectors, for example Hyper-Kamiokande, have a good chance of detecting the DSNB and these dips. If no dips are detected, this method serves as an independent constraint on the sterile neutrino self-interaction strength and mixing with active neutrinos. We show that relic sterile neutrino parameters that evade X-ray and structure bounds may nevertheless be testable by future detectors like TRISTAN, but may also produce dips in the DSNB which could be detectable. Such a detection would suggest the existence of a cosmologically-significant, strongly self-interacting sterile neutrino background, likely embedded in a richer dark sector.   
\end{abstract}

\maketitle

\section{Introduction}
\label{sec:Introduction}
Neutrino rest mass and associated physics demonstrates the existence of physics beyond the Standard Model (SM)~\cite{Kamiokande-II:1992hns, Super-Kamiokande:1998kpq, SNO:2002tuh}. 
The nature and origin of the neutrino masses remains an unsolved problem. Several proposed explanations rely on invoking extra neutrinos, which are singlets with respect to the SM gauge group and therefore deemed sterile. In the simplest predictions, for example seesaw type-I~\cite{Fritzsch:1975sr, Minkowski:1977sc, Wyler:1982dd, Foot:1988aq}, the masses of sterile neutrinos can be substantially larger than other observed particles, up to $10^{16}$~GeV~\cite{Maltoni:2000iq, deGouvea:2016qpx}. These large masses suppress sterile neutrino mixing with the active neutrinos (see, e.g., Refs.~\cite{King:2003jb, Altarelli:2004za, Mohapatra:2005wg, King:2015aea, deGouvea:2016qpx, Xing:2020ald} for recent reviews on neutrino mass models).

The mixing parameters of sterile neutrinos, which dictate their observational implications, are often motivated by solving a particular problem or anomaly. The~eV-mass range has gained interest in the past due to the LSND~\cite{LSND:1996ubh, LSND:1997vun}, MiniBooNE~\cite{MiniBooNE:2007uho, MiniBooNE:2010idf}, MicroBooNE~\cite{Denton:2021czb, Arguelles:2021meu} the reactor and short baseline neutrino experiments~\cite{Barinov:2021asz, Berryman:2021yan} anomalies/tensions, their potential impact on nucleosynthesis in core-collapse supernovae~\cite{Qian:1993dg, Caldwell:1999zk, McLaughlin:1999pd, Fetter:2002xx, Tamborra:2011is, Wu:2013gxa, Pllumbi:2014saa, Xiong:2019nvw}, and alleviating tensions~\cite{Planck:2018vyg, Riess:2019cxk} in the measurements of some of the cosmological parameters~\cite{Kreisch:2019yzn, Hannestad:2013ana, Dasgupta:2013zpn, Mirizzi:2014ama}.
The keV-mass range is interesting because of its potential to explain dark matter~\cite{Dodelson:1993je, Shi:1998km, Abazajian:2001nj, Dolgov:2001nz, Kusenko:2009up, Adhikari:2016upu} and plausible significant impact on the core-collapse supernova evolution~\cite{Kainulainen:1990bn,  Raffelt:1992bs, Shi:1993ee, Nunokawa:1997ct, Fuller:2003gy, Hidaka:2006sg, Hidaka:2007se, Raffelt:2011nc, Warren:2016slz, Arguelles:2016uwb, Suliga:2019bsq, Syvolap:2019dat, Suliga:2020vpz, Ray:2023gtu}.

A signal from the radiative decay of a keV-mass sterile neutrino is discernible in a variety of existing and upcoming X-ray telescopes~\cite{Abazajian:2001vt, Boyarsky:2007ge, Horiuchi:2013noa, Roach:2019ctw, Ando:2021fhj, Foster:2021ngm, Malyshev:2020hcc, Roach:2022lgo, Gerbino:2022nvz}. X-ray astronomy and large scale structure considerations comprise our best probe of this putative sector of particle physics. However, the nuclear decay searches for keV-mass sterile neutrinos with terrestrial detectors such as TRISTAN~\cite{KATRIN:2018oow, KATRIN:2022spi} and HUNTER~\cite{Smith:2016vku, Martoff:2021vxp} can serve as complementary probes with additional discovery potential~\cite{Abazajian:2023reo}.

Several of the aforementioned sterile neutrino scenarios, while solving some of the anomalies or problems, are in tension with cosmological and astrophysical bounds. This is the case for the $\mathcal{O}(1)$~eV-mass sterile neutrinos. It is also the case for the simplest active neutrino scattering-induced de-coherence production channel \cite{Dodelson:1993je} for dark matter keV-mass sterile neutrinos ~\cite{Adhikari:2016upu, Acero:2022wqg}. However, both of these scenarios may still be viable if sterile or active neutrinos interact with themselves by an exchange of a new dark sector mediator~\cite{Kreisch:2019yzn, Hannestad:2013ana, Dasgupta:2013zpn, Mirizzi:2014ama, Chu:2015ipa, Hansen:2017rxr, Jeong:2018yts, Johns:2019cwc, DeGouvea:2019wpf, RoyChoudhury:2020dmd, Kelly:2020pcy, Bringmann:2022aim, Davoudiasl:2023uiq, Astros:2023xhe, An:2023mkf, Spisak:2023xxx}.

As a consequence of their weak interactions, neutrinos can be excellent probes of physics in distant and extreme environments. Its their ability to preserve information about their production site that distinguishes them from other messengers. So far, neutrinos have revealed insights for only a couple of distant astrophysical point sources: the Sun~\cite{Davis:1968cp, Cleveland:1998nv, Fukuda:2001nj, Araki:2004mb, Bellini:2008mr} and SN1987A~\cite{Hirata:1989zj, Bionta:1987qt, ALEXEYEV1988209}~\footnote{Also potentially Blazar TXS 0506+056~\cite{IceCube:2018dnn} and Starbust Galaxy NGC 1068~\cite{IceCube:2022der}.}. These observations significantly deepen the understanding of these sources. In addition, they also enable limits on new physics connected with neutrinos such as the active neutrino lifetimes~\cite{Frieman:1987as, Beacom:2002cb, Funcke:2019grs, Ivanez-Ballesteros:2023lqa}, magnetic moments~\cite{Balantekin:2004tk}, and absolute masses~\cite{Spergel:1987ex}. 
Other guaranteed sources of neutrinos, even of cosmological origin, exist. These are especially appealing for probing new neutrino interactions with themselves or other dark relics. 
One of the guaranteed neutrino fluxes is the diffuse supernova neutrino background, which is an isotropic and stationary flux of neutrinos coming from the integrated contribution of all the past core-collapse events in the Universe (see, e.g. Refs~\cite{Horiuchi:2008jz, Beacom:2010kk, Lunardini:2010ab, Vitagliano:2019yzm, Kresse:2020nto, Horiuchi:2020jnc, Ando:2023fcc, Suliga:2022ica} for recent reviews). 

Here we show how future detection of the diffuse supernova neutrino background (DSNB) could facilitate probes of the physics of a relic background of $\mathcal{O} (1)$ keV-mass self-interacting sterile neutrinos. The DSNB neutrinos travel cosmological distances, allowing them ample opportunity to interact with the sterile relics if there exists a non-zero vacuum mixing between the active and sterile sector neutrinos. Even in the case of negligibly small mixing with the active sector, the long baselines, high number density of sterile states, and strong self-coupling enable sterile neutrinos to encode the information about their existence in the active neutrino DSNB signal. 

An exciting development would be that a future terrestrial sterile neutrino experiment, such as TRISTAN~\cite{KATRIN:2018oow}, finds evidence for a $\mathcal{O}(1)$ keV-mass sterile state that mixed with active species {\it and} a correlated dip signal is seen in the DSNB spectrum. This would be a clear signature of a strongly self-interacting sterile neutrino dark matter component. Such a development could also be construed as hinting at the existence of a rich dark sector. Likewise, if the laboratory experiments find evidence of these sterile states, but no corresponding dip signature in the DSNB is detected, then we can constrain self-interaction and vacuum mixing parameters of the sterile neutrinos. That, in turn, could narrow the range of possibilities for dark matter sterile neutrino production mechanisms.  

The following discussion is organized as follows: In Sec.~\ref{sec:nus-SI} we briefly review the theory of sterile neutrino self-interactions. In Sec.~\ref{sec:DSNB-modeling} we outline models for the standard DSNB, discuss astrophysical uncertainties affecting it, and we describe the implementation of the modifications to the DSNB from sterile neutrino self-interactions. In Sec.~\ref{sec:Results} we calculate the event rates in the Hyper-Kamiokande detector and derive the sensitivity limits on the sterile neutrino mixing and self-coupling parameters. Finally, we summarize and conclude in Sec.~\ref{sec:Discussion-and-Conclusions}.

\section{Sterile Neutrino Self-Interactions}
\label{sec:nus-SI}
In this section, we discuss the class of sterile neutrino self-interaction models considered in this study.

The keV-mass sterile neutrinos with self-interactions are viable dark matter candidates, see Refs.~\cite{Hannestad:2013ana, Hansen:2017rxr, Johns:2019cwc, Bringmann:2022aim, Astros:2023xhe, An:2023mkf, Spisak:2023xxx}. In those studies, self-interacting sterile neutrinos also mix with active neutrinos, enabling the sterile neutrino self-interactions to affect the active neutrino fluxes. Here, we focus on the plausible interactions between the DSNB neutrinos with the self-interacting sterile neutrino dark matter, which might imprint a detectable dip signal in the observable DSNB spectrum. Similar considerations were introduced for the DSNB active neutrino self-interactions with the cosmic active neutrino background~\cite{Goldberg:2005yw, Baker:2006gm} and MeV mass dark matter~\cite{Farzan:2014gza}. The dips may originate from the resonant interaction of the following form
\begin{equation}
\label{eq:s-channel-scattering}
\nu_s + \nu_s \to \phi \to \nu_s + \nu_s \, .
\end{equation}

We specifically consider cosmologically relevant interactions between sterile neutrinos ($\nu_s$) with the interaction Lagrangian of the form
\begin{equation}
\label{eq:Lagrangian-nus-phi}
\mathcal{L} = g_{s} \phi \nu_s \nu_s  \,,
\end{equation}
where we assume that the scalar-particle $\phi$ mediates the interaction and $g_s$ denotes the self-coupling between sterile neutrinos.

The cross section for a DSNB neutrino with energy $E_\nu$ in the MeV-energy range resonantly scattering off the keV-mass sterile background neutrinos, with momenta much smaller than the sterile neutrino rest mass $m_s$, can be calculated with the following Breit-Wigner formula
\begin{equation}
\label{eq:cross-section}
\sigma (E_\nu) = \frac{g_s^4}{4\pi} \frac{s}{(s-m_\phi^2)^2 +m_\phi^{2}\Gamma_\phi^2} \, ,
\end{equation}
where the momentum transfer is $s = 2 E_\nu m_s$, $m_\phi$ is the mediator mass with a decay width $\Gamma_\phi = g_s^2 m_\phi/4\pi$.
If the resolution at which the impact of this interaction can be registered in the detected flux is weaker than the decay width of the mediator, the formula in Eq.~\eqref{eq:cross-section} reduces to delta function of the form
\begin{equation}
\label{eq:reduced-cross-section}
\sigma (E_\nu) \approx \frac{\pi g_s^2}{m_\phi^2} E_\nu \delta (E_R - E_\nu) = \sigma_R E_\nu \delta (E_R - E_\nu) \, ,
\end{equation}
where the resonance energy is $E_R = m_\phi^2 / 2m_s$. 
In our work, we focus on the DSNB detection in future water Cherenkov detectors such as Hyper-Kamiokande~\cite{Hyper-Kamiokande:2018ofw}. This limits the range of $E_R$ to Hyper-Kamiokande's DSNB detection window which is expected to be approximately 12-25~MeV. 

Hyper-Kamiokande's energy resolution should be at least comparable to Super-Kamiokande's, which we conservatively assume to be of the order of $\Delta E_e = 2$~MeV for electron energies in the range $\sim$ 10-20~MeV~\cite{Super-Kamiokande:2005wtt}. Comparing the $\Delta E_e$ to the width of the resonance $2E_R\Gamma_\phi/m_\phi$, we arrive at a condition on the sterile neutrino self-coupling $g_s \lesssim 0.5 \:\Delta E_e / E_R$~\cite{Creque-Sarbinowski:2020qhz}, where we take the uncertainty in the neutrino energy to be comparable to the electron or positron energy uncertainty in the detector. For resonance energies in the range 15 - 25~MeV this leads to $g_s \lesssim 0.7 - 0.9$. Self-coupled dark matter bounds~\cite{Randall:2008ppe, Tulin:2017ara} , when applied to keV-mass sterile neutrinos with mediator masses $\mathcal{O}(40-200)$ times larger than the $m_s$, limit the sterile neutrino self-coupling to $g_s \lesssim 0.1-0.2$. As the latter limits are stronger (see Secs.~\ref{sec:Sensitivity-limits} and \ref{sec:Discussion-and-Conclusions} for more details), we will assume that it is safe to use the delta function approximation for the cross section.

\section{Diffuse Supernova Neutrino Background}
\label{sec:DSNB-modeling}

In this section, we review how to calculate the standard DSNB and briefly discuss the astrophysical uncertainties affecting this flux in Sec.~\ref{sec:DSNB-Standard}. In Sec.~\ref{sec:DSNB-vSI}, we discuss how to implement the modifications to the DSNB that steam from sterile neutrino self-interactions.

\subsection{Calculation of the Standard DSNB Signal}
\label{sec:DSNB-Standard}

In the most basic and minimal calculation the ingredients entering the DSNB calculation are: (1) The cosmological core-collapse supernova (CCSNe) rate $R_\mathrm{SN}$; (2) The neutrino flux $F_\mathrm{SN}$ with redshifted energies $E_{\nu}(1+z)$ emitted from an average CCSN; and (3) The Hubble parameter $H(z)$, where $z$ is the cosmological redshift. The final expression then yields the flux arriving at the Earth per a single neutrino flavor
\begin{equation}
\label{eq:DSNB-standard}
\phi_\alpha (E_\nu) = \int_0^{z_\mathrm{max}} dz \frac{1}{H(z)} \; R_\mathrm{SN} (z) \; F_\mathrm{SN} \left(E_\nu(1+z)\right) \,,
\end{equation}
where the neutrino energy detected at the Earth is $E_\nu$, the Hubble parameter in the standard $\Lambda\mathrm{CDM}$ model of cosmology is $H(z) = H_0 \sqrt{(1+z)^3\Omega_m + (1-\Omega_m)}$, and where the Hubble constant $H_0$ and the fraction of the matter density of the Universe $\Omega_m$ have been assumed to be their accepted standard values~\cite{ParticleDataGroup:2020ssz}.

We calculate the supernova rate $R_\mathrm{SN}$ assuming that it follows the redshift evolution of the star-formation rate density $\dot\rho_\star(z)$, which has been found to be well parametrized with~\cite{Yuksel:2008cu}
\begin{equation}
\label{eq:sfr}
\dot{\rho}_\star(z) \propto \left[(1+z)^{a\eta} + \left(\frac{1+z}{B}\right)^{b\eta}+\left(\frac{1+z}{C}\right)^{c\eta}\right]^{1/\eta} \, ,
\end{equation}
where the best fit values for all the parameters are $a = 3.4$, $b = -0.3$, $c = -3.5$, $\eta = -10$, $B = (1+z_1)^{1-a/b}$, and $C = (1+z_1)^{(b-a)/c}(1+z_2)^{1-b/c}$ with the breaks in the function at redshifts of $z_1 = 1$, and $z_2 = 4$. The star-formation rate density has units of [$M_\odot\;\mathrm{Mpc}^{-3} \:\mathrm{yr}^{-1}$].
This leads to the supernova rate 
\begin{equation}
\label{eq:SN_Rate}
R_\mathrm{SN} (z) = \frac{ \int_{8\;M_\odot}^{125\:M_\odot} dM (dN/dM) \dot \rho_{\star}(z)}{\int_{0.1M_\odot}^{125M_\odot} dM \; M \; (dN/dM)} \, , 
\end{equation}
where the $dN/dM$ is the initial mass function which we take to be a Salpeter form, $dN/dM \propto M^{-2.35}$~\cite{Salpeter:1955it}. We assume a local supernova rate $R_\mathrm{SN}(z=0) = 1.25 \times 10^{-4}\;\mathrm{Mpc}^{-3}\;\mathrm{yr}^{-1}$. A~conservative evaluation of values obtained from various electromagnetic observations limits this to $R_\mathrm{SN}(z=0) = 0.75 - 1.75 \times 10^{-4}\;\mathrm{Mpc}^{-3}\;\mathrm{yr}^{-1}$~\cite{Smartt:2008zd, Li:2010kc, Horiuchi:2011zz, Botticella:2011nd, Mattila:2012, Taylor:2014rlo, Strolger:2015kra}. 
As new observatories with larger fields of view, for example the Vera Rubin Observatory's LSST survey~\cite{LSST:2008ijt, 2009JCAP...01..047L} (expected to be fully operational in 2024), start to monitor large portions of the sky, the estimates of the supernova rate and its local value should become significantly more precise.

We model the single flavor emission-time-integrated neutrino spectrum coming from an average core-collapse supernova with a Fermi-Dirac distribution
\begin{equation}
    F_\mathrm{SN} (E_\nu) = \frac{E_\mathrm{tot}}{6} \frac{120}{7\pi^4}\frac{E_\nu^2}{T_\nu^4} \frac{1}{\exp(E_\nu/T_\nu) + 1} \, ,
\end{equation}
where $T_\nu$ is the neutrino temperature at the location (the neutrino sphere) of decoupling from the matter in the supernova core. This is connected to the mean energy by roughly $E_\nu = 3.15 T_\nu$. Most of the recent 1D and 3D supernova simulations suggest $T_\nu$ between 4-8~MeV depending on the type of the progenitor and neutrino flavor~\cite{Lentz:2011aa, Sukhbold:2015wba, Takiwaki:2013cqa, OConnor:2018sti, Burrows:2019zce}. In addition, the DSNB flux of a particular neutrino flavor will depend on all of the initial neutrino flavors' fluxes as a consequence of neutrino flavor mixing~\cite{Duan:2010bg, Chakraborty:2016yeg, Tamborra:2020cul, Patwardhan:2022mxg, Volpe:2023met}. In our work, we neglect this and assume that the final DSNB flux can be described just by a single $T_\nu$.  However, in projecting the future experimental sensitivity to  new physics we include a generous uncertainty on the DSNB normalization and shape (see Sec.~\ref{sec:Sensitivity-limits} for more details). For more detailed and sophisticated DSNB predictions that incorporate the emission spectrum from a collection of several to tens of numerical SN models see, e.g., Refs.~\cite{Kresse:2020nto, Horiuchi:2020jnc}.

\subsection{Modifications of the DSNB due to Sterile Neutrino Self-Interactions}
\label{sec:DSNB-vSI}

Sterile neutrino self-interactions ($\nu_s$SI) can impact the DSNB if there exists a non-zero mixing between the sterile and active neutrinos and the sterile neutrinos have a non-zero self-coupling.  
In this case, the DNSB flux for a particular active neutrino flavor $\nu_\alpha$ is modified to 
\begin{equation}
\label{eq:DSNB-vsSI}
\begin{split}
\phi_\alpha (E_\nu) \backsimeq & \sum_{i=1}^{3} |U_{\alpha i}|^2 \int_0^{z_\mathrm{max}} dz \;  \frac{P_i(E_\nu, z)}{H(z)} \times \; \\
& R_\mathrm{SN} (z) \; F_\mathrm{SN}^{i} \left(E_\nu(1+z)\right) \,, 
\end{split}
\end{equation}
where the probability of the $i$-th light mass state neutrino interacting with the fourth (mostly sterile) mass state depends on the optical depth $\tau_i(E_\nu, z)$ and is given by $P_i(E_\nu, z) = e^{-\tau_i(E_\nu, z)}$. This approach neglects the reemergence of the down-scattered DSNB neutrinos which should be a good approximation for small $|U_{\alpha 4}|^2$.

The expression for the optical depth using the small resonance width approximation, i.e., cross section for the $s$-channel interaction of the form in Eq.~\eqref{eq:reduced-cross-section}, takes a compact form~\cite{Jeong:2018yts, Creque-Sarbinowski:2020qhz}
\begin{equation}
\label{eq:probability-of-interaction}
   \tau_i(E_\nu, z) \backsimeq  \tau_R \Theta(z-z_R)   = \frac{\Gamma_R (z_R)}{(1+z_R) H(z_R)} \Theta(z-z_R) \, ,
\end{equation}
where $z_R = {E_R}/{E_\nu} - 1$ and the scattering rate reduces to $\Gamma_R (z_R)  {\backsimeq |U_{si}|^2} n_{\nu_s} (z_R) \sigma_R$ with $n_{\nu_s} (z_R) = n_{\nu_s} (1+z_R)^3$. The number density of sterile neutrinos is taken to be the cosmological average dark matter (DM) density $n_{\nu_s} \approx 1.26\: (1~\mathrm{keV}/m_s)$~cm$^{-3}$~\cite{ParticleDataGroup:2020ssz}. We note that this is a conservative approach, as any DM overdensities, especially at low redshifts $z \approx 1$, could lead to an increased rate of the self-interactions and stronger limits.

We calculate the $\bar\nu_e$ DSNB fluxes, for the standard and $\nu_s$SI cases, at the Earth assuming that all of the emission-time-integrated (mostly active) neutrino mass state fluxes emitted from an average supernova are similar. In that case, approximate unitarity allows us to obtain the sum in Eq.~\eqref{eq:DSNB-vsSI} simply. We use two different neutrino temperatures $T_\nu = \{4, 8\}$~MeV to calculate the standard DSNB and $\nu_s$SI affected DSNB. As mentioned in the Sec.~\ref{sec:DSNB-Standard}, however, we include a marginalization over the shape and normalization of the DSNB flux to obtain future sensitivity limits  (see Sec.~\ref{sec:Sensitivity-limits} for more details).

\section{Projected sensitivity}
\label{sec:Results}

 In Sec.~\ref{sec:DSNB-current-limits} below we summarize the status of the experimental limits and the detection prospects for the DSNB. Following this discussion, in Sec.~\ref{sec:HK} we show the expected DSNB calculated event rates in the Hyper-Kamiokande detector with and without inclusion of $\nu_s$SI. In Sec.~\ref{sec:Sensitivity-limits}, we present our sensitivity limits on the sterile neutrino parameters.

\subsection{Current Limits and Detection Perspectives for the DSNB}
\label{sec:DSNB-current-limits}

The DSNB has not been detected yet. The most stringent current limits on the $\bar\nu_e$ component come from Super-Kamiokande (SK)~\cite{Super-Kamiokande:2021jaq, Super-Kamiokande:2023xup}, $\nu_e$ from SNO~\cite{SNO:2020gqd}, and the heavy lepton flavors also from SK~\cite{Lunardini:2008xd} but can be significantly improved with the dark matter direct detection experiments~\cite{Suliga:2021hek}. In the last few years significant efforts have been made by the SK 
Collaboration to mitigate backgrounds. This can allow that detector to operate with lower thresholds for a DSNB detection scheme utilizing gadolinium in the water~\cite{Beacom:2003nk, Super-Kamiokande:2023xup}. In addition, the currently being constructed and planned detectors such as Jiangmen Underground Neutrino Observatory (JUNO)~\cite{JUNO:2022lpc}, Hyper-Kamiokande (HK)~\cite{Hyper-Kamiokande:2018ofw} and the Deep Underground Neutrino Experiment (DUNE)~\cite{Zhu:2018rwc} have the potential to not only detect $\bar\nu_e$ and $\nu_e$ components of the DSNB but also measure this flux with relatively good precision~\cite{Moller:2018kpn}. In the future, with 10~kton\;yr exposures the DM direct detection experiments potentially may be able to detect the DSNB at the $\sim$3-4~$\sigma$ level~\cite{Suliga:2021hek, Zhuang:2023dzd}.

\subsection{Detection in the Hyper-Kamiokande}
\label{sec:HK}

As an example, here we focus on the predictions for 3740 kton\;yr exposures in the HK experiment~\cite{Hyper-Kamiokande:2018ofw} enriched with 0.1\% Gadolinium. HK is a water Cherenkov neutrino detector currently under construction in Japan. It is expected to start taking data in 2027.  The nominal fiducial volume for DSNB detection of a single tank is 187~kton~\cite{Hyper-Kamiokande:2018ofw}. The main detection channel for DSNB is the Inverse Beta Decay (IBD) $\bar\nu_e + \mathrm{p} \rightarrow e^+ + \mathrm{n}$. This is because both reaction products can be identified with high confidence either by the produced Cherenkov ring or the $\gamma$-rays from the nucleus excited by the thermal capture of a neutron.

The positron energy differential event IBD rate can be calculated with the following formula
\begin{equation}
\label{eq:event-rate}
\frac{dN}{dE_e} =  \varepsilon N_{t} T \int dE_{\nu} \sigma_\mathrm{IBD}(E_e, E_{\nu}) \; {\phi}_{\bar\nu_e}(E_{\nu}) \ ,
\end{equation}
where $\varepsilon = 67\%$ is the efficiency of the detection, $N_t = 1.25 \times 10^{34}$ is the number of hydrogen targets in the fiducial volume of the detector, $T = 20\:\mathrm{yr}$ is the time of data taking\footnote{Note that if the two tank configuration is built, 10~yr of data taking should suffice to give the same results.}, and $\sigma_{i}(E,E_{\nu}) $ is the IBD differential cross section~\cite{Strumia:2003zx, Ricciardi:2022pru}.

We limit the defined energy window to 12 - 30~MeV. The irreducible backgrounds present in this window are $\nu_e$ atmospheric charged-current events, invisible muon decays,
${}^9$Li spallation, and $\bar\nu_e$ from nuclear reactors. The chosen lower energy window bound cuts the reactor neutrino background and significantly reduces the events from ${}^9$Li spallation, while the rising invisible muon decay background dictates the upper bound. 
In addition, we bin the calculated event rates in 2~MeV bins based on the energy resolution of SK $\Delta E_e = 2$~MeV, for electron energies in the range $\sim$10 - 20~MeV~\cite{Super-Kamiokande:2005wtt}.

\begin{figure}[t]
 	\includegraphics[width=0.99\columnwidth]{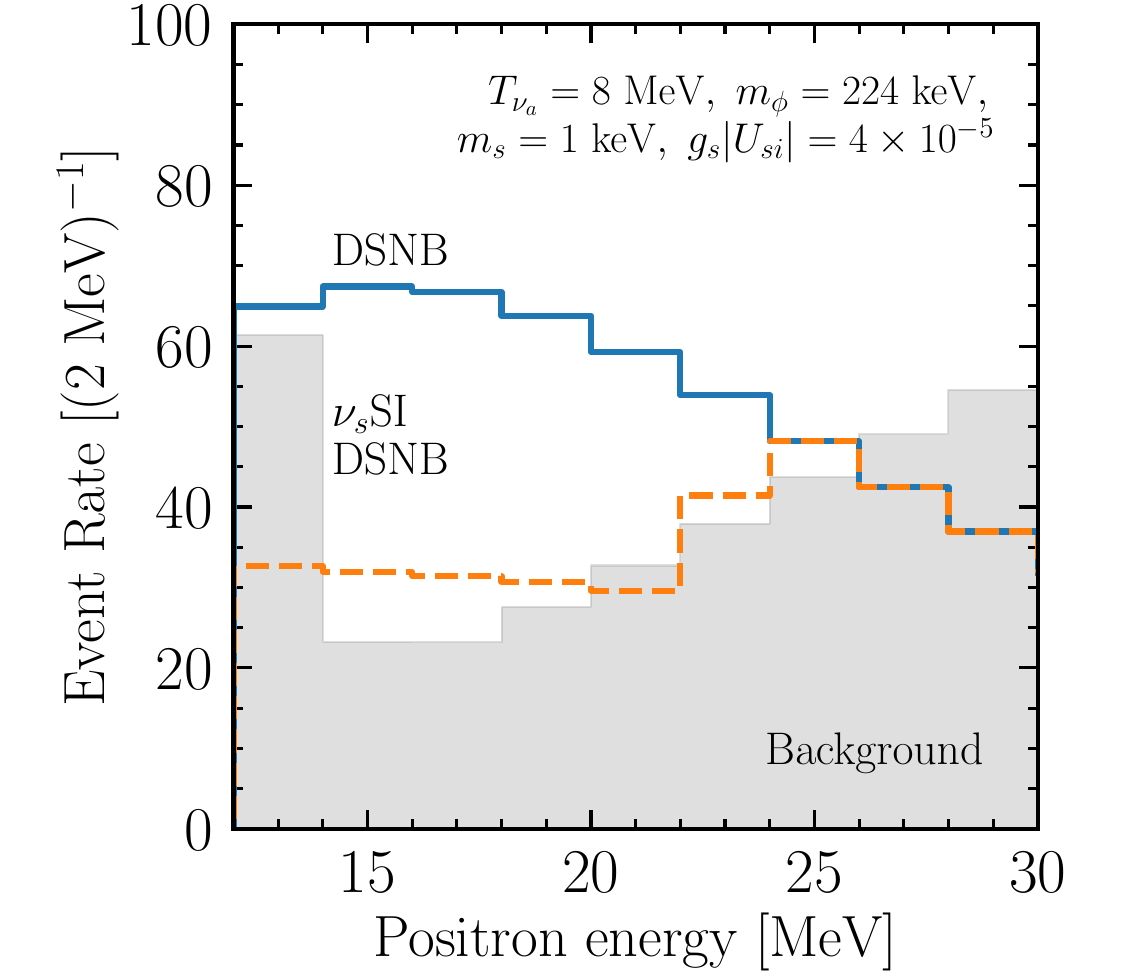}
	\caption{The differential DSNB event rate predictions as a function of the positron energy with $T_\nu = 8~\mathrm{MeV}$ for the standard DSNB model (solid blue line) and $\nu_s$SI DSNB with $m_\phi = 224~\mathrm{keV}$, $m_s = 1~\mathrm{keV}$, and $g_s|U_\mathrm{si}| = 4 \times 10^{-5}$ (dashed orange line), and the irreducible backgrounds (gray region) for HK with a 3740 kton$\;$yr exposure, 67\% detection efficiency, and binned into 2~MeV bins. The $\nu_s$SI DSNB event rate differs significantly from the standard DSNB prediction only for positrons with energies smaller or equal to the resonance energy $E_R$ dictated by the sterile neutrino and $\nu_s$SI mediator masses.}
	\label{Fig:Rate}
\end{figure}

Figure~\ref{Fig:Rate} shows the standard (solid blue line) and $\nu_s$SI (dashed-orange line) DSNB event rates in a 3740 kton$\;$yr exposure for HK, together with the detector backgrounds (gray region). Positrons with energies corresponding to neutrino energies below $E_R = 25$~MeV in the $\nu_s$SI DSNB event rate are depleted due to $\nu_s$SI. This feature makes it feasible to break the degeneracy between the astrophysical uncertainties in the standard DSNB prediction, which are expected to induce only smooth changes, and the presence of self-interactions between the sterile neutrino component of the light mass eigenstates in the DSNB neutrinos and the dark relic background (a sterile component of the heavy mass eigenstate). The DSNB detection should be the most sensitive to configurations of sterile neutrino and mediator masses that lead to dips in the middle of the DSNB detection window. In such cases, while half of the events will look as expected with the standard DSNB prediction, the dip in the middle suppresses the lower energy part of the window where most of the standard DSNB events would have been observed.

\subsection{Limits on the Sterile Neutrino Parameters}
\label{sec:Sensitivity-limits}

We calculate the 95\% confidence level (C.L.) limits on the sterile neutrino vacuum mixing angle with active species using a simple $\Delta\chi^2$ test 
\begin{equation}
\label{eq:chi2}
    \Delta \chi^2 = \min_{x, y} \sum_i \frac{(N_i^\mathrm{SM} - N_i^\mathrm{\nu_sSI})^2}{N_i^\mathrm{SM}+B_i} + \left(\frac{x}{\sigma_x}\right)^2 \, .
\end{equation}
Here the $i$-th index represents the bin number, $N_i^\mathrm{SM} \equiv N_i^\mathrm{SM}(T_\nu)$ is the number of DSNB events in the $i$-th bin without self-interactions, where we assume a particular supernova neutrino temperature $T_\nu$, and where $B_i$ is the number of background events, and $N_i^\mathrm{\nu_sSI} \equiv (1+x) N_i^\mathrm{\nu_sSi}(g_s|U_{si}|, m_s, T_\nu=y)$ is the number of DSNB events in the $i$-th bin with self-interactions. In this last expression, taking a range $x = [-1, 1]$ allows for changes in the normalization of the DSNB spectrum with a one sigma uncertainty $\sigma_x$ at a level of 50\%, and similarly for its spectral shape alteration for $y = [4, 8~\mathrm{MeV}]$.

To find the minimal $g_s|U_{si}|$ satisfying the condition $\Delta\chi^2 \geq \chi_\mathrm{95C.L}^2$ we use the $\Delta\chi^2$ minimized over $x$ and $y$ to account for the astrophysical uncertainties affecting the DSNB.

\begin{figure*}[t]
    \includegraphics[width=0.99\columnwidth]{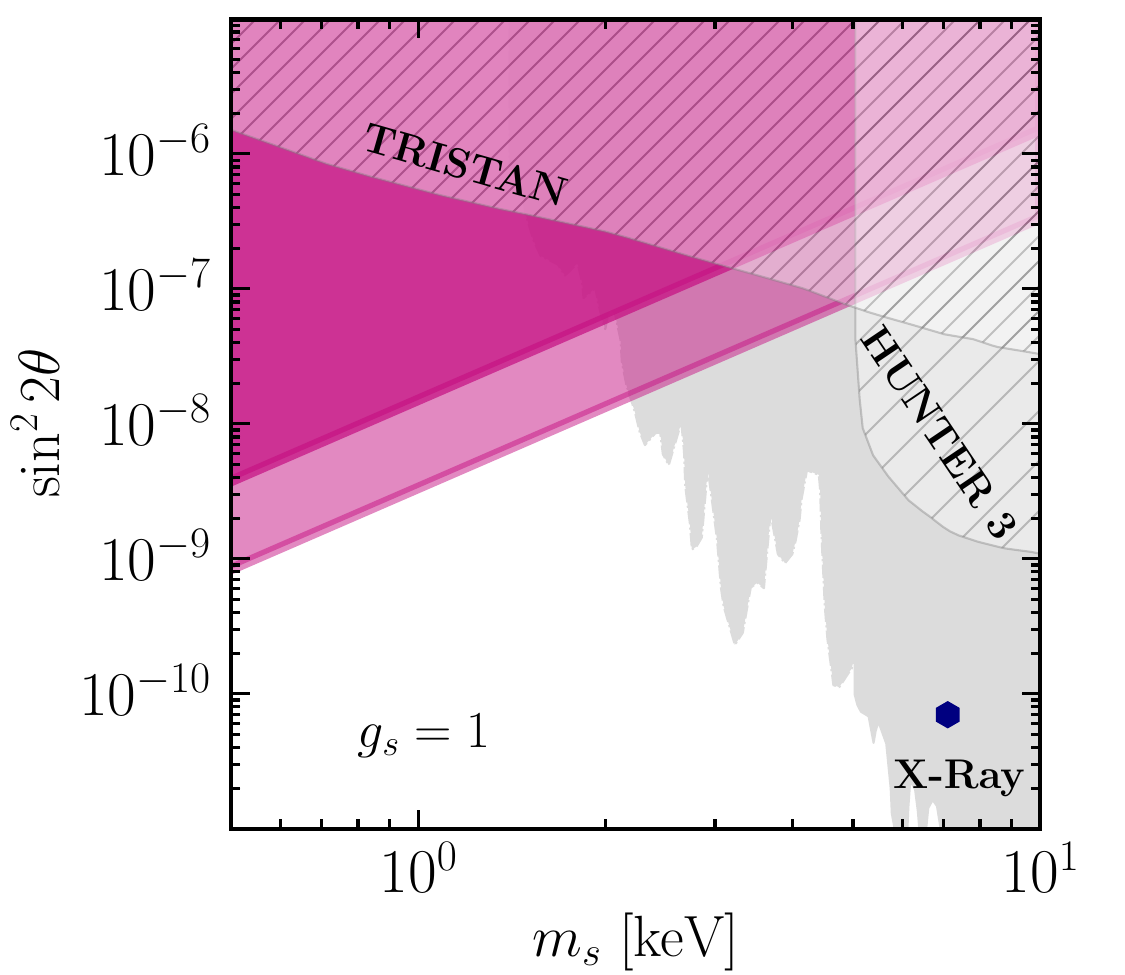}
 	\includegraphics[width=0.99\columnwidth]{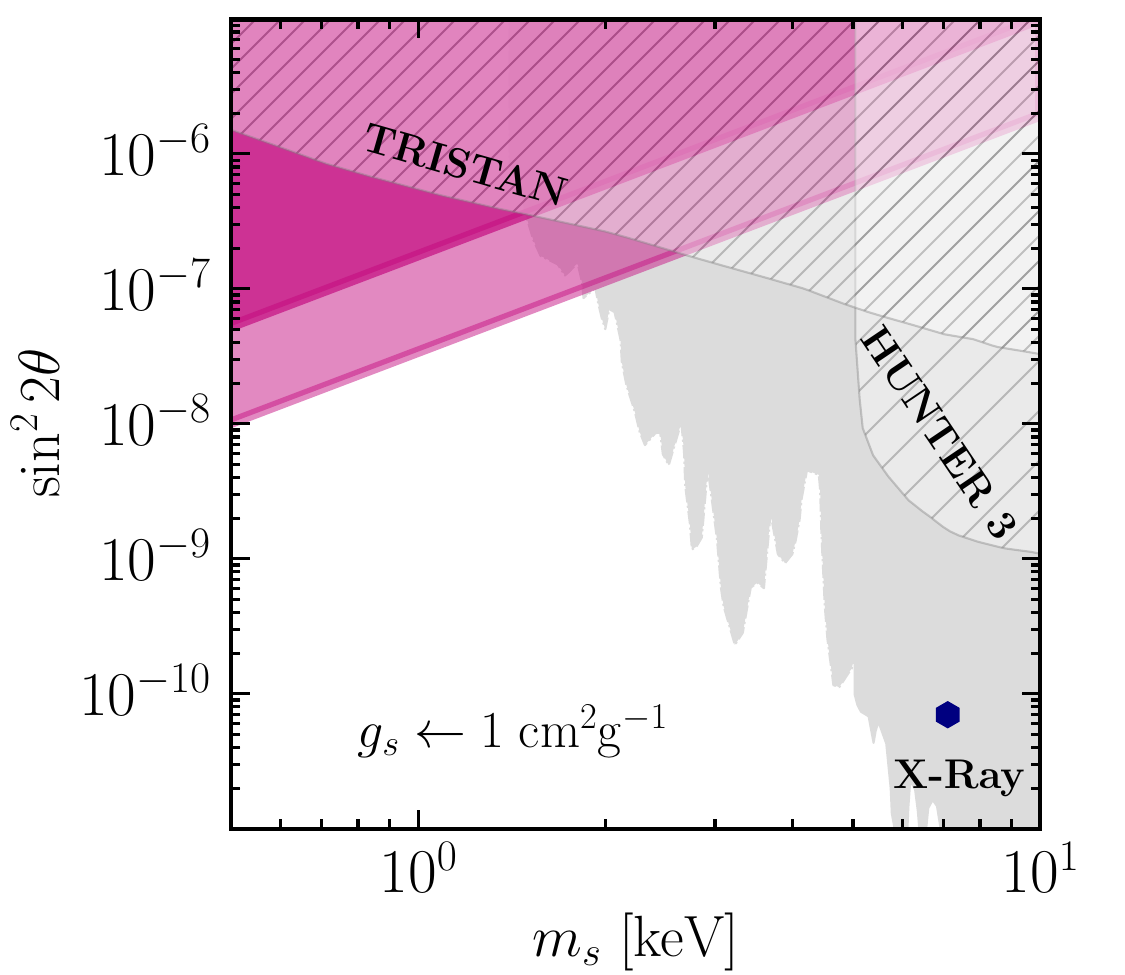}
	\caption{The 95\% C.L. sensitivity limits on the mass $m_s$, and the effective vacuum mixing angle with active species, $\theta$ of the self-interacting sterile neutrino dark matter. The mediator mass dictates the resonance energy $E_R$ in the DSNB observation window. We consider two cases where the indicated parameters give detectable dips: $T_\nu = 4$~MeV and $E_R = 15$~MeV (dark{er} pink regions); and $T_\nu = 8$~MeV and $E_R = 20$~MeV (light{er} pink regions). 
 On the left panel, the sterile neutrino self-coupling is set to $g_s = 1$ and on the right panel $g_s \approx 0.1-0.2$. We also show the X-ray limits from radiative decay of DM (gray regions)~\cite{Horiuchi:2013noa, Foster:2021ngm, Roach:2022lgo}, future sensitivities of TRISTAN~\cite{KATRIN:2018oow} (densely hatched regions) and HUNTER~3~\cite{Smith:2016vku, Martoff:2021vxp} ({sparsely} hatched regions), and the 3.5 keV line sterile neutrino hint (blue octagons)~\cite{Bulbul:2014sua, Boyarsky:2014jta}. The calculated sensitivities are stronger for smaller masses of the sterile neutrinos because they have a higher number density.}
	\label{Fig:limits}
\end{figure*}

Figure~\ref{Fig:limits} shows the calculated 95\% C.L. limits on the self-interacting sterile neutrinos' mixing angle and mass for $E_R = 15$~MeV and $T_\nu = 4$~MeV (dark pink regions) and $E_R = 20$~MeV and $T_\nu = 8$~MeV (light pink regions). The sterile neutrino self-coupling in the left panel is $g_s = 1$. In the right panel, the $g_s$ value corresponds to the limit on the dark matter self-coupling cross section, which is approximately $1\;\mathrm{cm}^2\:\mathrm{g}^{-1}$, from several of the late time astrophysical observations~\cite{Tulin:2017ara} (right panel).
Assuming that the self-scattering cross section outside of the $s$-channel resonance can be approximated with $\sigma_T \approx g_s^4 m_s^2 / 4 \pi m_\phi^4$~\cite{Spergel:1999mh} we find that the limit on the self-coupling is $g_s \lesssim 0.2 (\frac{\sigma_T}{m_s}/ 1~\mathrm{cm}^2\;\mathrm{g}^{-1})^{1/4} (m_s/10~\mathrm{keV})^{1/4} (E_R/25~\mathrm{MeV})^{1/2}$. For the values of sterile neutrino and mediator masses considered here, the resonance energy $E_R$ is 15 and 20 MeV, which translates to a limit on the self-coupling of the order of $g_s \lesssim 0.1 - 0.2$.

We find that our sensitivity limits can probe sterile neutrino masses $m_s \lesssim 2$~keV, which are not ruled out by the X-ray radiative decay constraints~\cite{Horiuchi:2013noa, Foster:2021ngm, Roach:2022lgo}, but can be probed by TRISTAN~\cite{KATRIN:2018oow} (statistical reach) in the future. If TRISTAN identifies a signal in that region of the sterile neutrino parameter space, and the detected DSNB event rate suggests the presence of spectral dips, that would point to compelling evidence for strongly self-interacting sterile neutrino dark matter. However, structure based constraints may well limit production schemes for a dark matter relic density comprised of sterile neutrinos with these relatively low masses (see Sec.~\ref{sec:Discussion-and-Conclusions} for more details).

{

TRISTAN project~\cite{KATRIN:2018oow} is a currently under development~\cite{KATRIN:2022spi} extension of the Karlsruhe Tritium Neutrino (KATRIN) experiment~\cite{KATRIN:2021dfa}. KATRIN is a high resolution spectrometer looking at the endpoint of the electron spectrum from tritium decay from an extremely powerful tritium source. The aim of that experiment is to measure the deviation from the maximum possible electron endpoint energy of $E_0 = 18.6$~keV to measure the electron antineutrino mass. KATRIN has already set the best terrestrial limits on that parameter $m_{\bar\nu_e} = 0.8$~eV~\cite{KATRIN:2019yun, KATRIN:2021fgc, KATRIN:2021uub}.

Nominally, the KATRIN energy window extends to 40 - 100 eV below the $E_0$~\cite{KATRIN:2022spi}. 
TRISTAN would be a novel detector system with an extended energy window to look for heavy sterile neutrinos up to $m_s = E_0$. The massive sterile neutrino can leave a feature in the spectrum of the electrons up to an energy $E_0 - m_s$~\cite{KATRIN:2018oow}. KATRIN already runs in a low luminosity mode allowing it to extend the energy window down to 0.01 - 1.6~keV below the endpoint energy for the tritium beta decay spectrum~\cite{KATRIN:2022spi}. For those sterile neutrino masses it lead to best terrestrial experimental limits on the active-sterile mixing for the tested heavy neutrino masses.

The Heavy Unseen Neutrinos from Total Energy-momentum Reconstruction (HUNTER)~\cite{Smith:2016vku, Martoff:2021vxp} experiment relies on the complete kinematical reconstruction of the electron-capture decay (K-capture) of the $^{131}$Cs isotope. The observable products of the decay are the recoil of the nucleus followed by an atomic X-ray photon and a single Auger electron (or sometimes two electrons); the invisible measure is the neutrino energy. Precision measurement of the momenta of all the visible particles from each decay will allow the experimentalist to reconstruct the momentum and mass of the neutrino.
The advantage of the complete kinematical reconstruction over the searches focusing on measuring the $\beta$ decay electron spectrum is that events involving a massive neutrino would appear as a distinct group rather than a kink in the spectrum. HUNTER will go after higher sterile neutrino masses than TRISTAN in a range from approximately from a few keV to a few hundreds of keV~\cite{Smith:2016vku, Martoff:2021vxp}.
}

In the parameter region where the X-ray limits are more constraining, the sensitivity limits estimated here can serve as complementary probes. In addition, note that the sensitivity limits presented here are only mildly affected by the astrophysical uncertainties. Better knowledge of the normalization and the DSNB spectral shape would improve the projected sensitivities but not by a significant factor.

\section{Discussion and Conclusions}
\label{sec:Discussion-and-Conclusions}

The keV-mass sterile neutrinos have been deemed \lq\lq natural\rq\rq\ dark matter candidates for a long time~\cite{Adhikari:2016upu}. The minimal Dodelson-Widrow collisional decoherence production mechanism~\cite{Dodelson:1993je}, however, is struggling to retain any viable parameter space for producing a relic density of dark matter that is compatible with the X-ray constraints from sterile neutrino radiative decay~\cite{Boyarsky:2007ge, Horiuchi:2013noa, Roach:2019ctw, Ando:2021fhj, Foster:2021ngm, Malyshev:2020hcc, Gerbino:2022nvz}. Some additional feature has to be added to this simplest mechanism to make it compatible with the data. One attempt at this is the Shi-Fuller mechanism~\cite{Shi:1993ee}, which reduces the required value of the mixing angle to produce the $\nu_s$ dark matter. This mechanism, however, requires a (significant) non-zero lepton number, and is nevertheless in tension with large scale structure considerations. Another viable alternative may be provided by sterile and active neutrino self-interactions~\cite{Hansen:2017rxr, DeGouvea:2019wpf, Bringmann:2022aim, Astros:2023xhe, Spisak:2023xxx}.

In our work, we have shown that sterile keV-mass neutrino self-interactions can imprint distinct features in the diffuse supernova neutrino background. We find that to circumvent the X-ray radiative decay bounds, the self-coupling needs to be substantial, i.e., above approximately $g_s \gtrsim 10^{-2}$. Such values are still consistent with the constraints from the bullet clusters~\cite{Randall:2008ppe} and other late-times astrophysical observations~\cite{Tulin:2017ara}. These self coupling limits are of the order of $g_s \lesssim 0.1-0.2$ for the values of $m_s$ and $m_\phi$ considered here.

The self-interacting sterile neutrinos with strong self-coupling overproduce the dark matter density in models with a single sterile neutrino and a single mediator~\cite{Hansen:2017rxr, Johns:2019cwc, Bringmann:2022aim, Astros:2023xhe}. Ways to ameliorate this issue could be a dilution of the sterile neutrino density by out-of-equilibrium heavy particle decays~\cite{Asaka:2006ek, Patwardhan:2015kga} or late-time vacuum phase transitions which suppress the mixing with the active sector at early times~\cite{Fuller:1990mq, Patwardhan:2014iha}.

Sterile neutrino dark matter, like all dark matter, is also subject to various phase space constraints, for example the Tremaine-Gunn bound for collision-less dark matter. Another phase space limit that applies to fermionic dark matter particles is the Pauli exclusion principle. 
Comparing the Fermi velocity with the escape velocity of a given galaxy~\cite{Tremaine:1979we}, this method yields a limit of approximately 25~eV for the Milky Way~\cite{Adhikari:2016upu} and around 250~eV for dwarf galaxies~\cite{Boyarsky:2008ju, Adhikari:2016upu, DiPaolo:2017geq, Savchenko:2019qnn}. More restrictive limits stem from the \emph{Liouville theorem} for collisionless particles, i.e., the Tremaine-Gunn bound. This states that the phase-space density of dark matter does not change in the course of its \emph{disipationless} and \emph{collisionless} evolution~\cite{Tremaine:1979we}. For thermally produced sterile neutrinos this bound restricts their masses to be below approximately 6~keV~\cite{Adhikari:2016upu}. However, changes in the initially produced velocity distribution of the dark matter, for example through dissipation in the dark sector, could in principle relax that limit. The physics in this case could be complex and difficult to 
analyze~\cite{Das:2010ts}. Here we remain agnostic as to how one could produce the relic density of the self-interacting sterile neutrinos. Our considerations here might give an important bound on such speculative model building.

The keV-mass sterile neutrinos can also impact the shape of the linear matter power spectrum. This would lead to observable changes in the Lyman-$\alpha$ forest of distant quasars~\cite{Meiksin:2007rz, Adhikari:2016upu, Zelko:2022tgf}. The sterile neutrinos imprint their signatures on small-scale structures through their free streaming length. This means that structure formation is only sensitive to the velocity distribution of dark matter particles. For a trivial thermal velocity distribution, this restricts the sterile neutrino masses to be below few keV. However, a more complex evolution of the velocity distribution function, i.e., a non-thermal velocity profile, could result in altered limits~\cite{Dienes:2021cxp}.

We note that the strongly self-coupling sterile neutrinos we suggest could be probed and constrained by future DSNB observations and also could be consistent with the required self-coupling for dark matter self-interactions invoked to explain an assortment of small-scale issues. These issues include, for example, the cusp-core problem for dwarf galaxies~\cite{Spergel:1999mh, Tulin:2017ara}.

We also note that the existence of strongly interacting sterile neutrinos may have nontrivial consequences for the core-collapse supernovae evolution. To fully understand the impact and feedback effects on supernovae evolution, further investigation is needed and beyond the scope of this paper.

An exciting prospect could develop along the following lines:  Terrestrial detectors, such as TRISTAN, see a signal of a sterile neutrino with mass $\mathcal{O}(1)\:\mathrm{keV}$; {\it and} active neutrino detectors see the dips in the DSNB that we discuss here. That outcome would hint at the existence of self-interacting sterile neutrino dark matter most likely emerging from a non-trivial dark sector.
\vspace{-0.05 cm}
\begin{acknowledgments}
We are grateful for helpful discussions with {Luk\'{a}\v{s} Gr\'{a}f}, Amol Patwardhan, Manibrata Sen, and Jacob Spisak.
This work was supported in part by the National Science Foundation Grants  No. PHY-2209578 at UCSD, NSF Grant No. PHY-2108339 at the University of Wisconsin,
and the NSF N3AS Physics Frontier Center, NSF Grant No. PHY-2020275, and the Heising-Simons Foundation (2017-228). 
Authors acknowledge partial support from the Institute for Nuclear Theory at the University of Washington for its kind hospitality and stimulating research environment. This research was supported in part by the INT's U.S. Department of Energy Grant No. DE-FG02-00ER41132. A.~B.~B., A.~R., and A.~M.~S. are also grateful to the Mainz Institute for Theoretical Physics (MITP) of the Cluster of Excellence PRISMA$^{+}$ (Project ID 39083149). Part of this research was completed while at the Center for Theoretical Underground Physics and Related Areas (CETUP*).
\end{acknowledgments}

\phantom{i}

\bibliography{ref}

\begin{thebibliography}{168}%
\makeatletter
\providecommand \@ifxundefined [1]{%
 \@ifx{#1\undefined}
}%
\providecommand \@ifnum [1]{%
 \ifnum #1\expandafter \@firstoftwo
 \else \expandafter \@secondoftwo
 \fi
}%
\providecommand \@ifx [1]{%
 \ifx #1\expandafter \@firstoftwo
 \else \expandafter \@secondoftwo
 \fi
}%
\providecommand \natexlab [1]{#1}%
\providecommand \enquote  [1]{``#1''}%
\providecommand \bibnamefont  [1]{#1}%
\providecommand \bibfnamefont [1]{#1}%
\providecommand \citenamefont [1]{#1}%
\providecommand \href@noop [0]{\@secondoftwo}%
\providecommand \href [0]{\begingroup \@sanitize@url \@href}%
\providecommand \@href[1]{\@@startlink{#1}\@@href}%
\providecommand \@@href[1]{\endgroup#1\@@endlink}%
\providecommand \@sanitize@url [0]{\catcode `\\12\catcode `\$12\catcode
  `\&12\catcode `\#12\catcode `\^12\catcode `\_12\catcode `\%12\relax}%
\providecommand \@@startlink[1]{}%
\providecommand \@@endlink[0]{}%
\providecommand \url  [0]{\begingroup\@sanitize@url \@url }%
\providecommand \@url [1]{\endgroup\@href {#1}{\urlprefix }}%
\providecommand \urlprefix  [0]{URL }%
\providecommand \Eprint [0]{\href }%
\providecommand \doibase [0]{http://dx.doi.org/}%
\providecommand \selectlanguage [0]{\@gobble}%
\providecommand \bibinfo  [0]{\@secondoftwo}%
\providecommand \bibfield  [0]{\@secondoftwo}%
\providecommand \translation [1]{[#1]}%
\providecommand \BibitemOpen [0]{}%
\providecommand \bibitemStop [0]{}%
\providecommand \bibitemNoStop [0]{.\EOS\space}%
\providecommand \EOS [0]{\spacefactor3000\relax}%
\providecommand \BibitemShut  [1]{\csname bibitem#1\endcsname}%
\let\auto@bib@innerbib\@empty
\bibitem [{\citenamefont {Hirata}\ \emph {et~al.}(1992)\citenamefont {Hirata}
  \emph {et~al.}}]{Kamiokande-II:1992hns}%
  \BibitemOpen
  \bibfield  {author} {\bibinfo {author} {\bibfnamefont {K.~S.}\ \bibnamefont
  {Hirata}} \emph {et~al.} (\bibinfo {collaboration} {Kamiokande-II}),\ }\href
  {\doibase 10.1016/0370-2693(92)90788-6} {\bibfield  {journal} {\bibinfo
  {journal} {Phys. Lett. B}\ }\textbf {\bibinfo {volume} {280}},\ \bibinfo
  {pages} {146} (\bibinfo {year} {1992})}\BibitemShut {NoStop}%
\bibitem [{\citenamefont {Fukuda}\ \emph {et~al.}(1998)\citenamefont {Fukuda}
  \emph {et~al.}}]{Super-Kamiokande:1998kpq}%
  \BibitemOpen
  \bibfield  {author} {\bibinfo {author} {\bibfnamefont {Y.}~\bibnamefont
  {Fukuda}} \emph {et~al.} (\bibinfo {collaboration} {Super-Kamiokande}),\
  }\href {\doibase 10.1103/PhysRevLett.81.1562} {\bibfield  {journal} {\bibinfo
   {journal} {Phys. Rev. Lett.}\ }\textbf {\bibinfo {volume} {81}},\ \bibinfo
  {pages} {1562} (\bibinfo {year} {1998})},\ \Eprint
  {http://arxiv.org/abs/hep-ex/9807003} {arXiv:hep-ex/9807003} \BibitemShut
  {NoStop}%
\bibitem [{\citenamefont {Ahmad}\ \emph {et~al.}(2002)\citenamefont {Ahmad}
  \emph {et~al.}}]{SNO:2002tuh}%
  \BibitemOpen
  \bibfield  {author} {\bibinfo {author} {\bibfnamefont {Q.~R.}\ \bibnamefont
  {Ahmad}} \emph {et~al.} (\bibinfo {collaboration} {SNO}),\ }\href {\doibase
  10.1103/PhysRevLett.89.011301} {\bibfield  {journal} {\bibinfo  {journal}
  {Phys. Rev. Lett.}\ }\textbf {\bibinfo {volume} {89}},\ \bibinfo {pages}
  {011301} (\bibinfo {year} {2002})},\ \Eprint
  {http://arxiv.org/abs/nucl-ex/0204008} {arXiv:nucl-ex/0204008} \BibitemShut
  {NoStop}%
\bibitem [{\citenamefont {Fritzsch}\ \emph {et~al.}(1975)\citenamefont
  {Fritzsch}, \citenamefont {Gell-Mann},\ and\ \citenamefont
  {Minkowski}}]{Fritzsch:1975sr}%
  \BibitemOpen
  \bibfield  {author} {\bibinfo {author} {\bibfnamefont {H.}~\bibnamefont
  {Fritzsch}}, \bibinfo {author} {\bibfnamefont {M.}~\bibnamefont {Gell-Mann}},
  \ and\ \bibinfo {author} {\bibfnamefont {P.}~\bibnamefont {Minkowski}},\
  }\href {\doibase 10.1016/0370-2693(75)90040-4} {\bibfield  {journal}
  {\bibinfo  {journal} {Phys. Lett. B}\ }\textbf {\bibinfo {volume} {59}},\
  \bibinfo {pages} {256} (\bibinfo {year} {1975})}\BibitemShut {NoStop}%
\bibitem [{\citenamefont {Minkowski}(1977)}]{Minkowski:1977sc}%
  \BibitemOpen
  \bibfield  {author} {\bibinfo {author} {\bibfnamefont {P.}~\bibnamefont
  {Minkowski}},\ }\href {\doibase 10.1016/0370-2693(77)90435-X} {\bibfield
  {journal} {\bibinfo  {journal} {Phys. Lett. B}\ }\textbf {\bibinfo {volume}
  {67}},\ \bibinfo {pages} {421} (\bibinfo {year} {1977})}\BibitemShut
  {NoStop}%
\bibitem [{\citenamefont {Wyler}\ and\ \citenamefont
  {Wolfenstein}(1983)}]{Wyler:1982dd}%
  \BibitemOpen
  \bibfield  {author} {\bibinfo {author} {\bibfnamefont {D.}~\bibnamefont
  {Wyler}}\ and\ \bibinfo {author} {\bibfnamefont {L.}~\bibnamefont
  {Wolfenstein}},\ }\href {\doibase 10.1016/0550-3213(83)90482-0} {\bibfield
  {journal} {\bibinfo  {journal} {Nucl. Phys. B}\ }\textbf {\bibinfo {volume}
  {218}},\ \bibinfo {pages} {205} (\bibinfo {year} {1983})}\BibitemShut
  {NoStop}%
\bibitem [{\citenamefont {Foot}\ \emph {et~al.}(1989)\citenamefont {Foot},
  \citenamefont {Lew}, \citenamefont {He},\ and\ \citenamefont
  {Joshi}}]{Foot:1988aq}%
  \BibitemOpen
  \bibfield  {author} {\bibinfo {author} {\bibfnamefont {R.}~\bibnamefont
  {Foot}}, \bibinfo {author} {\bibfnamefont {H.}~\bibnamefont {Lew}}, \bibinfo
  {author} {\bibfnamefont {X.~G.}\ \bibnamefont {He}}, \ and\ \bibinfo {author}
  {\bibfnamefont {G.~C.}\ \bibnamefont {Joshi}},\ }\href {\doibase
  10.1007/BF01415558} {\bibfield  {journal} {\bibinfo  {journal} {Z. Phys. C}\
  }\textbf {\bibinfo {volume} {44}},\ \bibinfo {pages} {441} (\bibinfo {year}
  {1989})}\BibitemShut {NoStop}%
\bibitem [{\citenamefont {Maltoni}\ \emph {et~al.}(2001)\citenamefont
  {Maltoni}, \citenamefont {Niczyporuk},\ and\ \citenamefont
  {Willenbrock}}]{Maltoni:2000iq}%
  \BibitemOpen
  \bibfield  {author} {\bibinfo {author} {\bibfnamefont {F.}~\bibnamefont
  {Maltoni}}, \bibinfo {author} {\bibfnamefont {J.~M.}\ \bibnamefont
  {Niczyporuk}}, \ and\ \bibinfo {author} {\bibfnamefont {S.}~\bibnamefont
  {Willenbrock}},\ }\href {\doibase 10.1103/PhysRevLett.86.212} {\bibfield
  {journal} {\bibinfo  {journal} {Phys. Rev. Lett.}\ }\textbf {\bibinfo
  {volume} {86}},\ \bibinfo {pages} {212} (\bibinfo {year} {2001})},\ \Eprint
  {http://arxiv.org/abs/hep-ph/0006358} {arXiv:hep-ph/0006358} \BibitemShut
  {NoStop}%
\bibitem [{\citenamefont {de~Gouv\^ea}(2016)}]{deGouvea:2016qpx}%
  \BibitemOpen
  \bibfield  {author} {\bibinfo {author} {\bibfnamefont {A.}~\bibnamefont
  {de~Gouv\^ea}},\ }\href {\doibase 10.1146/annurev-nucl-102115-044600}
  {\bibfield  {journal} {\bibinfo  {journal} {Ann. Rev. Nucl. Part. Sci.}\
  }\textbf {\bibinfo {volume} {66}},\ \bibinfo {pages} {197} (\bibinfo {year}
  {2016})}\BibitemShut {NoStop}%
\bibitem [{\citenamefont {King}(2004)}]{King:2003jb}%
  \BibitemOpen
  \bibfield  {author} {\bibinfo {author} {\bibfnamefont {S.~F.}\ \bibnamefont
  {King}},\ }\href {\doibase 10.1088/0034-4885/67/2/R01} {\bibfield  {journal}
  {\bibinfo  {journal} {Rept. Prog. Phys.}\ }\textbf {\bibinfo {volume} {67}},\
  \bibinfo {pages} {107} (\bibinfo {year} {2004})},\ \Eprint
  {http://arxiv.org/abs/hep-ph/0310204} {arXiv:hep-ph/0310204} \BibitemShut
  {NoStop}%
\bibitem [{\citenamefont {Altarelli}\ and\ \citenamefont
  {Feruglio}(2004)}]{Altarelli:2004za}%
  \BibitemOpen
  \bibfield  {author} {\bibinfo {author} {\bibfnamefont {G.}~\bibnamefont
  {Altarelli}}\ and\ \bibinfo {author} {\bibfnamefont {F.}~\bibnamefont
  {Feruglio}},\ }\href {\doibase 10.1088/1367-2630/6/1/106} {\bibfield
  {journal} {\bibinfo  {journal} {New J. Phys.}\ }\textbf {\bibinfo {volume}
  {6}},\ \bibinfo {pages} {106} (\bibinfo {year} {2004})},\ \Eprint
  {http://arxiv.org/abs/hep-ph/0405048} {arXiv:hep-ph/0405048} \BibitemShut
  {NoStop}%
\bibitem [{\citenamefont {Mohapatra}\ \emph {et~al.}(2007)\citenamefont
  {Mohapatra} \emph {et~al.}}]{Mohapatra:2005wg}%
  \BibitemOpen
  \bibfield  {author} {\bibinfo {author} {\bibfnamefont {R.~N.}\ \bibnamefont
  {Mohapatra}} \emph {et~al.},\ }\href {\doibase 10.1088/0034-4885/70/11/R02}
  {\bibfield  {journal} {\bibinfo  {journal} {Rept. Prog. Phys.}\ }\textbf
  {\bibinfo {volume} {70}},\ \bibinfo {pages} {1757} (\bibinfo {year}
  {2007})},\ \Eprint {http://arxiv.org/abs/hep-ph/0510213}
  {arXiv:hep-ph/0510213} \BibitemShut {NoStop}%
\bibitem [{\citenamefont {King}(2015)}]{King:2015aea}%
  \BibitemOpen
  \bibfield  {author} {\bibinfo {author} {\bibfnamefont {S.~F.}\ \bibnamefont
  {King}},\ }\href {\doibase 10.1088/0954-3899/42/12/123001} {\bibfield
  {journal} {\bibinfo  {journal} {J. Phys. G}\ }\textbf {\bibinfo {volume}
  {42}},\ \bibinfo {pages} {123001} (\bibinfo {year} {2015})},\ \Eprint
  {http://arxiv.org/abs/1510.02091} {arXiv:1510.02091 [hep-ph]} \BibitemShut
  {NoStop}%
\bibitem [{\citenamefont {Xing}\ and\ \citenamefont
  {Zhao}(2021)}]{Xing:2020ald}%
  \BibitemOpen
  \bibfield  {author} {\bibinfo {author} {\bibfnamefont {Z.-z.}\ \bibnamefont
  {Xing}}\ and\ \bibinfo {author} {\bibfnamefont {Z.-h.}\ \bibnamefont
  {Zhao}},\ }\href {\doibase 10.1088/1361-6633/abf086} {\bibfield  {journal}
  {\bibinfo  {journal} {Rept. Prog. Phys.}\ }\textbf {\bibinfo {volume} {84}},\
  \bibinfo {pages} {066201} (\bibinfo {year} {2021})},\ \Eprint
  {http://arxiv.org/abs/2008.12090} {arXiv:2008.12090 [hep-ph]} \BibitemShut
  {NoStop}%
\bibitem [{\citenamefont {Athanassopoulos}\ \emph {et~al.}(1996)\citenamefont
  {Athanassopoulos} \emph {et~al.}}]{LSND:1996ubh}%
  \BibitemOpen
  \bibfield  {author} {\bibinfo {author} {\bibfnamefont {C.}~\bibnamefont
  {Athanassopoulos}} \emph {et~al.} (\bibinfo {collaboration} {LSND}),\ }\href
  {\doibase 10.1103/PhysRevLett.77.3082} {\bibfield  {journal} {\bibinfo
  {journal} {Phys. Rev. Lett.}\ }\textbf {\bibinfo {volume} {77}},\ \bibinfo
  {pages} {3082} (\bibinfo {year} {1996})},\ \Eprint
  {http://arxiv.org/abs/nucl-ex/9605003} {arXiv:nucl-ex/9605003} \BibitemShut
  {NoStop}%
\bibitem [{\citenamefont {Athanassopoulos}\ \emph {et~al.}(1998)\citenamefont
  {Athanassopoulos} \emph {et~al.}}]{LSND:1997vun}%
  \BibitemOpen
  \bibfield  {author} {\bibinfo {author} {\bibfnamefont {C.}~\bibnamefont
  {Athanassopoulos}} \emph {et~al.} (\bibinfo {collaboration} {LSND}),\ }\href
  {\doibase 10.1103/PhysRevLett.81.1774} {\bibfield  {journal} {\bibinfo
  {journal} {Phys. Rev. Lett.}\ }\textbf {\bibinfo {volume} {81}},\ \bibinfo
  {pages} {1774} (\bibinfo {year} {1998})},\ \Eprint
  {http://arxiv.org/abs/nucl-ex/9709006} {arXiv:nucl-ex/9709006} \BibitemShut
  {NoStop}%
\bibitem [{\citenamefont {Aguilar-Arevalo}\ \emph {et~al.}(2007)\citenamefont
  {Aguilar-Arevalo} \emph {et~al.}}]{MiniBooNE:2007uho}%
  \BibitemOpen
  \bibfield  {author} {\bibinfo {author} {\bibfnamefont {A.~A.}\ \bibnamefont
  {Aguilar-Arevalo}} \emph {et~al.} (\bibinfo {collaboration} {MiniBooNE}),\
  }\href {\doibase 10.1103/PhysRevLett.98.231801} {\bibfield  {journal}
  {\bibinfo  {journal} {Phys. Rev. Lett.}\ }\textbf {\bibinfo {volume} {98}},\
  \bibinfo {pages} {231801} (\bibinfo {year} {2007})},\ \Eprint
  {http://arxiv.org/abs/0704.1500} {arXiv:0704.1500 [hep-ex]} \BibitemShut
  {NoStop}%
\bibitem [{\citenamefont {Aguilar-Arevalo}\ \emph {et~al.}(2010)\citenamefont
  {Aguilar-Arevalo} \emph {et~al.}}]{MiniBooNE:2010idf}%
  \BibitemOpen
  \bibfield  {author} {\bibinfo {author} {\bibfnamefont {A.~A.}\ \bibnamefont
  {Aguilar-Arevalo}} \emph {et~al.} (\bibinfo {collaboration} {MiniBooNE}),\
  }\href {\doibase 10.1103/PhysRevLett.105.181801} {\bibfield  {journal}
  {\bibinfo  {journal} {Phys. Rev. Lett.}\ }\textbf {\bibinfo {volume} {105}},\
  \bibinfo {pages} {181801} (\bibinfo {year} {2010})},\ \Eprint
  {http://arxiv.org/abs/1007.1150} {arXiv:1007.1150 [hep-ex]} \BibitemShut
  {NoStop}%
\bibitem [{\citenamefont {Denton}(2022)}]{Denton:2021czb}%
  \BibitemOpen
  \bibfield  {author} {\bibinfo {author} {\bibfnamefont {P.~B.}\ \bibnamefont
  {Denton}},\ }\href {\doibase 10.1103/PhysRevLett.129.061801} {\bibfield
  {journal} {\bibinfo  {journal} {Phys. Rev. Lett.}\ }\textbf {\bibinfo
  {volume} {129}},\ \bibinfo {pages} {061801} (\bibinfo {year} {2022})},\
  \Eprint {http://arxiv.org/abs/2111.05793} {arXiv:2111.05793 [hep-ph]}
  \BibitemShut {NoStop}%
\bibitem [{\citenamefont {Arg\"uelles}\ \emph {et~al.}(2022)\citenamefont
  {Arg\"uelles}, \citenamefont {Esteban}, \citenamefont {Hostert},
  \citenamefont {Kelly}, \citenamefont {Kopp}, \citenamefont {Machado},
  \citenamefont {Martinez-Soler},\ and\ \citenamefont
  {Perez-Gonzalez}}]{Arguelles:2021meu}%
  \BibitemOpen
  \bibfield  {author} {\bibinfo {author} {\bibfnamefont {C.~A.}\ \bibnamefont
  {Arg\"uelles}}, \bibinfo {author} {\bibfnamefont {I.}~\bibnamefont
  {Esteban}}, \bibinfo {author} {\bibfnamefont {M.}~\bibnamefont {Hostert}},
  \bibinfo {author} {\bibfnamefont {K.~J.}\ \bibnamefont {Kelly}}, \bibinfo
  {author} {\bibfnamefont {J.}~\bibnamefont {Kopp}}, \bibinfo {author}
  {\bibfnamefont {P.~A.~N.}\ \bibnamefont {Machado}}, \bibinfo {author}
  {\bibfnamefont {I.}~\bibnamefont {Martinez-Soler}}, \ and\ \bibinfo {author}
  {\bibfnamefont {Y.~F.}\ \bibnamefont {Perez-Gonzalez}},\ }\href {\doibase
  10.1103/PhysRevLett.128.241802} {\bibfield  {journal} {\bibinfo  {journal}
  {Phys. Rev. Lett.}\ }\textbf {\bibinfo {volume} {128}},\ \bibinfo {pages}
  {241802} (\bibinfo {year} {2022})},\ \Eprint
  {http://arxiv.org/abs/2111.10359} {arXiv:2111.10359 [hep-ph]} \BibitemShut
  {NoStop}%
\bibitem [{\citenamefont {Barinov}\ \emph {et~al.}(2022)\citenamefont {Barinov}
  \emph {et~al.}}]{Barinov:2021asz}%
  \BibitemOpen
  \bibfield  {author} {\bibinfo {author} {\bibfnamefont {V.~V.}\ \bibnamefont
  {Barinov}} \emph {et~al.},\ }\href {\doibase 10.1103/PhysRevLett.128.232501}
  {\bibfield  {journal} {\bibinfo  {journal} {Phys. Rev. Lett.}\ }\textbf
  {\bibinfo {volume} {128}},\ \bibinfo {pages} {232501} (\bibinfo {year}
  {2022})},\ \Eprint {http://arxiv.org/abs/2109.11482} {arXiv:2109.11482
  [nucl-ex]} \BibitemShut {NoStop}%
\bibitem [{\citenamefont {Berryman}\ \emph {et~al.}(2022)\citenamefont
  {Berryman}, \citenamefont {Coloma}, \citenamefont {Huber}, \citenamefont
  {Schwetz},\ and\ \citenamefont {Zhou}}]{Berryman:2021yan}%
  \BibitemOpen
  \bibfield  {author} {\bibinfo {author} {\bibfnamefont {J.~M.}\ \bibnamefont
  {Berryman}}, \bibinfo {author} {\bibfnamefont {P.}~\bibnamefont {Coloma}},
  \bibinfo {author} {\bibfnamefont {P.}~\bibnamefont {Huber}}, \bibinfo
  {author} {\bibfnamefont {T.}~\bibnamefont {Schwetz}}, \ and\ \bibinfo
  {author} {\bibfnamefont {A.}~\bibnamefont {Zhou}},\ }\href {\doibase
  10.1007/JHEP02(2022)055} {\bibfield  {journal} {\bibinfo  {journal} {JHEP}\
  }\textbf {\bibinfo {volume} {02}},\ \bibinfo {pages} {055} (\bibinfo {year}
  {2022})},\ \Eprint {http://arxiv.org/abs/2111.12530} {arXiv:2111.12530
  [hep-ph]} \BibitemShut {NoStop}%
\bibitem [{\citenamefont {Qian}\ \emph {et~al.}(1993)\citenamefont {Qian},
  \citenamefont {Fuller}, \citenamefont {Mathews}, \citenamefont {Mayle},
  \citenamefont {Wilson},\ and\ \citenamefont {Woosley}}]{Qian:1993dg}%
  \BibitemOpen
  \bibfield  {author} {\bibinfo {author} {\bibfnamefont {Y.-Z.}\ \bibnamefont
  {Qian}}, \bibinfo {author} {\bibfnamefont {G.~M.}\ \bibnamefont {Fuller}},
  \bibinfo {author} {\bibfnamefont {G.~J.}\ \bibnamefont {Mathews}}, \bibinfo
  {author} {\bibfnamefont {R.}~\bibnamefont {Mayle}}, \bibinfo {author}
  {\bibfnamefont {J.~R.}\ \bibnamefont {Wilson}}, \ and\ \bibinfo {author}
  {\bibfnamefont {S.~E.}\ \bibnamefont {Woosley}},\ }\href {\doibase
  10.1103/PhysRevLett.71.1965} {\bibfield  {journal} {\bibinfo  {journal}
  {Phys. Rev. Lett.}\ }\textbf {\bibinfo {volume} {71}},\ \bibinfo {pages}
  {1965} (\bibinfo {year} {1993})}\BibitemShut {NoStop}%
\bibitem [{\citenamefont {Caldwell}\ \emph {et~al.}(2000)\citenamefont
  {Caldwell}, \citenamefont {Fuller},\ and\ \citenamefont
  {Qian}}]{Caldwell:1999zk}%
  \BibitemOpen
  \bibfield  {author} {\bibinfo {author} {\bibfnamefont {D.~O.}\ \bibnamefont
  {Caldwell}}, \bibinfo {author} {\bibfnamefont {G.~M.}\ \bibnamefont
  {Fuller}}, \ and\ \bibinfo {author} {\bibfnamefont {Y.-Z.}\ \bibnamefont
  {Qian}},\ }\href {\doibase 10.1103/PhysRevD.61.123005} {\bibfield  {journal}
  {\bibinfo  {journal} {Phys. Rev. D}\ }\textbf {\bibinfo {volume} {61}},\
  \bibinfo {pages} {123005} (\bibinfo {year} {2000})},\ \Eprint
  {http://arxiv.org/abs/astro-ph/9910175} {arXiv:astro-ph/9910175} \BibitemShut
  {NoStop}%
\bibitem [{\citenamefont {McLaughlin}\ \emph {et~al.}(1999)\citenamefont
  {McLaughlin}, \citenamefont {Fetter}, \citenamefont {Balantekin},\ and\
  \citenamefont {Fuller}}]{McLaughlin:1999pd}%
  \BibitemOpen
  \bibfield  {author} {\bibinfo {author} {\bibfnamefont {G.~C.}\ \bibnamefont
  {McLaughlin}}, \bibinfo {author} {\bibfnamefont {J.~M.}\ \bibnamefont
  {Fetter}}, \bibinfo {author} {\bibfnamefont {A.~B.}\ \bibnamefont
  {Balantekin}}, \ and\ \bibinfo {author} {\bibfnamefont {G.~M.}\ \bibnamefont
  {Fuller}},\ }\href {\doibase 10.1103/PhysRevC.59.2873} {\bibfield  {journal}
  {\bibinfo  {journal} {Phys. Rev. C}\ }\textbf {\bibinfo {volume} {59}},\
  \bibinfo {pages} {2873} (\bibinfo {year} {1999})},\ \Eprint
  {http://arxiv.org/abs/astro-ph/9902106} {arXiv:astro-ph/9902106} \BibitemShut
  {NoStop}%
\bibitem [{\citenamefont {Fetter}\ \emph {et~al.}(2003)\citenamefont {Fetter},
  \citenamefont {McLaughlin}, \citenamefont {Balantekin},\ and\ \citenamefont
  {Fuller}}]{Fetter:2002xx}%
  \BibitemOpen
  \bibfield  {author} {\bibinfo {author} {\bibfnamefont {J.}~\bibnamefont
  {Fetter}}, \bibinfo {author} {\bibfnamefont {G.~C.}\ \bibnamefont
  {McLaughlin}}, \bibinfo {author} {\bibfnamefont {A.~B.}\ \bibnamefont
  {Balantekin}}, \ and\ \bibinfo {author} {\bibfnamefont {G.~M.}\ \bibnamefont
  {Fuller}},\ }\href {\doibase 10.1016/S0927-6505(02)00156-1} {\bibfield
  {journal} {\bibinfo  {journal} {Astropart. Phys.}\ }\textbf {\bibinfo
  {volume} {18}},\ \bibinfo {pages} {433} (\bibinfo {year} {2003})},\ \Eprint
  {http://arxiv.org/abs/hep-ph/0205029} {arXiv:hep-ph/0205029} \BibitemShut
  {NoStop}%
\bibitem [{\citenamefont {Tamborra}\ \emph {et~al.}(2012)\citenamefont
  {Tamborra}, \citenamefont {Raffelt}, \citenamefont {Hudepohl},\ and\
  \citenamefont {Janka}}]{Tamborra:2011is}%
  \BibitemOpen
  \bibfield  {author} {\bibinfo {author} {\bibfnamefont {I.}~\bibnamefont
  {Tamborra}}, \bibinfo {author} {\bibfnamefont {G.~G.}\ \bibnamefont
  {Raffelt}}, \bibinfo {author} {\bibfnamefont {L.}~\bibnamefont {Hudepohl}}, \
  and\ \bibinfo {author} {\bibfnamefont {H.-T.}\ \bibnamefont {Janka}},\ }\href
  {\doibase 10.1088/1475-7516/2012/01/013} {\bibfield  {journal} {\bibinfo
  {journal} {JCAP}\ }\textbf {\bibinfo {volume} {01}},\ \bibinfo {pages} {013}
  (\bibinfo {year} {2012})},\ \Eprint {http://arxiv.org/abs/1110.2104}
  {arXiv:1110.2104 [astro-ph.SR]} \BibitemShut {NoStop}%
\bibitem [{\citenamefont {Wu}\ \emph {et~al.}(2014)\citenamefont {Wu},
  \citenamefont {Fischer}, \citenamefont {Huther}, \citenamefont
  {Mart\'\i{}nez-Pinedo},\ and\ \citenamefont {Qian}}]{Wu:2013gxa}%
  \BibitemOpen
  \bibfield  {author} {\bibinfo {author} {\bibfnamefont {M.-R.}\ \bibnamefont
  {Wu}}, \bibinfo {author} {\bibfnamefont {T.}~\bibnamefont {Fischer}},
  \bibinfo {author} {\bibfnamefont {L.}~\bibnamefont {Huther}}, \bibinfo
  {author} {\bibfnamefont {G.}~\bibnamefont {Mart\'\i{}nez-Pinedo}}, \ and\
  \bibinfo {author} {\bibfnamefont {Y.-Z.}\ \bibnamefont {Qian}},\ }\href
  {\doibase 10.1103/PhysRevD.89.061303} {\bibfield  {journal} {\bibinfo
  {journal} {Phys. Rev. D}\ }\textbf {\bibinfo {volume} {89}},\ \bibinfo
  {pages} {061303} (\bibinfo {year} {2014})},\ \Eprint
  {http://arxiv.org/abs/1305.2382} {arXiv:1305.2382 [astro-ph.HE]} \BibitemShut
  {NoStop}%
\bibitem [{\citenamefont {Pllumbi}\ \emph {et~al.}(2015)\citenamefont
  {Pllumbi}, \citenamefont {Tamborra}, \citenamefont {Wanajo}, \citenamefont
  {Janka},\ and\ \citenamefont {H\"udepohl}}]{Pllumbi:2014saa}%
  \BibitemOpen
  \bibfield  {author} {\bibinfo {author} {\bibfnamefont {E.}~\bibnamefont
  {Pllumbi}}, \bibinfo {author} {\bibfnamefont {I.}~\bibnamefont {Tamborra}},
  \bibinfo {author} {\bibfnamefont {S.}~\bibnamefont {Wanajo}}, \bibinfo
  {author} {\bibfnamefont {H.-T.}\ \bibnamefont {Janka}}, \ and\ \bibinfo
  {author} {\bibfnamefont {L.}~\bibnamefont {H\"udepohl}},\ }\href {\doibase
  10.1088/0004-637X/808/2/188} {\bibfield  {journal} {\bibinfo  {journal}
  {Astrophys. J.}\ }\textbf {\bibinfo {volume} {808}},\ \bibinfo {pages} {188}
  (\bibinfo {year} {2015})},\ \Eprint {http://arxiv.org/abs/1406.2596}
  {arXiv:1406.2596 [astro-ph.SR]} \BibitemShut {NoStop}%
\bibitem [{\citenamefont {Xiong}\ \emph {et~al.}(2019)\citenamefont {Xiong},
  \citenamefont {Wu},\ and\ \citenamefont {Qian}}]{Xiong:2019nvw}%
  \BibitemOpen
  \bibfield  {author} {\bibinfo {author} {\bibfnamefont {Z.}~\bibnamefont
  {Xiong}}, \bibinfo {author} {\bibfnamefont {M.-R.}\ \bibnamefont {Wu}}, \
  and\ \bibinfo {author} {\bibfnamefont {Y.-Z.}\ \bibnamefont {Qian}},\ }\href
  {\doibase 10.3847/1538-4357/ab2870} {\  (\bibinfo {year} {2019}),\
  10.3847/1538-4357/ab2870},\ \Eprint {http://arxiv.org/abs/1904.09371}
  {arXiv:1904.09371 [astro-ph.HE]} \BibitemShut {NoStop}%
\bibitem [{\citenamefont {Aghanim}\ \emph {et~al.}(2020)\citenamefont {Aghanim}
  \emph {et~al.}}]{Planck:2018vyg}%
  \BibitemOpen
  \bibfield  {author} {\bibinfo {author} {\bibfnamefont {N.}~\bibnamefont
  {Aghanim}} \emph {et~al.} (\bibinfo {collaboration} {Planck}),\ }\href
  {\doibase 10.1051/0004-6361/201833910} {\bibfield  {journal} {\bibinfo
  {journal} {Astron. Astrophys.}\ }\textbf {\bibinfo {volume} {641}},\ \bibinfo
  {pages} {A6} (\bibinfo {year} {2020})},\ \bibinfo {note} {[Erratum:
  Astron.Astrophys. 652, C4 (2021)]},\ \Eprint
  {http://arxiv.org/abs/1807.06209} {arXiv:1807.06209 [astro-ph.CO]}
  \BibitemShut {NoStop}%
\bibitem [{\citenamefont {Riess}\ \emph {et~al.}(2019)\citenamefont {Riess},
  \citenamefont {Casertano}, \citenamefont {Yuan}, \citenamefont {Macri},\ and\
  \citenamefont {Scolnic}}]{Riess:2019cxk}%
  \BibitemOpen
  \bibfield  {author} {\bibinfo {author} {\bibfnamefont {A.~G.}\ \bibnamefont
  {Riess}}, \bibinfo {author} {\bibfnamefont {S.}~\bibnamefont {Casertano}},
  \bibinfo {author} {\bibfnamefont {W.}~\bibnamefont {Yuan}}, \bibinfo {author}
  {\bibfnamefont {L.~M.}\ \bibnamefont {Macri}}, \ and\ \bibinfo {author}
  {\bibfnamefont {D.}~\bibnamefont {Scolnic}},\ }\href {\doibase
  10.3847/1538-4357/ab1422} {\bibfield  {journal} {\bibinfo  {journal}
  {Astrophys. J.}\ }\textbf {\bibinfo {volume} {876}},\ \bibinfo {pages} {85}
  (\bibinfo {year} {2019})},\ \Eprint {http://arxiv.org/abs/1903.07603}
  {arXiv:1903.07603 [astro-ph.CO]} \BibitemShut {NoStop}%
\bibitem [{\citenamefont {Kreisch}\ \emph {et~al.}(2020)\citenamefont
  {Kreisch}, \citenamefont {Cyr-Racine},\ and\ \citenamefont
  {Dor\'e}}]{Kreisch:2019yzn}%
  \BibitemOpen
  \bibfield  {author} {\bibinfo {author} {\bibfnamefont {C.~D.}\ \bibnamefont
  {Kreisch}}, \bibinfo {author} {\bibfnamefont {F.-Y.}\ \bibnamefont
  {Cyr-Racine}}, \ and\ \bibinfo {author} {\bibfnamefont {O.}~\bibnamefont
  {Dor\'e}},\ }\href {\doibase 10.1103/PhysRevD.101.123505} {\bibfield
  {journal} {\bibinfo  {journal} {Phys. Rev. D}\ }\textbf {\bibinfo {volume}
  {101}},\ \bibinfo {pages} {123505} (\bibinfo {year} {2020})},\ \Eprint
  {http://arxiv.org/abs/1902.00534} {arXiv:1902.00534 [astro-ph.CO]}
  \BibitemShut {NoStop}%
\bibitem [{\citenamefont {Hannestad}\ \emph {et~al.}(2014)\citenamefont
  {Hannestad}, \citenamefont {Hansen},\ and\ \citenamefont
  {Tram}}]{Hannestad:2013ana}%
  \BibitemOpen
  \bibfield  {author} {\bibinfo {author} {\bibfnamefont {S.}~\bibnamefont
  {Hannestad}}, \bibinfo {author} {\bibfnamefont {R.~S.}\ \bibnamefont
  {Hansen}}, \ and\ \bibinfo {author} {\bibfnamefont {T.}~\bibnamefont
  {Tram}},\ }\href {\doibase 10.1103/PhysRevLett.112.031802} {\bibfield
  {journal} {\bibinfo  {journal} {Phys. Rev. Lett.}\ }\textbf {\bibinfo
  {volume} {112}},\ \bibinfo {pages} {031802} (\bibinfo {year} {2014})},\
  \Eprint {http://arxiv.org/abs/1310.5926} {arXiv:1310.5926 [astro-ph.CO]}
  \BibitemShut {NoStop}%
\bibitem [{\citenamefont {Dasgupta}\ and\ \citenamefont
  {Kopp}(2014)}]{Dasgupta:2013zpn}%
  \BibitemOpen
  \bibfield  {author} {\bibinfo {author} {\bibfnamefont {B.}~\bibnamefont
  {Dasgupta}}\ and\ \bibinfo {author} {\bibfnamefont {J.}~\bibnamefont
  {Kopp}},\ }\href {\doibase 10.1103/PhysRevLett.112.031803} {\bibfield
  {journal} {\bibinfo  {journal} {Phys. Rev. Lett.}\ }\textbf {\bibinfo
  {volume} {112}},\ \bibinfo {pages} {031803} (\bibinfo {year} {2014})},\
  \Eprint {http://arxiv.org/abs/1310.6337} {arXiv:1310.6337 [hep-ph]}
  \BibitemShut {NoStop}%
\bibitem [{\citenamefont {Mirizzi}\ \emph {et~al.}(2015)\citenamefont
  {Mirizzi}, \citenamefont {Mangano}, \citenamefont {Pisanti},\ and\
  \citenamefont {Saviano}}]{Mirizzi:2014ama}%
  \BibitemOpen
  \bibfield  {author} {\bibinfo {author} {\bibfnamefont {A.}~\bibnamefont
  {Mirizzi}}, \bibinfo {author} {\bibfnamefont {G.}~\bibnamefont {Mangano}},
  \bibinfo {author} {\bibfnamefont {O.}~\bibnamefont {Pisanti}}, \ and\
  \bibinfo {author} {\bibfnamefont {N.}~\bibnamefont {Saviano}},\ }\href
  {\doibase 10.1103/PhysRevD.91.025019} {\bibfield  {journal} {\bibinfo
  {journal} {Phys. Rev. D}\ }\textbf {\bibinfo {volume} {91}},\ \bibinfo
  {pages} {025019} (\bibinfo {year} {2015})},\ \Eprint
  {http://arxiv.org/abs/1410.1385} {arXiv:1410.1385 [hep-ph]} \BibitemShut
  {NoStop}%
\bibitem [{\citenamefont {Dodelson}\ and\ \citenamefont
  {Widrow}(1994)}]{Dodelson:1993je}%
  \BibitemOpen
  \bibfield  {author} {\bibinfo {author} {\bibfnamefont {S.}~\bibnamefont
  {Dodelson}}\ and\ \bibinfo {author} {\bibfnamefont {L.~M.}\ \bibnamefont
  {Widrow}},\ }\href {\doibase 10.1103/PhysRevLett.72.17} {\bibfield  {journal}
  {\bibinfo  {journal} {Phys. Rev. Lett.}\ }\textbf {\bibinfo {volume} {72}},\
  \bibinfo {pages} {17} (\bibinfo {year} {1994})},\ \Eprint
  {http://arxiv.org/abs/hep-ph/9303287} {arXiv:hep-ph/9303287} \BibitemShut
  {NoStop}%
\bibitem [{\citenamefont {Shi}\ and\ \citenamefont
  {Fuller}(1999)}]{Shi:1998km}%
  \BibitemOpen
  \bibfield  {author} {\bibinfo {author} {\bibfnamefont {X.-D.}\ \bibnamefont
  {Shi}}\ and\ \bibinfo {author} {\bibfnamefont {G.~M.}\ \bibnamefont
  {Fuller}},\ }\href {\doibase 10.1103/PhysRevLett.82.2832} {\bibfield
  {journal} {\bibinfo  {journal} {Phys. Rev. Lett.}\ }\textbf {\bibinfo
  {volume} {82}},\ \bibinfo {pages} {2832} (\bibinfo {year} {1999})},\ \Eprint
  {http://arxiv.org/abs/astro-ph/9810076} {arXiv:astro-ph/9810076} \BibitemShut
  {NoStop}%
\bibitem [{\citenamefont {Abazajian}\ \emph
  {et~al.}(2001{\natexlab{a}})\citenamefont {Abazajian}, \citenamefont
  {Fuller},\ and\ \citenamefont {Patel}}]{Abazajian:2001nj}%
  \BibitemOpen
  \bibfield  {author} {\bibinfo {author} {\bibfnamefont {K.}~\bibnamefont
  {Abazajian}}, \bibinfo {author} {\bibfnamefont {G.~M.}\ \bibnamefont
  {Fuller}}, \ and\ \bibinfo {author} {\bibfnamefont {M.}~\bibnamefont
  {Patel}},\ }\href {\doibase 10.1103/PhysRevD.64.023501} {\bibfield  {journal}
  {\bibinfo  {journal} {Phys. Rev. D}\ }\textbf {\bibinfo {volume} {64}},\
  \bibinfo {pages} {023501} (\bibinfo {year} {2001}{\natexlab{a}})},\ \Eprint
  {http://arxiv.org/abs/astro-ph/0101524} {arXiv:astro-ph/0101524} \BibitemShut
  {NoStop}%
\bibitem [{\citenamefont {Dolgov}\ and\ \citenamefont
  {Hansen}(2001)}]{Dolgov:2001nz}%
  \BibitemOpen
  \bibfield  {author} {\bibinfo {author} {\bibfnamefont {A.~D.}\ \bibnamefont
  {Dolgov}}\ and\ \bibinfo {author} {\bibfnamefont {S.~H.}\ \bibnamefont
  {Hansen}},\ }\href@noop {} {\  (\bibinfo {year} {2001})},\ \Eprint
  {http://arxiv.org/abs/hep-ph/0103118} {arXiv:hep-ph/0103118} \BibitemShut
  {NoStop}%
\bibitem [{\citenamefont {Kusenko}(2009)}]{Kusenko:2009up}%
  \BibitemOpen
  \bibfield  {author} {\bibinfo {author} {\bibfnamefont {A.}~\bibnamefont
  {Kusenko}},\ }\href {\doibase 10.1016/j.physrep.2009.07.004} {\bibfield
  {journal} {\bibinfo  {journal} {Phys. Rept.}\ }\textbf {\bibinfo {volume}
  {481}},\ \bibinfo {pages} {1} (\bibinfo {year} {2009})},\ \Eprint
  {http://arxiv.org/abs/0906.2968} {arXiv:0906.2968 [hep-ph]} \BibitemShut
  {NoStop}%
\bibitem [{\citenamefont {Adhikari}\ \emph {et~al.}(2017)\citenamefont
  {Adhikari} \emph {et~al.}}]{Adhikari:2016upu}%
  \BibitemOpen
  \bibfield  {author} {\bibinfo {author} {\bibfnamefont {R.}~\bibnamefont
  {Adhikari}} \emph {et~al.},\ }\href {\doibase 10.1088/1475-7516/2017/01/025}
  {\bibfield  {journal} {\bibinfo  {journal} {JCAP}\ }\textbf {\bibinfo
  {volume} {01}},\ \bibinfo {pages} {025} (\bibinfo {year} {2017})},\ \Eprint
  {http://arxiv.org/abs/1602.04816} {arXiv:1602.04816 [hep-ph]} \BibitemShut
  {NoStop}%
\bibitem [{\citenamefont {Kainulainen}\ \emph {et~al.}(1991)\citenamefont
  {Kainulainen}, \citenamefont {Maalampi},\ and\ \citenamefont
  {Peltoniemi}}]{Kainulainen:1990bn}%
  \BibitemOpen
  \bibfield  {author} {\bibinfo {author} {\bibfnamefont {K.}~\bibnamefont
  {Kainulainen}}, \bibinfo {author} {\bibfnamefont {J.}~\bibnamefont
  {Maalampi}}, \ and\ \bibinfo {author} {\bibfnamefont {J.~T.}\ \bibnamefont
  {Peltoniemi}},\ }\href {\doibase 10.1016/0550-3213(91)90354-Z} {\bibfield
  {journal} {\bibinfo  {journal} {Nucl. Phys. B}\ }\textbf {\bibinfo {volume}
  {358}},\ \bibinfo {pages} {435} (\bibinfo {year} {1991})}\BibitemShut
  {NoStop}%
\bibitem [{\citenamefont {Raffelt}\ and\ \citenamefont
  {Sigl}(1993)}]{Raffelt:1992bs}%
  \BibitemOpen
  \bibfield  {author} {\bibinfo {author} {\bibfnamefont {G.}~\bibnamefont
  {Raffelt}}\ and\ \bibinfo {author} {\bibfnamefont {G.}~\bibnamefont {Sigl}},\
  }\href {\doibase 10.1016/0927-6505(93)90020-E} {\bibfield  {journal}
  {\bibinfo  {journal} {Astropart. Phys.}\ }\textbf {\bibinfo {volume} {1}},\
  \bibinfo {pages} {165} (\bibinfo {year} {1993})},\ \Eprint
  {http://arxiv.org/abs/astro-ph/9209005} {arXiv:astro-ph/9209005} \BibitemShut
  {NoStop}%
\bibitem [{\citenamefont {Shi}\ and\ \citenamefont {Sigl}(1994)}]{Shi:1993ee}%
  \BibitemOpen
  \bibfield  {author} {\bibinfo {author} {\bibfnamefont {X.}~\bibnamefont
  {Shi}}\ and\ \bibinfo {author} {\bibfnamefont {G.}~\bibnamefont {Sigl}},\
  }\href {\doibase 10.1016/0370-2693(94)91232-7} {\bibfield  {journal}
  {\bibinfo  {journal} {Phys. Lett. B}\ }\textbf {\bibinfo {volume} {323}},\
  \bibinfo {pages} {360} (\bibinfo {year} {1994})},\ \bibinfo {note} {[Erratum:
  Phys.Lett.B 324, 516--516 (1994)]},\ \Eprint
  {http://arxiv.org/abs/hep-ph/9312247} {arXiv:hep-ph/9312247} \BibitemShut
  {NoStop}%
\bibitem [{\citenamefont {Nunokawa}\ \emph {et~al.}(1997)\citenamefont
  {Nunokawa}, \citenamefont {Peltoniemi}, \citenamefont {Rossi},\ and\
  \citenamefont {Valle}}]{Nunokawa:1997ct}%
  \BibitemOpen
  \bibfield  {author} {\bibinfo {author} {\bibfnamefont {H.}~\bibnamefont
  {Nunokawa}}, \bibinfo {author} {\bibfnamefont {J.~T.}\ \bibnamefont
  {Peltoniemi}}, \bibinfo {author} {\bibfnamefont {A.}~\bibnamefont {Rossi}}, \
  and\ \bibinfo {author} {\bibfnamefont {J.~W.~F.}\ \bibnamefont {Valle}},\
  }\href {\doibase 10.1103/PhysRevD.56.1704} {\bibfield  {journal} {\bibinfo
  {journal} {Phys. Rev. D}\ }\textbf {\bibinfo {volume} {56}},\ \bibinfo
  {pages} {1704} (\bibinfo {year} {1997})},\ \Eprint
  {http://arxiv.org/abs/hep-ph/9702372} {arXiv:hep-ph/9702372} \BibitemShut
  {NoStop}%
\bibitem [{\citenamefont {Fuller}\ \emph {et~al.}(2003)\citenamefont {Fuller},
  \citenamefont {Kusenko}, \citenamefont {Mocioiu},\ and\ \citenamefont
  {Pascoli}}]{Fuller:2003gy}%
  \BibitemOpen
  \bibfield  {author} {\bibinfo {author} {\bibfnamefont {G.~M.}\ \bibnamefont
  {Fuller}}, \bibinfo {author} {\bibfnamefont {A.}~\bibnamefont {Kusenko}},
  \bibinfo {author} {\bibfnamefont {I.}~\bibnamefont {Mocioiu}}, \ and\
  \bibinfo {author} {\bibfnamefont {S.}~\bibnamefont {Pascoli}},\ }\href
  {\doibase 10.1103/PhysRevD.68.103002} {\bibfield  {journal} {\bibinfo
  {journal} {Phys. Rev. D}\ }\textbf {\bibinfo {volume} {68}},\ \bibinfo
  {pages} {103002} (\bibinfo {year} {2003})},\ \Eprint
  {http://arxiv.org/abs/astro-ph/0307267} {arXiv:astro-ph/0307267} \BibitemShut
  {NoStop}%
\bibitem [{\citenamefont {Hidaka}\ and\ \citenamefont
  {Fuller}(2006)}]{Hidaka:2006sg}%
  \BibitemOpen
  \bibfield  {author} {\bibinfo {author} {\bibfnamefont {J.}~\bibnamefont
  {Hidaka}}\ and\ \bibinfo {author} {\bibfnamefont {G.~M.}\ \bibnamefont
  {Fuller}},\ }\href {\doibase 10.1103/PhysRevD.74.125015} {\bibfield
  {journal} {\bibinfo  {journal} {Phys. Rev. D}\ }\textbf {\bibinfo {volume}
  {74}},\ \bibinfo {pages} {125015} (\bibinfo {year} {2006})},\ \Eprint
  {http://arxiv.org/abs/astro-ph/0609425} {arXiv:astro-ph/0609425} \BibitemShut
  {NoStop}%
\bibitem [{\citenamefont {Hidaka}\ and\ \citenamefont
  {Fuller}(2007)}]{Hidaka:2007se}%
  \BibitemOpen
  \bibfield  {author} {\bibinfo {author} {\bibfnamefont {J.}~\bibnamefont
  {Hidaka}}\ and\ \bibinfo {author} {\bibfnamefont {G.~M.}\ \bibnamefont
  {Fuller}},\ }\href {\doibase 10.1103/PhysRevD.76.083516} {\bibfield
  {journal} {\bibinfo  {journal} {Phys. Rev. D}\ }\textbf {\bibinfo {volume}
  {76}},\ \bibinfo {pages} {083516} (\bibinfo {year} {2007})},\ \Eprint
  {http://arxiv.org/abs/0706.3886} {arXiv:0706.3886 [astro-ph]} \BibitemShut
  {NoStop}%
\bibitem [{\citenamefont {Raffelt}\ and\ \citenamefont
  {Zhou}(2011)}]{Raffelt:2011nc}%
  \BibitemOpen
  \bibfield  {author} {\bibinfo {author} {\bibfnamefont {G.~G.}\ \bibnamefont
  {Raffelt}}\ and\ \bibinfo {author} {\bibfnamefont {S.}~\bibnamefont {Zhou}},\
  }\href {\doibase 10.1103/PhysRevD.83.093014} {\bibfield  {journal} {\bibinfo
  {journal} {Phys. Rev. D}\ }\textbf {\bibinfo {volume} {83}},\ \bibinfo
  {pages} {093014} (\bibinfo {year} {2011})},\ \Eprint
  {http://arxiv.org/abs/1102.5124} {arXiv:1102.5124 [hep-ph]} \BibitemShut
  {NoStop}%
\bibitem [{\citenamefont {Warren}\ \emph {et~al.}(2016)\citenamefont {Warren},
  \citenamefont {Mathews}, \citenamefont {Meixner}, \citenamefont {Hidaka},\
  and\ \citenamefont {Kajino}}]{Warren:2016slz}%
  \BibitemOpen
  \bibfield  {author} {\bibinfo {author} {\bibfnamefont {M.}~\bibnamefont
  {Warren}}, \bibinfo {author} {\bibfnamefont {G.~J.}\ \bibnamefont {Mathews}},
  \bibinfo {author} {\bibfnamefont {M.}~\bibnamefont {Meixner}}, \bibinfo
  {author} {\bibfnamefont {J.}~\bibnamefont {Hidaka}}, \ and\ \bibinfo {author}
  {\bibfnamefont {T.}~\bibnamefont {Kajino}},\ }\href {\doibase
  10.1142/S0217751X16501372} {\bibfield  {journal} {\bibinfo  {journal} {Int.
  J. Mod. Phys. A}\ }\textbf {\bibinfo {volume} {31}},\ \bibinfo {pages}
  {1650137} (\bibinfo {year} {2016})},\ \Eprint
  {http://arxiv.org/abs/1603.05503} {arXiv:1603.05503 [astro-ph.HE]}
  \BibitemShut {NoStop}%
\bibitem [{\citenamefont {Arg\"uelles}\ \emph {et~al.}(2019)\citenamefont
  {Arg\"uelles}, \citenamefont {Brdar},\ and\ \citenamefont
  {Kopp}}]{Arguelles:2016uwb}%
  \BibitemOpen
  \bibfield  {author} {\bibinfo {author} {\bibfnamefont {C.~A.}\ \bibnamefont
  {Arg\"uelles}}, \bibinfo {author} {\bibfnamefont {V.}~\bibnamefont {Brdar}},
  \ and\ \bibinfo {author} {\bibfnamefont {J.}~\bibnamefont {Kopp}},\ }\href
  {\doibase 10.1103/PhysRevD.99.043012} {\bibfield  {journal} {\bibinfo
  {journal} {Phys. Rev. D}\ }\textbf {\bibinfo {volume} {99}},\ \bibinfo
  {pages} {043012} (\bibinfo {year} {2019})},\ \Eprint
  {http://arxiv.org/abs/1605.00654} {arXiv:1605.00654 [hep-ph]} \BibitemShut
  {NoStop}%
\bibitem [{\citenamefont {Suliga}\ \emph {et~al.}(2019)\citenamefont {Suliga},
  \citenamefont {Tamborra},\ and\ \citenamefont {Wu}}]{Suliga:2019bsq}%
  \BibitemOpen
  \bibfield  {author} {\bibinfo {author} {\bibfnamefont {A.~M.}\ \bibnamefont
  {Suliga}}, \bibinfo {author} {\bibfnamefont {I.}~\bibnamefont {Tamborra}}, \
  and\ \bibinfo {author} {\bibfnamefont {M.-R.}\ \bibnamefont {Wu}},\ }\href
  {\doibase 10.1088/1475-7516/2019/12/019} {\bibfield  {journal} {\bibinfo
  {journal} {JCAP}\ }\textbf {\bibinfo {volume} {12}},\ \bibinfo {pages} {019}
  (\bibinfo {year} {2019})},\ \Eprint {http://arxiv.org/abs/1908.11382}
  {arXiv:1908.11382 [astro-ph.HE]} \BibitemShut {NoStop}%
\bibitem [{\citenamefont {Syvolap}\ \emph {et~al.}(2022)\citenamefont
  {Syvolap}, \citenamefont {Ruchayskiy},\ and\ \citenamefont
  {Boyarsky}}]{Syvolap:2019dat}%
  \BibitemOpen
  \bibfield  {author} {\bibinfo {author} {\bibfnamefont {V.}~\bibnamefont
  {Syvolap}}, \bibinfo {author} {\bibfnamefont {O.}~\bibnamefont {Ruchayskiy}},
  \ and\ \bibinfo {author} {\bibfnamefont {A.}~\bibnamefont {Boyarsky}},\
  }\href {\doibase 10.1103/PhysRevD.106.015017} {\bibfield  {journal} {\bibinfo
   {journal} {Phys. Rev. D}\ }\textbf {\bibinfo {volume} {106}},\ \bibinfo
  {pages} {015017} (\bibinfo {year} {2022})},\ \Eprint
  {http://arxiv.org/abs/1909.06320} {arXiv:1909.06320 [hep-ph]} \BibitemShut
  {NoStop}%
\bibitem [{\citenamefont {Suliga}\ \emph {et~al.}(2020)\citenamefont {Suliga},
  \citenamefont {Tamborra},\ and\ \citenamefont {Wu}}]{Suliga:2020vpz}%
  \BibitemOpen
  \bibfield  {author} {\bibinfo {author} {\bibfnamefont {A.~M.}\ \bibnamefont
  {Suliga}}, \bibinfo {author} {\bibfnamefont {I.}~\bibnamefont {Tamborra}}, \
  and\ \bibinfo {author} {\bibfnamefont {M.-R.}\ \bibnamefont {Wu}},\ }\href
  {\doibase 10.1088/1475-7516/2020/08/018} {\bibfield  {journal} {\bibinfo
  {journal} {JCAP}\ }\textbf {\bibinfo {volume} {08}},\ \bibinfo {pages} {018}
  (\bibinfo {year} {2020})},\ \Eprint {http://arxiv.org/abs/2004.11389}
  {arXiv:2004.11389 [astro-ph.HE]} \BibitemShut {NoStop}%
\bibitem [{\citenamefont {Ray}\ and\ \citenamefont {Qian}(2023)}]{Ray:2023gtu}%
  \BibitemOpen
  \bibfield  {author} {\bibinfo {author} {\bibfnamefont {A.}~\bibnamefont
  {Ray}}\ and\ \bibinfo {author} {\bibfnamefont {Y.-Z.}\ \bibnamefont {Qian}},\
  }\href {\doibase 10.1103/PhysRevD.108.063025} {\bibfield  {journal} {\bibinfo
   {journal} {Phys. Rev. D}\ }\textbf {\bibinfo {volume} {108}},\ \bibinfo
  {pages} {063025} (\bibinfo {year} {2023})},\ \Eprint
  {http://arxiv.org/abs/2306.08209} {arXiv:2306.08209 [hep-ph]} \BibitemShut
  {NoStop}%
\bibitem [{\citenamefont {Abazajian}\ \emph
  {et~al.}(2001{\natexlab{b}})\citenamefont {Abazajian}, \citenamefont
  {Fuller},\ and\ \citenamefont {Tucker}}]{Abazajian:2001vt}%
  \BibitemOpen
  \bibfield  {author} {\bibinfo {author} {\bibfnamefont {K.}~\bibnamefont
  {Abazajian}}, \bibinfo {author} {\bibfnamefont {G.~M.}\ \bibnamefont
  {Fuller}}, \ and\ \bibinfo {author} {\bibfnamefont {W.~H.}\ \bibnamefont
  {Tucker}},\ }\href {\doibase 10.1086/323867} {\bibfield  {journal} {\bibinfo
  {journal} {Astrophys. J.}\ }\textbf {\bibinfo {volume} {562}},\ \bibinfo
  {pages} {593} (\bibinfo {year} {2001}{\natexlab{b}})},\ \Eprint
  {http://arxiv.org/abs/astro-ph/0106002} {arXiv:astro-ph/0106002} \BibitemShut
  {NoStop}%
\bibitem [{\citenamefont {Boyarsky}\ \emph {et~al.}(2008)\citenamefont
  {Boyarsky}, \citenamefont {Malyshev}, \citenamefont {Neronov},\ and\
  \citenamefont {Ruchayskiy}}]{Boyarsky:2007ge}%
  \BibitemOpen
  \bibfield  {author} {\bibinfo {author} {\bibfnamefont {A.}~\bibnamefont
  {Boyarsky}}, \bibinfo {author} {\bibfnamefont {D.}~\bibnamefont {Malyshev}},
  \bibinfo {author} {\bibfnamefont {A.}~\bibnamefont {Neronov}}, \ and\
  \bibinfo {author} {\bibfnamefont {O.}~\bibnamefont {Ruchayskiy}},\ }\href
  {\doibase 10.1111/j.1365-2966.2008.13003.x} {\bibfield  {journal} {\bibinfo
  {journal} {Mon. Not. Roy. Astron. Soc.}\ }\textbf {\bibinfo {volume} {387}},\
  \bibinfo {pages} {1345} (\bibinfo {year} {2008})},\ \Eprint
  {http://arxiv.org/abs/0710.4922} {arXiv:0710.4922 [astro-ph]} \BibitemShut
  {NoStop}%
\bibitem [{\citenamefont {Horiuchi}\ \emph {et~al.}(2014)\citenamefont
  {Horiuchi}, \citenamefont {Humphrey}, \citenamefont {Onorbe}, \citenamefont
  {Abazajian}, \citenamefont {Kaplinghat},\ and\ \citenamefont
  {Garrison-Kimmel}}]{Horiuchi:2013noa}%
  \BibitemOpen
  \bibfield  {author} {\bibinfo {author} {\bibfnamefont {S.}~\bibnamefont
  {Horiuchi}}, \bibinfo {author} {\bibfnamefont {P.~J.}\ \bibnamefont
  {Humphrey}}, \bibinfo {author} {\bibfnamefont {J.}~\bibnamefont {Onorbe}},
  \bibinfo {author} {\bibfnamefont {K.~N.}\ \bibnamefont {Abazajian}}, \bibinfo
  {author} {\bibfnamefont {M.}~\bibnamefont {Kaplinghat}}, \ and\ \bibinfo
  {author} {\bibfnamefont {S.}~\bibnamefont {Garrison-Kimmel}},\ }\href
  {\doibase 10.1103/PhysRevD.89.025017} {\bibfield  {journal} {\bibinfo
  {journal} {Phys. Rev. D}\ }\textbf {\bibinfo {volume} {89}},\ \bibinfo
  {pages} {025017} (\bibinfo {year} {2014})},\ \Eprint
  {http://arxiv.org/abs/1311.0282} {arXiv:1311.0282 [astro-ph.CO]} \BibitemShut
  {NoStop}%
\bibitem [{\citenamefont {Roach}\ \emph {et~al.}(2020)\citenamefont {Roach},
  \citenamefont {Ng}, \citenamefont {Perez}, \citenamefont {Beacom},
  \citenamefont {Horiuchi}, \citenamefont {Krivonos},\ and\ \citenamefont
  {Wik}}]{Roach:2019ctw}%
  \BibitemOpen
  \bibfield  {author} {\bibinfo {author} {\bibfnamefont {B.~M.}\ \bibnamefont
  {Roach}}, \bibinfo {author} {\bibfnamefont {K.~C.~Y.}\ \bibnamefont {Ng}},
  \bibinfo {author} {\bibfnamefont {K.}~\bibnamefont {Perez}}, \bibinfo
  {author} {\bibfnamefont {J.~F.}\ \bibnamefont {Beacom}}, \bibinfo {author}
  {\bibfnamefont {S.}~\bibnamefont {Horiuchi}}, \bibinfo {author}
  {\bibfnamefont {R.}~\bibnamefont {Krivonos}}, \ and\ \bibinfo {author}
  {\bibfnamefont {D.~R.}\ \bibnamefont {Wik}},\ }\href {\doibase
  10.1103/PhysRevD.101.103011} {\bibfield  {journal} {\bibinfo  {journal}
  {Phys. Rev. D}\ }\textbf {\bibinfo {volume} {101}},\ \bibinfo {pages}
  {103011} (\bibinfo {year} {2020})},\ \Eprint
  {http://arxiv.org/abs/1908.09037} {arXiv:1908.09037 [astro-ph.HE]}
  \BibitemShut {NoStop}%
\bibitem [{\citenamefont {Ando}\ \emph {et~al.}(2021)\citenamefont {Ando} \emph
  {et~al.}}]{Ando:2021fhj}%
  \BibitemOpen
  \bibfield  {author} {\bibinfo {author} {\bibfnamefont {S.}~\bibnamefont
  {Ando}} \emph {et~al.},\ }\href {\doibase 10.1103/PhysRevD.104.023022}
  {\bibfield  {journal} {\bibinfo  {journal} {Phys. Rev. D}\ }\textbf {\bibinfo
  {volume} {104}},\ \bibinfo {pages} {023022} (\bibinfo {year} {2021})},\
  \Eprint {http://arxiv.org/abs/2103.13242} {arXiv:2103.13242 [astro-ph.HE]}
  \BibitemShut {NoStop}%
\bibitem [{\citenamefont {Foster}\ \emph {et~al.}(2021)\citenamefont {Foster},
  \citenamefont {Kongsore}, \citenamefont {Dessert}, \citenamefont {Park},
  \citenamefont {Rodd}, \citenamefont {Cranmer},\ and\ \citenamefont
  {Safdi}}]{Foster:2021ngm}%
  \BibitemOpen
  \bibfield  {author} {\bibinfo {author} {\bibfnamefont {J.~W.}\ \bibnamefont
  {Foster}}, \bibinfo {author} {\bibfnamefont {M.}~\bibnamefont {Kongsore}},
  \bibinfo {author} {\bibfnamefont {C.}~\bibnamefont {Dessert}}, \bibinfo
  {author} {\bibfnamefont {Y.}~\bibnamefont {Park}}, \bibinfo {author}
  {\bibfnamefont {N.~L.}\ \bibnamefont {Rodd}}, \bibinfo {author}
  {\bibfnamefont {K.}~\bibnamefont {Cranmer}}, \ and\ \bibinfo {author}
  {\bibfnamefont {B.~R.}\ \bibnamefont {Safdi}},\ }\href {\doibase
  10.1103/PhysRevLett.127.051101} {\bibfield  {journal} {\bibinfo  {journal}
  {Phys. Rev. Lett.}\ }\textbf {\bibinfo {volume} {127}},\ \bibinfo {pages}
  {051101} (\bibinfo {year} {2021})},\ \Eprint
  {http://arxiv.org/abs/2102.02207} {arXiv:2102.02207 [astro-ph.CO]}
  \BibitemShut {NoStop}%
\bibitem [{\citenamefont {Malyshev}\ \emph {et~al.}(2020)\citenamefont
  {Malyshev}, \citenamefont {Thorpe-Morgan}, \citenamefont {Santangelo},
  \citenamefont {Jochum},\ and\ \citenamefont {Zhang}}]{Malyshev:2020hcc}%
  \BibitemOpen
  \bibfield  {author} {\bibinfo {author} {\bibfnamefont {D.}~\bibnamefont
  {Malyshev}}, \bibinfo {author} {\bibfnamefont {C.}~\bibnamefont
  {Thorpe-Morgan}}, \bibinfo {author} {\bibfnamefont {A.}~\bibnamefont
  {Santangelo}}, \bibinfo {author} {\bibfnamefont {J.}~\bibnamefont {Jochum}},
  \ and\ \bibinfo {author} {\bibfnamefont {S.-N.}\ \bibnamefont {Zhang}},\
  }\href {\doibase 10.1103/PhysRevD.101.123009} {\bibfield  {journal} {\bibinfo
   {journal} {Phys. Rev. D}\ }\textbf {\bibinfo {volume} {101}},\ \bibinfo
  {pages} {123009} (\bibinfo {year} {2020})},\ \Eprint
  {http://arxiv.org/abs/2001.07014} {arXiv:2001.07014 [astro-ph.HE]}
  \BibitemShut {NoStop}%
\bibitem [{\citenamefont {Roach}\ \emph {et~al.}(2023)\citenamefont {Roach},
  \citenamefont {Rossland}, \citenamefont {Ng}, \citenamefont {Perez},
  \citenamefont {Beacom}, \citenamefont {Grefenstette}, \citenamefont
  {Horiuchi}, \citenamefont {Krivonos},\ and\ \citenamefont
  {Wik}}]{Roach:2022lgo}%
  \BibitemOpen
  \bibfield  {author} {\bibinfo {author} {\bibfnamefont {B.~M.}\ \bibnamefont
  {Roach}}, \bibinfo {author} {\bibfnamefont {S.}~\bibnamefont {Rossland}},
  \bibinfo {author} {\bibfnamefont {K.~C.~Y.}\ \bibnamefont {Ng}}, \bibinfo
  {author} {\bibfnamefont {K.}~\bibnamefont {Perez}}, \bibinfo {author}
  {\bibfnamefont {J.~F.}\ \bibnamefont {Beacom}}, \bibinfo {author}
  {\bibfnamefont {B.~W.}\ \bibnamefont {Grefenstette}}, \bibinfo {author}
  {\bibfnamefont {S.}~\bibnamefont {Horiuchi}}, \bibinfo {author}
  {\bibfnamefont {R.}~\bibnamefont {Krivonos}}, \ and\ \bibinfo {author}
  {\bibfnamefont {D.~R.}\ \bibnamefont {Wik}},\ }\href {\doibase
  10.1103/PhysRevD.107.023009} {\bibfield  {journal} {\bibinfo  {journal}
  {Phys. Rev. D}\ }\textbf {\bibinfo {volume} {107}},\ \bibinfo {pages}
  {023009} (\bibinfo {year} {2023})},\ \Eprint
  {http://arxiv.org/abs/2207.04572} {arXiv:2207.04572 [astro-ph.HE]}
  \BibitemShut {NoStop}%
\bibitem [{\citenamefont {Gerbino}\ \emph {et~al.}(2023)\citenamefont {Gerbino}
  \emph {et~al.}}]{Gerbino:2022nvz}%
  \BibitemOpen
  \bibfield  {author} {\bibinfo {author} {\bibfnamefont {M.}~\bibnamefont
  {Gerbino}} \emph {et~al.},\ }\href {\doibase 10.1016/j.dark.2023.101333}
  {\bibfield  {journal} {\bibinfo  {journal} {Phys. Dark Univ.}\ }\textbf
  {\bibinfo {volume} {42}},\ \bibinfo {pages} {101333} (\bibinfo {year}
  {2023})},\ \Eprint {http://arxiv.org/abs/2203.07377} {arXiv:2203.07377
  [hep-ph]} \BibitemShut {NoStop}%
\bibitem [{\citenamefont {Mertens}\ \emph {et~al.}(2019)\citenamefont {Mertens}
  \emph {et~al.}}]{KATRIN:2018oow}%
  \BibitemOpen
  \bibfield  {author} {\bibinfo {author} {\bibfnamefont {S.}~\bibnamefont
  {Mertens}} \emph {et~al.} (\bibinfo {collaboration} {KATRIN}),\ }\href
  {\doibase 10.1088/1361-6471/ab12fe} {\bibfield  {journal} {\bibinfo
  {journal} {J. Phys. G}\ }\textbf {\bibinfo {volume} {46}},\ \bibinfo {pages}
  {065203} (\bibinfo {year} {2019})},\ \Eprint
  {http://arxiv.org/abs/1810.06711} {arXiv:1810.06711 [physics.ins-det]}
  \BibitemShut {NoStop}%
\bibitem [{\citenamefont {Aker}\ \emph {et~al.}(2023)\citenamefont {Aker} \emph
  {et~al.}}]{KATRIN:2022spi}%
  \BibitemOpen
  \bibfield  {author} {\bibinfo {author} {\bibfnamefont {M.}~\bibnamefont
  {Aker}} \emph {et~al.} (\bibinfo {collaboration} {KATRIN}),\ }\href {\doibase
  10.1140/epjc/s10052-023-11818-y} {\bibfield  {journal} {\bibinfo  {journal}
  {Eur. Phys. J. C}\ }\textbf {\bibinfo {volume} {83}},\ \bibinfo {pages} {763}
  (\bibinfo {year} {2023})},\ \Eprint {http://arxiv.org/abs/2207.06337}
  {arXiv:2207.06337 [nucl-ex]} \BibitemShut {NoStop}%
\bibitem [{\citenamefont {Smith}(2019)}]{Smith:2016vku}%
  \BibitemOpen
  \bibfield  {author} {\bibinfo {author} {\bibfnamefont {P.~F.}\ \bibnamefont
  {Smith}},\ }\href {\doibase 10.1088/1367-2630/ab1502} {\bibfield  {journal}
  {\bibinfo  {journal} {New J. Phys.}\ }\textbf {\bibinfo {volume} {21}},\
  \bibinfo {pages} {053022} (\bibinfo {year} {2019})},\ \Eprint
  {http://arxiv.org/abs/1607.06876} {arXiv:1607.06876 [physics.ins-det]}
  \BibitemShut {NoStop}%
\bibitem [{\citenamefont {Martoff}\ \emph {et~al.}(2021)\citenamefont {Martoff}
  \emph {et~al.}}]{Martoff:2021vxp}%
  \BibitemOpen
  \bibfield  {author} {\bibinfo {author} {\bibfnamefont {C.~J.}\ \bibnamefont
  {Martoff}} \emph {et~al.},\ }\href {\doibase 10.1088/2058-9565/abdb9b}
  {\bibfield  {journal} {\bibinfo  {journal} {Quantum Sci. Technol.}\ }\textbf
  {\bibinfo {volume} {6}},\ \bibinfo {pages} {024008} (\bibinfo {year}
  {2021})}\BibitemShut {NoStop}%
\bibitem [{\citenamefont {Abazajian}\ and\ \citenamefont
  {Escudero}(2023)}]{Abazajian:2023reo}%
  \BibitemOpen
  \bibfield  {author} {\bibinfo {author} {\bibfnamefont {K.~N.}\ \bibnamefont
  {Abazajian}}\ and\ \bibinfo {author} {\bibfnamefont {H.~G.}\ \bibnamefont
  {Escudero}},\ }\href@noop {} {\  (\bibinfo {year} {2023})},\ \Eprint
  {http://arxiv.org/abs/2309.11492} {arXiv:2309.11492 [hep-ph]} \BibitemShut
  {NoStop}%
\bibitem [{\citenamefont {Acero}\ \emph {et~al.}(2022)\citenamefont {Acero}
  \emph {et~al.}}]{Acero:2022wqg}%
  \BibitemOpen
  \bibfield  {author} {\bibinfo {author} {\bibfnamefont {M.~A.}\ \bibnamefont
  {Acero}} \emph {et~al.},\ }\href@noop {} {\  (\bibinfo {year} {2022})},\
  \Eprint {http://arxiv.org/abs/2203.07323} {arXiv:2203.07323 [hep-ex]}
  \BibitemShut {NoStop}%
\bibitem [{\citenamefont {Chu}\ \emph {et~al.}(2015)\citenamefont {Chu},
  \citenamefont {Dasgupta},\ and\ \citenamefont {Kopp}}]{Chu:2015ipa}%
  \BibitemOpen
  \bibfield  {author} {\bibinfo {author} {\bibfnamefont {X.}~\bibnamefont
  {Chu}}, \bibinfo {author} {\bibfnamefont {B.}~\bibnamefont {Dasgupta}}, \
  and\ \bibinfo {author} {\bibfnamefont {J.}~\bibnamefont {Kopp}},\ }\href
  {\doibase 10.1088/1475-7516/2015/10/011} {\bibfield  {journal} {\bibinfo
  {journal} {JCAP}\ }\textbf {\bibinfo {volume} {10}},\ \bibinfo {pages} {011}
  (\bibinfo {year} {2015})},\ \Eprint {http://arxiv.org/abs/1505.02795}
  {arXiv:1505.02795 [hep-ph]} \BibitemShut {NoStop}%
\bibitem [{\citenamefont {Hansen}\ and\ \citenamefont
  {Vogl}(2017)}]{Hansen:2017rxr}%
  \BibitemOpen
  \bibfield  {author} {\bibinfo {author} {\bibfnamefont {R.~S.~L.}\
  \bibnamefont {Hansen}}\ and\ \bibinfo {author} {\bibfnamefont
  {S.}~\bibnamefont {Vogl}},\ }\href {\doibase 10.1103/PhysRevLett.119.251305}
  {\bibfield  {journal} {\bibinfo  {journal} {Phys. Rev. Lett.}\ }\textbf
  {\bibinfo {volume} {119}},\ \bibinfo {pages} {251305} (\bibinfo {year}
  {2017})},\ \Eprint {http://arxiv.org/abs/1706.02707} {arXiv:1706.02707
  [hep-ph]} \BibitemShut {NoStop}%
\bibitem [{\citenamefont {Jeong}\ \emph {et~al.}(2018)\citenamefont {Jeong},
  \citenamefont {Palomares-Ruiz}, \citenamefont {Reno},\ and\ \citenamefont
  {Sarcevic}}]{Jeong:2018yts}%
  \BibitemOpen
  \bibfield  {author} {\bibinfo {author} {\bibfnamefont {Y.~S.}\ \bibnamefont
  {Jeong}}, \bibinfo {author} {\bibfnamefont {S.}~\bibnamefont
  {Palomares-Ruiz}}, \bibinfo {author} {\bibfnamefont {M.~H.}\ \bibnamefont
  {Reno}}, \ and\ \bibinfo {author} {\bibfnamefont {I.}~\bibnamefont
  {Sarcevic}},\ }\href {\doibase 10.1088/1475-7516/2018/06/019} {\bibfield
  {journal} {\bibinfo  {journal} {JCAP}\ }\textbf {\bibinfo {volume} {06}},\
  \bibinfo {pages} {019} (\bibinfo {year} {2018})},\ \Eprint
  {http://arxiv.org/abs/1803.04541} {arXiv:1803.04541 [hep-ph]} \BibitemShut
  {NoStop}%
\bibitem [{\citenamefont {Johns}\ and\ \citenamefont
  {Fuller}(2019)}]{Johns:2019cwc}%
  \BibitemOpen
  \bibfield  {author} {\bibinfo {author} {\bibfnamefont {L.}~\bibnamefont
  {Johns}}\ and\ \bibinfo {author} {\bibfnamefont {G.~M.}\ \bibnamefont
  {Fuller}},\ }\href {\doibase 10.1103/PhysRevD.100.023533} {\bibfield
  {journal} {\bibinfo  {journal} {Phys. Rev. D}\ }\textbf {\bibinfo {volume}
  {100}},\ \bibinfo {pages} {023533} (\bibinfo {year} {2019})},\ \Eprint
  {http://arxiv.org/abs/1903.08296} {arXiv:1903.08296 [hep-ph]} \BibitemShut
  {NoStop}%
\bibitem [{\citenamefont {De~Gouv\^ea}\ \emph {et~al.}(2020)\citenamefont
  {De~Gouv\^ea}, \citenamefont {Sen}, \citenamefont {Tangarife},\ and\
  \citenamefont {Zhang}}]{DeGouvea:2019wpf}%
  \BibitemOpen
  \bibfield  {author} {\bibinfo {author} {\bibfnamefont {A.}~\bibnamefont
  {De~Gouv\^ea}}, \bibinfo {author} {\bibfnamefont {M.}~\bibnamefont {Sen}},
  \bibinfo {author} {\bibfnamefont {W.}~\bibnamefont {Tangarife}}, \ and\
  \bibinfo {author} {\bibfnamefont {Y.}~\bibnamefont {Zhang}},\ }\href
  {\doibase 10.1103/PhysRevLett.124.081802} {\bibfield  {journal} {\bibinfo
  {journal} {Phys. Rev. Lett.}\ }\textbf {\bibinfo {volume} {124}},\ \bibinfo
  {pages} {081802} (\bibinfo {year} {2020})},\ \Eprint
  {http://arxiv.org/abs/1910.04901} {arXiv:1910.04901 [hep-ph]} \BibitemShut
  {NoStop}%
\bibitem [{\citenamefont {Roy~Choudhury}\ \emph {et~al.}(2021)\citenamefont
  {Roy~Choudhury}, \citenamefont {Hannestad},\ and\ \citenamefont
  {Tram}}]{RoyChoudhury:2020dmd}%
  \BibitemOpen
  \bibfield  {author} {\bibinfo {author} {\bibfnamefont {S.}~\bibnamefont
  {Roy~Choudhury}}, \bibinfo {author} {\bibfnamefont {S.}~\bibnamefont
  {Hannestad}}, \ and\ \bibinfo {author} {\bibfnamefont {T.}~\bibnamefont
  {Tram}},\ }\href {\doibase 10.1088/1475-7516/2021/03/084} {\bibfield
  {journal} {\bibinfo  {journal} {JCAP}\ }\textbf {\bibinfo {volume} {03}},\
  \bibinfo {pages} {084} (\bibinfo {year} {2021})},\ \Eprint
  {http://arxiv.org/abs/2012.07519} {arXiv:2012.07519 [astro-ph.CO]}
  \BibitemShut {NoStop}%
\bibitem [{\citenamefont {Kelly}\ \emph {et~al.}(2020)\citenamefont {Kelly},
  \citenamefont {Sen}, \citenamefont {Tangarife},\ and\ \citenamefont
  {Zhang}}]{Kelly:2020pcy}%
  \BibitemOpen
  \bibfield  {author} {\bibinfo {author} {\bibfnamefont {K.~J.}\ \bibnamefont
  {Kelly}}, \bibinfo {author} {\bibfnamefont {M.}~\bibnamefont {Sen}}, \bibinfo
  {author} {\bibfnamefont {W.}~\bibnamefont {Tangarife}}, \ and\ \bibinfo
  {author} {\bibfnamefont {Y.}~\bibnamefont {Zhang}},\ }\href {\doibase
  10.1103/PhysRevD.101.115031} {\bibfield  {journal} {\bibinfo  {journal}
  {Phys. Rev. D}\ }\textbf {\bibinfo {volume} {101}},\ \bibinfo {pages}
  {115031} (\bibinfo {year} {2020})},\ \Eprint
  {http://arxiv.org/abs/2005.03681} {arXiv:2005.03681 [hep-ph]} \BibitemShut
  {NoStop}%
\bibitem [{\citenamefont {Bringmann}\ \emph {et~al.}(2023)\citenamefont
  {Bringmann}, \citenamefont {Depta}, \citenamefont {Hufnagel}, \citenamefont
  {Kersten}, \citenamefont {Ruderman},\ and\ \citenamefont
  {Schmidt-Hoberg}}]{Bringmann:2022aim}%
  \BibitemOpen
  \bibfield  {author} {\bibinfo {author} {\bibfnamefont {T.}~\bibnamefont
  {Bringmann}}, \bibinfo {author} {\bibfnamefont {P.~F.}\ \bibnamefont
  {Depta}}, \bibinfo {author} {\bibfnamefont {M.}~\bibnamefont {Hufnagel}},
  \bibinfo {author} {\bibfnamefont {J.}~\bibnamefont {Kersten}}, \bibinfo
  {author} {\bibfnamefont {J.~T.}\ \bibnamefont {Ruderman}}, \ and\ \bibinfo
  {author} {\bibfnamefont {K.}~\bibnamefont {Schmidt-Hoberg}},\ }\href
  {\doibase 10.1103/PhysRevD.107.L071702} {\bibfield  {journal} {\bibinfo
  {journal} {Phys. Rev. D}\ }\textbf {\bibinfo {volume} {107}},\ \bibinfo
  {pages} {L071702} (\bibinfo {year} {2023})},\ \Eprint
  {http://arxiv.org/abs/2206.10630} {arXiv:2206.10630 [hep-ph]} \BibitemShut
  {NoStop}%
\bibitem [{\citenamefont {Davoudiasl}\ and\ \citenamefont
  {Denton}(2023)}]{Davoudiasl:2023uiq}%
  \BibitemOpen
  \bibfield  {author} {\bibinfo {author} {\bibfnamefont {H.}~\bibnamefont
  {Davoudiasl}}\ and\ \bibinfo {author} {\bibfnamefont {P.~B.}\ \bibnamefont
  {Denton}},\ }\href {\doibase 10.1103/PhysRevD.108.035013} {\bibfield
  {journal} {\bibinfo  {journal} {Phys. Rev. D}\ }\textbf {\bibinfo {volume}
  {108}},\ \bibinfo {pages} {035013} (\bibinfo {year} {2023})},\ \Eprint
  {http://arxiv.org/abs/2301.09651} {arXiv:2301.09651 [hep-ph]} \BibitemShut
  {NoStop}%
\bibitem [{\citenamefont {Astros}\ and\ \citenamefont
  {Vogl}(2023)}]{Astros:2023xhe}%
  \BibitemOpen
  \bibfield  {author} {\bibinfo {author} {\bibfnamefont {M.~D.}\ \bibnamefont
  {Astros}}\ and\ \bibinfo {author} {\bibfnamefont {S.}~\bibnamefont {Vogl}},\
  }\href@noop {} {\  (\bibinfo {year} {2023})},\ \Eprint
  {http://arxiv.org/abs/2307.15565} {arXiv:2307.15565 [hep-ph]} \BibitemShut
  {NoStop}%
\bibitem [{\citenamefont {An}\ \emph {et~al.}(2023)\citenamefont {An},
  \citenamefont {Gluscevic}, \citenamefont {Nadler},\ and\ \citenamefont
  {Zhang}}]{An:2023mkf}%
  \BibitemOpen
  \bibfield  {author} {\bibinfo {author} {\bibfnamefont {R.}~\bibnamefont
  {An}}, \bibinfo {author} {\bibfnamefont {V.}~\bibnamefont {Gluscevic}},
  \bibinfo {author} {\bibfnamefont {E.~O.}\ \bibnamefont {Nadler}}, \ and\
  \bibinfo {author} {\bibfnamefont {Y.}~\bibnamefont {Zhang}},\ }\href
  {\doibase 10.3847/2041-8213/acf049} {\bibfield  {journal} {\bibinfo
  {journal} {Astrophys. J. Lett.}\ }\textbf {\bibinfo {volume} {954}},\
  \bibinfo {pages} {L18} (\bibinfo {year} {2023})},\ \Eprint
  {http://arxiv.org/abs/2301.08299} {arXiv:2301.08299 [astro-ph.CO]}
  \BibitemShut {NoStop}%
\bibitem [{\citenamefont {Spisak}\ \emph {et~al.}()\citenamefont {Spisak},
  \citenamefont {Graf}, \citenamefont {Patwardhan},\ and\ \citenamefont
  {Fuller}}]{Spisak:2023xxx}%
  \BibitemOpen
  \bibfield  {author} {\bibinfo {author} {\bibfnamefont {J.}~\bibnamefont
  {Spisak}}, \bibinfo {author} {\bibfnamefont {L.}~\bibnamefont {Graf}},
  \bibinfo {author} {\bibfnamefont {A.}~\bibnamefont {Patwardhan}}, \ and\
  \bibinfo {author} {\bibfnamefont {G.}~\bibnamefont {Fuller}},\ }\href@noop {}
  {\ }\Eprint {http://arxiv.org/abs/(to be published)} {(to be published)}
  \BibitemShut {NoStop}%
\bibitem [{\citenamefont {Davis~Jr.}(1968)}]{Davis:1968cp}%
  \BibitemOpen
  \bibfield  {author} {\bibinfo {author} {\bibfnamefont {R.}~\bibnamefont
  {Davis~Jr.}},\ }\href {\doibase 10.1103/PhysRevLett.20.1205} {\bibfield
  {journal} {\bibinfo  {journal} {Phys. Rev. Lett.}\ }\textbf {\bibinfo
  {volume} {20}},\ \bibinfo {pages} {1205} (\bibinfo {year} {1968})},\ \Eprint
  {http://arxiv.org/abs/Solar Neutrinos. II: Experimental} {arXiv:Solar
  Neutrinos. II: Experimental [physics.gen-ph]} \BibitemShut {NoStop}%
\bibitem [{\citenamefont {Cleveland}\ \emph {et~al.}(1998)\citenamefont
  {Cleveland}, \citenamefont {Daily}, \citenamefont {Davis}, \citenamefont
  {Distel}, \citenamefont {Lande}, \citenamefont {Lee}, \citenamefont
  {Wildenhain},\ and\ \citenamefont {Ullman}}]{Cleveland:1998nv}%
  \BibitemOpen
  \bibfield  {author} {\bibinfo {author} {\bibfnamefont {B.~T.}\ \bibnamefont
  {Cleveland}}, \bibinfo {author} {\bibfnamefont {T.}~\bibnamefont {Daily}},
  \bibinfo {author} {\bibfnamefont {R.}~\bibnamefont {Davis}, \bibfnamefont
  {Jr.}}, \bibinfo {author} {\bibfnamefont {J.~R.}\ \bibnamefont {Distel}},
  \bibinfo {author} {\bibfnamefont {K.}~\bibnamefont {Lande}}, \bibinfo
  {author} {\bibfnamefont {C.~K.}\ \bibnamefont {Lee}}, \bibinfo {author}
  {\bibfnamefont {P.~S.}\ \bibnamefont {Wildenhain}}, \ and\ \bibinfo {author}
  {\bibfnamefont {J.}~\bibnamefont {Ullman}},\ }\href {\doibase 10.1086/305343}
  {\bibfield  {journal} {\bibinfo  {journal} {Astrophys. J.}\ }\textbf
  {\bibinfo {volume} {496}},\ \bibinfo {pages} {505} (\bibinfo {year}
  {1998})}\BibitemShut {NoStop}%
\bibitem [{\citenamefont {Fukuda}\ \emph {et~al.}(2001)\citenamefont {Fukuda}
  \emph {et~al.}}]{Fukuda:2001nj}%
  \BibitemOpen
  \bibfield  {author} {\bibinfo {author} {\bibfnamefont {S.}~\bibnamefont
  {Fukuda}} \emph {et~al.} (\bibinfo {collaboration} {Super-Kamiokande
  Collaboration}),\ }\href {\doibase 10.1103/PhysRevLett.86.5651} {\bibfield
  {journal} {\bibinfo  {journal} {Phys. Rev. Lett.}\ }\textbf {\bibinfo
  {volume} {86}},\ \bibinfo {pages} {5651} (\bibinfo {year} {2001})},\ \Eprint
  {http://arxiv.org/abs/hep-ex/0103032} {arXiv:hep-ex/0103032 [hep-ex]}
  \BibitemShut {NoStop}%
\bibitem [{\citenamefont {Araki}\ \emph {et~al.}(2005)\citenamefont {Araki}
  \emph {et~al.}}]{Araki:2004mb}%
  \BibitemOpen
  \bibfield  {author} {\bibinfo {author} {\bibfnamefont {T.}~\bibnamefont
  {Araki}} \emph {et~al.} (\bibinfo {collaboration} {KamLAND}),\ }\href
  {\doibase 10.1103/PhysRevLett.94.081801} {\bibfield  {journal} {\bibinfo
  {journal} {Phys. Rev. Lett.}\ }\textbf {\bibinfo {volume} {94}},\ \bibinfo
  {pages} {081801} (\bibinfo {year} {2005})},\ \Eprint
  {http://arxiv.org/abs/hep-ex/0406035} {arXiv:hep-ex/0406035 [hep-ex]}
  \BibitemShut {NoStop}%
\bibitem [{\citenamefont {Bellini}\ \emph {et~al.}(2008)\citenamefont {Bellini}
  \emph {et~al.}}]{Bellini:2008mr}%
  \BibitemOpen
  \bibfield  {author} {\bibinfo {author} {\bibfnamefont {G.}~\bibnamefont
  {Bellini}} \emph {et~al.} (\bibinfo {collaboration} {Borexino}),\ }\href
  {\doibase 10.1103/PhysRevLett.101.091302} {\bibfield  {journal} {\bibinfo
  {journal} {Phys. Rev. Lett.}\ }\textbf {\bibinfo {volume} {101}},\ \bibinfo
  {pages} {091302} (\bibinfo {year} {2008})},\ \Eprint
  {http://arxiv.org/abs/0805.3843} {arXiv:0805.3843 [hep-ex]} \BibitemShut
  {NoStop}%
\bibitem [{\citenamefont {Hirata}\ \emph {et~al.}(1989)\citenamefont {Hirata},
  \citenamefont {Kajita}, \citenamefont {Koshiba}, \citenamefont {Nakahata},
  \citenamefont {Oyama}, \citenamefont {Suzuki}, \citenamefont {Totsuka},
  \citenamefont {Kifune}, \citenamefont {Suda}, \citenamefont {Takahashi},\
  and\ \citenamefont {et~al.}}]{Hirata:1989zj}%
  \BibitemOpen
  \bibfield  {author} {\bibinfo {author} {\bibfnamefont {K.}~\bibnamefont
  {Hirata}}, \bibinfo {author} {\bibfnamefont {T.}~\bibnamefont {Kajita}},
  \bibinfo {author} {\bibfnamefont {M.}~\bibnamefont {Koshiba}}, \bibinfo
  {author} {\bibfnamefont {M.}~\bibnamefont {Nakahata}}, \bibinfo {author}
  {\bibfnamefont {Y.}~\bibnamefont {Oyama}}, \bibinfo {author} {\bibfnamefont
  {A.}~\bibnamefont {Suzuki}}, \bibinfo {author} {\bibfnamefont
  {Y.}~\bibnamefont {Totsuka}}, \bibinfo {author} {\bibfnamefont
  {T.}~\bibnamefont {Kifune}}, \bibinfo {author} {\bibfnamefont
  {T.}~\bibnamefont {Suda}}, \bibinfo {author} {\bibfnamefont {M.}~\bibnamefont
  {Takahashi}}, \ and\ \bibinfo {author} {\bibnamefont {et~al.}},\ }\href
  {\doibase 10.1103/PhysRevLett.63.16} {\bibfield  {journal} {\bibinfo
  {journal} {Phys. Rev. Lett.}\ }\textbf {\bibinfo {volume} {63}},\ \bibinfo
  {pages} {16} (\bibinfo {year} {1989})}\BibitemShut {NoStop}%
\bibitem [{\citenamefont {Bionta}\ \emph {et~al.}(1987)\citenamefont {Bionta}
  \emph {et~al.}}]{Bionta:1987qt}%
  \BibitemOpen
  \bibfield  {author} {\bibinfo {author} {\bibfnamefont {R.~M.}\ \bibnamefont
  {Bionta}} \emph {et~al.},\ }\href {\doibase 10.1103/PhysRevLett.58.1494}
  {\bibfield  {journal} {\bibinfo  {journal} {Phys. Rev. Lett.}\ }\textbf
  {\bibinfo {volume} {58}},\ \bibinfo {pages} {1494} (\bibinfo {year}
  {1987})}\BibitemShut {NoStop}%
\bibitem [{\citenamefont {Alexeyev}\ \emph {et~al.}(1988)\citenamefont
  {Alexeyev}, \citenamefont {Alexeyeva}, \citenamefont {Krivosheina},\ and\
  \citenamefont {Volchenko}}]{ALEXEYEV1988209}%
  \BibitemOpen
  \bibfield  {author} {\bibinfo {author} {\bibfnamefont {E.}~\bibnamefont
  {Alexeyev}}, \bibinfo {author} {\bibfnamefont {L.}~\bibnamefont {Alexeyeva}},
  \bibinfo {author} {\bibfnamefont {I.}~\bibnamefont {Krivosheina}}, \ and\
  \bibinfo {author} {\bibfnamefont {V.}~\bibnamefont {Volchenko}},\ }\href
  {\doibase https://doi.org/10.1016/0370-2693(88)91651-6} {\bibfield  {journal}
  {\bibinfo  {journal} {Physics Letters B}\ }\textbf {\bibinfo {volume}
  {205}},\ \bibinfo {pages} {209} (\bibinfo {year} {1988})}\BibitemShut
  {NoStop}%
\bibitem [{\citenamefont {Collaboration}\ \emph {et~al.}(2018)\citenamefont
  {Collaboration} \emph {et~al.}}]{IceCube:2018dnn}%
  \BibitemOpen
  \bibfield  {author} {\bibinfo {author} {\bibfnamefont {I.}~\bibnamefont
  {Collaboration}} \emph {et~al.} (\bibinfo {collaboration} {IceCube}),\ }\href
  {\doibase 10.1126/science.aat2890} {\bibfield  {journal} {\bibinfo  {journal}
  {Science}\ }\textbf {\bibinfo {volume} {361}},\ \bibinfo {pages} {147}
  (\bibinfo {year} {2018})},\ \Eprint {http://arxiv.org/abs/1807.08794}
  {arXiv:1807.08794 [astro-ph.HE]} \BibitemShut {NoStop}%
\bibitem [{\citenamefont {Abbasi}\ \emph {et~al.}(2022)\citenamefont {Abbasi}
  \emph {et~al.}}]{IceCube:2022der}%
  \BibitemOpen
  \bibfield  {author} {\bibinfo {author} {\bibfnamefont {R.}~\bibnamefont
  {Abbasi}} \emph {et~al.} (\bibinfo {collaboration} {IceCube}),\ }\href
  {\doibase 10.1126/science.abg3395} {\bibfield  {journal} {\bibinfo  {journal}
  {Science}\ }\textbf {\bibinfo {volume} {378}},\ \bibinfo {pages} {538}
  (\bibinfo {year} {2022})},\ \Eprint {http://arxiv.org/abs/2211.09972}
  {arXiv:2211.09972 [astro-ph.HE]} \BibitemShut {NoStop}%
\bibitem [{\citenamefont {Frieman}\ \emph {et~al.}(1988)\citenamefont
  {Frieman}, \citenamefont {Haber},\ and\ \citenamefont
  {Freese}}]{Frieman:1987as}%
  \BibitemOpen
  \bibfield  {author} {\bibinfo {author} {\bibfnamefont {J.~A.}\ \bibnamefont
  {Frieman}}, \bibinfo {author} {\bibfnamefont {H.~E.}\ \bibnamefont {Haber}},
  \ and\ \bibinfo {author} {\bibfnamefont {K.}~\bibnamefont {Freese}},\ }\href
  {\doibase 10.1016/0370-2693(88)91120-3} {\bibfield  {journal} {\bibinfo
  {journal} {Phys. Lett. B}\ }\textbf {\bibinfo {volume} {200}},\ \bibinfo
  {pages} {115} (\bibinfo {year} {1988})}\BibitemShut {NoStop}%
\bibitem [{\citenamefont {Beacom}\ and\ \citenamefont
  {Bell}(2002)}]{Beacom:2002cb}%
  \BibitemOpen
  \bibfield  {author} {\bibinfo {author} {\bibfnamefont {J.~F.}\ \bibnamefont
  {Beacom}}\ and\ \bibinfo {author} {\bibfnamefont {N.~F.}\ \bibnamefont
  {Bell}},\ }\href {\doibase 10.1103/PhysRevD.65.113009} {\bibfield  {journal}
  {\bibinfo  {journal} {Phys. Rev. D}\ }\textbf {\bibinfo {volume} {65}},\
  \bibinfo {pages} {113009} (\bibinfo {year} {2002})},\ \Eprint
  {http://arxiv.org/abs/hep-ph/0204111} {hep-ph/0204111} \BibitemShut {NoStop}%
\bibitem [{\citenamefont {Funcke}\ \emph {et~al.}(2020)\citenamefont {Funcke},
  \citenamefont {Raffelt},\ and\ \citenamefont {Vitagliano}}]{Funcke:2019grs}%
  \BibitemOpen
  \bibfield  {author} {\bibinfo {author} {\bibfnamefont {L.}~\bibnamefont
  {Funcke}}, \bibinfo {author} {\bibfnamefont {G.}~\bibnamefont {Raffelt}}, \
  and\ \bibinfo {author} {\bibfnamefont {E.}~\bibnamefont {Vitagliano}},\
  }\href {\doibase 10.1103/PhysRevD.101.015025} {\bibfield  {journal} {\bibinfo
   {journal} {Phys. Rev. D}\ }\textbf {\bibinfo {volume} {101}},\ \bibinfo
  {pages} {015025} (\bibinfo {year} {2020})},\ \Eprint
  {http://arxiv.org/abs/1905.01264} {arXiv:1905.01264 [hep-ph]} \BibitemShut
  {NoStop}%
\bibitem [{\citenamefont {Iv\'a\~nez Ballesteros}\ and\ \citenamefont
  {Volpe}(2023)}]{Ivanez-Ballesteros:2023lqa}%
  \BibitemOpen
  \bibfield  {author} {\bibinfo {author} {\bibfnamefont {P.}~\bibnamefont
  {Iv\'a\~nez Ballesteros}}\ and\ \bibinfo {author} {\bibfnamefont {M.~C.}\
  \bibnamefont {Volpe}},\ }\href@noop {} {\  (\bibinfo {year} {2023})},\
  \Eprint {http://arxiv.org/abs/2307.03549} {arXiv:2307.03549 [hep-ph]}
  \BibitemShut {NoStop}%
\bibitem [{\citenamefont {Balantekin}\ and\ \citenamefont
  {Volpe}(2005)}]{Balantekin:2004tk}%
  \BibitemOpen
  \bibfield  {author} {\bibinfo {author} {\bibfnamefont {A.~B.}\ \bibnamefont
  {Balantekin}}\ and\ \bibinfo {author} {\bibfnamefont {C.}~\bibnamefont
  {Volpe}},\ }\href {\doibase 10.1103/PhysRevD.72.033008} {\bibfield  {journal}
  {\bibinfo  {journal} {Phys. Rev. D}\ }\textbf {\bibinfo {volume} {72}},\
  \bibinfo {pages} {033008} (\bibinfo {year} {2005})},\ \Eprint
  {http://arxiv.org/abs/hep-ph/0411148} {arXiv:hep-ph/0411148} \BibitemShut
  {NoStop}%
\bibitem [{\citenamefont {Spergel}\ and\ \citenamefont
  {Bahcall}(1988)}]{Spergel:1987ex}%
  \BibitemOpen
  \bibfield  {author} {\bibinfo {author} {\bibfnamefont {D.~N.}\ \bibnamefont
  {Spergel}}\ and\ \bibinfo {author} {\bibfnamefont {J.~N.}\ \bibnamefont
  {Bahcall}},\ }\href {\doibase 10.1016/0370-2693(88)90790-3} {\bibfield
  {journal} {\bibinfo  {journal} {Phys. Lett. B}\ }\textbf {\bibinfo {volume}
  {200}},\ \bibinfo {pages} {366} (\bibinfo {year} {1988})}\BibitemShut
  {NoStop}%
\bibitem [{\citenamefont {Horiuchi}\ \emph {et~al.}(2009)\citenamefont
  {Horiuchi}, \citenamefont {Beacom},\ and\ \citenamefont
  {Dwek}}]{Horiuchi:2008jz}%
  \BibitemOpen
  \bibfield  {author} {\bibinfo {author} {\bibfnamefont {S.}~\bibnamefont
  {Horiuchi}}, \bibinfo {author} {\bibfnamefont {J.~F.}\ \bibnamefont
  {Beacom}}, \ and\ \bibinfo {author} {\bibfnamefont {E.}~\bibnamefont
  {Dwek}},\ }\href {\doibase 10.1103/PhysRevD.79.083013} {\bibfield  {journal}
  {\bibinfo  {journal} {Phys. Rev. D}\ }\textbf {\bibinfo {volume} {79}},\
  \bibinfo {pages} {083013} (\bibinfo {year} {2009})},\ \Eprint
  {http://arxiv.org/abs/0812.3157} {arXiv:0812.3157 [astro-ph]} \BibitemShut
  {NoStop}%
\bibitem [{\citenamefont {Beacom}(2010)}]{Beacom:2010kk}%
  \BibitemOpen
  \bibfield  {author} {\bibinfo {author} {\bibfnamefont {J.~F.}\ \bibnamefont
  {Beacom}},\ }\href {\doibase 10.1146/annurev.nucl.010909.083331} {\bibfield
  {journal} {\bibinfo  {journal} {Ann. Rev. Nucl. Part. Sci.}\ }\textbf
  {\bibinfo {volume} {60}},\ \bibinfo {pages} {439} (\bibinfo {year} {2010})},\
  \Eprint {http://arxiv.org/abs/1004.3311} {arXiv:1004.3311 [astro-ph.HE]}
  \BibitemShut {NoStop}%
\bibitem [{\citenamefont {Lunardini}(2016)}]{Lunardini:2010ab}%
  \BibitemOpen
  \bibfield  {author} {\bibinfo {author} {\bibfnamefont {C.}~\bibnamefont
  {Lunardini}},\ }\href {\doibase 10.1016/j.astropartphys.2016.02.005}
  {\bibfield  {journal} {\bibinfo  {journal} {Astropart. Phys.}\ }\textbf
  {\bibinfo {volume} {79}},\ \bibinfo {pages} {49} (\bibinfo {year} {2016})},\
  \Eprint {http://arxiv.org/abs/1007.3252} {arXiv:1007.3252 [astro-ph.CO]}
  \BibitemShut {NoStop}%
\bibitem [{\citenamefont {Vitagliano}\ \emph {et~al.}(2020)\citenamefont
  {Vitagliano}, \citenamefont {Tamborra},\ and\ \citenamefont
  {Raffelt}}]{Vitagliano:2019yzm}%
  \BibitemOpen
  \bibfield  {author} {\bibinfo {author} {\bibfnamefont {E.}~\bibnamefont
  {Vitagliano}}, \bibinfo {author} {\bibfnamefont {I.}~\bibnamefont
  {Tamborra}}, \ and\ \bibinfo {author} {\bibfnamefont {G.}~\bibnamefont
  {Raffelt}},\ }\href {\doibase 10.1103/RevModPhys.92.045006} {\bibfield
  {journal} {\bibinfo  {journal} {Rev. Mod. Phys.}\ }\textbf {\bibinfo {volume}
  {92}},\ \bibinfo {pages} {45006} (\bibinfo {year} {2020})},\ \Eprint
  {http://arxiv.org/abs/1910.11878} {arXiv:1910.11878 [astro-ph.HE]}
  \BibitemShut {NoStop}%
\bibitem [{\citenamefont {Kresse}\ \emph {et~al.}(2021)\citenamefont {Kresse},
  \citenamefont {Ertl},\ and\ \citenamefont {Janka}}]{Kresse:2020nto}%
  \BibitemOpen
  \bibfield  {author} {\bibinfo {author} {\bibfnamefont {D.}~\bibnamefont
  {Kresse}}, \bibinfo {author} {\bibfnamefont {T.}~\bibnamefont {Ertl}}, \ and\
  \bibinfo {author} {\bibfnamefont {H.-T.}\ \bibnamefont {Janka}},\ }\href
  {\doibase 10.3847/1538-4357/abd54e} {\bibfield  {journal} {\bibinfo
  {journal} {Astrophys. J.}\ }\textbf {\bibinfo {volume} {909}},\ \bibinfo
  {pages} {169} (\bibinfo {year} {2021})},\ \Eprint
  {http://arxiv.org/abs/2010.04728} {arXiv:2010.04728 [astro-ph.HE]}
  \BibitemShut {NoStop}%
\bibitem [{\citenamefont {Horiuchi}\ \emph {et~al.}(2021)\citenamefont
  {Horiuchi}, \citenamefont {Kinugawa}, \citenamefont {Takiwaki}, \citenamefont
  {Takahashi},\ and\ \citenamefont {Kotake}}]{Horiuchi:2020jnc}%
  \BibitemOpen
  \bibfield  {author} {\bibinfo {author} {\bibfnamefont {S.}~\bibnamefont
  {Horiuchi}}, \bibinfo {author} {\bibfnamefont {T.}~\bibnamefont {Kinugawa}},
  \bibinfo {author} {\bibfnamefont {T.}~\bibnamefont {Takiwaki}}, \bibinfo
  {author} {\bibfnamefont {K.}~\bibnamefont {Takahashi}}, \ and\ \bibinfo
  {author} {\bibfnamefont {K.}~\bibnamefont {Kotake}},\ }\href {\doibase
  10.1103/PhysRevD.103.043003} {\bibfield  {journal} {\bibinfo  {journal}
  {Phys. Rev. D}\ }\textbf {\bibinfo {volume} {103}},\ \bibinfo {pages}
  {043003} (\bibinfo {year} {2021})},\ \Eprint
  {http://arxiv.org/abs/2012.08524} {arXiv:2012.08524 [astro-ph.HE]}
  \BibitemShut {NoStop}%
\bibitem [{\citenamefont {Ando}\ \emph {et~al.}(2023)\citenamefont {Ando},
  \citenamefont {Ekanger}, \citenamefont {Horiuchi},\ and\ \citenamefont
  {Koshio}}]{Ando:2023fcc}%
  \BibitemOpen
  \bibfield  {author} {\bibinfo {author} {\bibfnamefont {S.}~\bibnamefont
  {Ando}}, \bibinfo {author} {\bibfnamefont {N.}~\bibnamefont {Ekanger}},
  \bibinfo {author} {\bibfnamefont {S.}~\bibnamefont {Horiuchi}}, \ and\
  \bibinfo {author} {\bibfnamefont {Y.}~\bibnamefont {Koshio}}\ }(\bibinfo
  {year} {2023})\ \Eprint {http://arxiv.org/abs/2306.16076} {arXiv:2306.16076
  [astro-ph.HE]} \BibitemShut {NoStop}%
\bibitem [{\citenamefont {Suliga}(2022)}]{Suliga:2022ica}%
  \BibitemOpen
  \bibfield  {author} {\bibinfo {author} {\bibfnamefont {A.~M.}\ \bibnamefont
  {Suliga}},\ }\enquote {\bibinfo {title} {{Diffuse Supernova Neutrino
  Background}},}\ in\ \href {\doibase 10.1007/978-981-15-8818-1_129-1} {\emph
  {\bibinfo {booktitle} {{Handbook of Nuclear Physics}}}},\ \bibinfo {editor}
  {edited by\ \bibinfo {editor} {\bibfnamefont {I.}~\bibnamefont {Tanihata}},
  \bibinfo {editor} {\bibfnamefont {H.}~\bibnamefont {Toki}}, \ and\ \bibinfo
  {editor} {\bibfnamefont {T.}~\bibnamefont {Kajino}}}\ (\bibinfo {year}
  {2022})\ pp.\ \bibinfo {pages} {1--18},\ \Eprint
  {http://arxiv.org/abs/2207.09632} {arXiv:2207.09632 [astro-ph.HE]}
  \BibitemShut {NoStop}%
\bibitem [{\citenamefont {Goldberg}\ \emph {et~al.}(2006)\citenamefont
  {Goldberg}, \citenamefont {Perez},\ and\ \citenamefont
  {Sarcevic}}]{Goldberg:2005yw}%
  \BibitemOpen
  \bibfield  {author} {\bibinfo {author} {\bibfnamefont {H.}~\bibnamefont
  {Goldberg}}, \bibinfo {author} {\bibfnamefont {G.}~\bibnamefont {Perez}}, \
  and\ \bibinfo {author} {\bibfnamefont {I.}~\bibnamefont {Sarcevic}},\ }\href
  {\doibase 10.1088/1126-6708/2006/11/023} {\bibfield  {journal} {\bibinfo
  {journal} {JHEP}\ }\textbf {\bibinfo {volume} {11}},\ \bibinfo {pages} {023}
  (\bibinfo {year} {2006})},\ \Eprint {http://arxiv.org/abs/hep-ph/0505221}
  {arXiv:hep-ph/0505221} \BibitemShut {NoStop}%
\bibitem [{\citenamefont {Baker}\ \emph {et~al.}(2007)\citenamefont {Baker},
  \citenamefont {Goldberg}, \citenamefont {Perez},\ and\ \citenamefont
  {Sarcevic}}]{Baker:2006gm}%
  \BibitemOpen
  \bibfield  {author} {\bibinfo {author} {\bibfnamefont {J.}~\bibnamefont
  {Baker}}, \bibinfo {author} {\bibfnamefont {H.}~\bibnamefont {Goldberg}},
  \bibinfo {author} {\bibfnamefont {G.}~\bibnamefont {Perez}}, \ and\ \bibinfo
  {author} {\bibfnamefont {I.}~\bibnamefont {Sarcevic}},\ }\href {\doibase
  10.1103/PhysRevD.76.063004} {\bibfield  {journal} {\bibinfo  {journal} {Phys.
  Rev. D}\ }\textbf {\bibinfo {volume} {76}},\ \bibinfo {pages} {063004}
  (\bibinfo {year} {2007})},\ \Eprint {http://arxiv.org/abs/hep-ph/0607281}
  {arXiv:hep-ph/0607281} \BibitemShut {NoStop}%
\bibitem [{\citenamefont {Farzan}\ and\ \citenamefont
  {Palomares-Ruiz}(2014)}]{Farzan:2014gza}%
  \BibitemOpen
  \bibfield  {author} {\bibinfo {author} {\bibfnamefont {Y.}~\bibnamefont
  {Farzan}}\ and\ \bibinfo {author} {\bibfnamefont {S.}~\bibnamefont
  {Palomares-Ruiz}},\ }\href {\doibase 10.1088/1475-7516/2014/06/014}
  {\bibfield  {journal} {\bibinfo  {journal} {JCAP}\ }\textbf {\bibinfo
  {volume} {06}},\ \bibinfo {pages} {014} (\bibinfo {year} {2014})},\ \Eprint
  {http://arxiv.org/abs/1401.7019} {arXiv:1401.7019 [hep-ph]} \BibitemShut
  {NoStop}%
\bibitem [{\citenamefont {Abe}\ \emph {et~al.}(2018)\citenamefont {Abe} \emph
  {et~al.}}]{Hyper-Kamiokande:2018ofw}%
  \BibitemOpen
  \bibfield  {author} {\bibinfo {author} {\bibfnamefont {K.}~\bibnamefont
  {Abe}} \emph {et~al.} (\bibinfo {collaboration} {Hyper-Kamiokande}),\
  }\href@noop {} {\  (\bibinfo {year} {2018})},\ \Eprint
  {http://arxiv.org/abs/1805.04163} {arXiv:1805.04163 [physics.ins-det]}
  \BibitemShut {NoStop}%
\bibitem [{\citenamefont {Hosaka}\ \emph {et~al.}(2006)\citenamefont {Hosaka}
  \emph {et~al.}}]{Super-Kamiokande:2005wtt}%
  \BibitemOpen
  \bibfield  {author} {\bibinfo {author} {\bibfnamefont {J.}~\bibnamefont
  {Hosaka}} \emph {et~al.} (\bibinfo {collaboration} {Super-Kamiokande}),\
  }\href {\doibase 10.1103/PhysRevD.73.112001} {\bibfield  {journal} {\bibinfo
  {journal} {Phys. Rev. D}\ }\textbf {\bibinfo {volume} {73}},\ \bibinfo
  {pages} {112001} (\bibinfo {year} {2006})},\ \Eprint
  {http://arxiv.org/abs/hep-ex/0508053} {arXiv:hep-ex/0508053} \BibitemShut
  {NoStop}%
\bibitem [{\citenamefont {Creque-Sarbinowski}\ \emph
  {et~al.}(2021)\citenamefont {Creque-Sarbinowski}, \citenamefont {Hyde},\ and\
  \citenamefont {Kamionkowski}}]{Creque-Sarbinowski:2020qhz}%
  \BibitemOpen
  \bibfield  {author} {\bibinfo {author} {\bibfnamefont {C.}~\bibnamefont
  {Creque-Sarbinowski}}, \bibinfo {author} {\bibfnamefont {J.}~\bibnamefont
  {Hyde}}, \ and\ \bibinfo {author} {\bibfnamefont {M.}~\bibnamefont
  {Kamionkowski}},\ }\href {\doibase 10.1103/PhysRevD.103.023527} {\bibfield
  {journal} {\bibinfo  {journal} {Phys. Rev. D}\ }\textbf {\bibinfo {volume}
  {103}},\ \bibinfo {pages} {023527} (\bibinfo {year} {2021})},\ \Eprint
  {http://arxiv.org/abs/2005.05332} {arXiv:2005.05332 [hep-ph]} \BibitemShut
  {NoStop}%
\bibitem [{\citenamefont {Randall}\ \emph {et~al.}(2008)\citenamefont
  {Randall}, \citenamefont {Markevitch}, \citenamefont {Clowe}, \citenamefont
  {Gonzalez},\ and\ \citenamefont {Bradac}}]{Randall:2008ppe}%
  \BibitemOpen
  \bibfield  {author} {\bibinfo {author} {\bibfnamefont {S.~W.}\ \bibnamefont
  {Randall}}, \bibinfo {author} {\bibfnamefont {M.}~\bibnamefont {Markevitch}},
  \bibinfo {author} {\bibfnamefont {D.}~\bibnamefont {Clowe}}, \bibinfo
  {author} {\bibfnamefont {A.~H.}\ \bibnamefont {Gonzalez}}, \ and\ \bibinfo
  {author} {\bibfnamefont {M.}~\bibnamefont {Bradac}},\ }\href {\doibase
  10.1086/587859} {\bibfield  {journal} {\bibinfo  {journal} {Astrophys. J.}\
  }\textbf {\bibinfo {volume} {679}},\ \bibinfo {pages} {1173} (\bibinfo {year}
  {2008})},\ \Eprint {http://arxiv.org/abs/0704.0261} {arXiv:0704.0261
  [astro-ph]} \BibitemShut {NoStop}%
\bibitem [{\citenamefont {Tulin}\ and\ \citenamefont
  {Yu}(2018)}]{Tulin:2017ara}%
  \BibitemOpen
  \bibfield  {author} {\bibinfo {author} {\bibfnamefont {S.}~\bibnamefont
  {Tulin}}\ and\ \bibinfo {author} {\bibfnamefont {H.-B.}\ \bibnamefont {Yu}},\
  }\href {\doibase 10.1016/j.physrep.2017.11.004} {\bibfield  {journal}
  {\bibinfo  {journal} {Phys. Rept.}\ }\textbf {\bibinfo {volume} {730}},\
  \bibinfo {pages} {1} (\bibinfo {year} {2018})},\ \Eprint
  {http://arxiv.org/abs/1705.02358} {arXiv:1705.02358 [hep-ph]} \BibitemShut
  {NoStop}%
\bibitem [{\citenamefont {Zyla}\ \emph {et~al.}(2020)\citenamefont {Zyla} \emph
  {et~al.}}]{ParticleDataGroup:2020ssz}%
  \BibitemOpen
  \bibfield  {author} {\bibinfo {author} {\bibfnamefont {P.~A.}\ \bibnamefont
  {Zyla}} \emph {et~al.} (\bibinfo {collaboration} {Particle Data Group}),\
  }\href {\doibase 10.1093/ptep/ptaa104} {\bibfield  {journal} {\bibinfo
  {journal} {PTEP}\ }\textbf {\bibinfo {volume} {2020}},\ \bibinfo {pages}
  {083C01} (\bibinfo {year} {2020})}\BibitemShut {NoStop}%
\bibitem [{\citenamefont {Yuksel}\ \emph {et~al.}(2008)\citenamefont {Yuksel},
  \citenamefont {Kistler}, \citenamefont {Beacom},\ and\ \citenamefont
  {Hopkins}}]{Yuksel:2008cu}%
  \BibitemOpen
  \bibfield  {author} {\bibinfo {author} {\bibfnamefont {H.}~\bibnamefont
  {Yuksel}}, \bibinfo {author} {\bibfnamefont {M.~D.}\ \bibnamefont {Kistler}},
  \bibinfo {author} {\bibfnamefont {J.~F.}\ \bibnamefont {Beacom}}, \ and\
  \bibinfo {author} {\bibfnamefont {A.~M.}\ \bibnamefont {Hopkins}},\ }\href
  {\doibase 10.1086/591449} {\bibfield  {journal} {\bibinfo  {journal}
  {Astrophys. J. Lett.}\ }\textbf {\bibinfo {volume} {683}},\ \bibinfo {pages}
  {L5} (\bibinfo {year} {2008})},\ \Eprint {http://arxiv.org/abs/0804.4008}
  {arXiv:0804.4008 [astro-ph]} \BibitemShut {NoStop}%
\bibitem [{\citenamefont {Salpeter}(1955)}]{Salpeter:1955it}%
  \BibitemOpen
  \bibfield  {author} {\bibinfo {author} {\bibfnamefont {E.~E.}\ \bibnamefont
  {Salpeter}},\ }\href {\doibase 10.1086/145971} {\bibfield  {journal}
  {\bibinfo  {journal} {Astrophys. J.}\ }\textbf {\bibinfo {volume} {121}},\
  \bibinfo {pages} {161} (\bibinfo {year} {1955})}\BibitemShut {NoStop}%
\bibitem [{\citenamefont {Smartt}\ \emph {et~al.}(2009)\citenamefont {Smartt},
  \citenamefont {Eldridge}, \citenamefont {Crockett},\ and\ \citenamefont
  {Maund}}]{Smartt:2008zd}%
  \BibitemOpen
  \bibfield  {author} {\bibinfo {author} {\bibfnamefont {S.~J.}\ \bibnamefont
  {Smartt}}, \bibinfo {author} {\bibfnamefont {J.~J.}\ \bibnamefont
  {Eldridge}}, \bibinfo {author} {\bibfnamefont {R.~M.}\ \bibnamefont
  {Crockett}}, \ and\ \bibinfo {author} {\bibfnamefont {J.~R.}\ \bibnamefont
  {Maund}},\ }\href {\doibase 10.1111/j.1365-2966.2009.14506.x} {\bibfield
  {journal} {\bibinfo  {journal} {Mon. Not. Roy. Astron. Soc.}\ }\textbf
  {\bibinfo {volume} {395}},\ \bibinfo {pages} {1409} (\bibinfo {year}
  {2009})},\ \Eprint {http://arxiv.org/abs/0809.0403} {arXiv:0809.0403
  [astro-ph]} \BibitemShut {NoStop}%
\bibitem [{\citenamefont {Li}\ \emph {et~al.}(2011)\citenamefont {Li} \emph
  {et~al.}}]{Li:2010kc}%
  \BibitemOpen
  \bibfield  {author} {\bibinfo {author} {\bibfnamefont {W.}~\bibnamefont {Li}}
  \emph {et~al.},\ }\href {\doibase 10.1111/j.1365-2966.2011.18160.x}
  {\bibfield  {journal} {\bibinfo  {journal} {Mon. Not. Roy. Astron. Soc.}\
  }\textbf {\bibinfo {volume} {412}},\ \bibinfo {pages} {1441} (\bibinfo {year}
  {2011})},\ \Eprint {http://arxiv.org/abs/1006.4612} {arXiv:1006.4612
  [astro-ph.SR]} \BibitemShut {NoStop}%
\bibitem [{\citenamefont {Horiuchi}\ \emph {et~al.}(2011)\citenamefont
  {Horiuchi}, \citenamefont {Beacom}, \citenamefont {Kochanek}, \citenamefont
  {Prieto}, \citenamefont {Stanek},\ and\ \citenamefont
  {Thompson}}]{Horiuchi:2011zz}%
  \BibitemOpen
  \bibfield  {author} {\bibinfo {author} {\bibfnamefont {S.}~\bibnamefont
  {Horiuchi}}, \bibinfo {author} {\bibfnamefont {J.~F.}\ \bibnamefont
  {Beacom}}, \bibinfo {author} {\bibfnamefont {C.~S.}\ \bibnamefont
  {Kochanek}}, \bibinfo {author} {\bibfnamefont {J.~L.}\ \bibnamefont
  {Prieto}}, \bibinfo {author} {\bibfnamefont {K.~Z.}\ \bibnamefont {Stanek}},
  \ and\ \bibinfo {author} {\bibfnamefont {T.~A.}\ \bibnamefont {Thompson}},\
  }\href {\doibase 10.1088/0004-637X/738/2/154} {\bibfield  {journal} {\bibinfo
   {journal} {Astrophys. J.}\ }\textbf {\bibinfo {volume} {738}},\ \bibinfo
  {pages} {154} (\bibinfo {year} {2011})},\ \Eprint
  {http://arxiv.org/abs/1102.1977} {arXiv:1102.1977 [astro-ph.CO]} \BibitemShut
  {NoStop}%
\bibitem [{\citenamefont {Botticella}\ \emph {et~al.}(2012)\citenamefont
  {Botticella}, \citenamefont {Smartt}, \citenamefont {Kennicutt},
  \citenamefont {Cappellaro}, \citenamefont {Sereno},\ and\ \citenamefont
  {Lee}}]{Botticella:2011nd}%
  \BibitemOpen
  \bibfield  {author} {\bibinfo {author} {\bibfnamefont {M.~T.}\ \bibnamefont
  {Botticella}}, \bibinfo {author} {\bibfnamefont {S.~J.}\ \bibnamefont
  {Smartt}}, \bibinfo {author} {\bibfnamefont {R.~C.}\ \bibnamefont
  {Kennicutt}, \bibfnamefont {Jr.}}, \bibinfo {author} {\bibfnamefont
  {E.}~\bibnamefont {Cappellaro}}, \bibinfo {author} {\bibfnamefont
  {M.}~\bibnamefont {Sereno}}, \ and\ \bibinfo {author} {\bibfnamefont {J.~C.}\
  \bibnamefont {Lee}},\ }\href {\doibase 10.1051/0004-6361/201117343}
  {\bibfield  {journal} {\bibinfo  {journal} {Astron. Astrophys.}\ }\textbf
  {\bibinfo {volume} {537}},\ \bibinfo {pages} {A132} (\bibinfo {year}
  {2012})},\ \Eprint {http://arxiv.org/abs/1111.1692} {arXiv:1111.1692
  [astro-ph.CO]} \BibitemShut {NoStop}%
\bibitem [{\citenamefont {Mattila}\ \emph {et~al.}(2012)\citenamefont {Mattila}
  \emph {et~al.}}]{Mattila:2012}%
  \BibitemOpen
  \bibfield  {author} {\bibinfo {author} {\bibfnamefont {S.}~\bibnamefont
  {Mattila}} \emph {et~al.},\ }\href {\doibase 10.1088/0004-637X/756/2/111}
  {\bibfield  {journal} {\bibinfo  {journal} {Astrophys. J.}\ }\textbf
  {\bibinfo {volume} {756}},\ \bibinfo {pages} {111} (\bibinfo {year}
  {2012})},\ \Eprint {http://arxiv.org/abs/1206.1314} {arXiv:1206.1314
  [astro-ph.CO]} \BibitemShut {NoStop}%
\bibitem [{\citenamefont {Taylor}\ \emph {et~al.}(2014)\citenamefont {Taylor}
  \emph {et~al.}}]{Taylor:2014rlo}%
  \BibitemOpen
  \bibfield  {author} {\bibinfo {author} {\bibfnamefont {M.}~\bibnamefont
  {Taylor}} \emph {et~al.},\ }\href {\doibase 10.1088/0004-637X/792/2/135}
  {\bibfield  {journal} {\bibinfo  {journal} {Astrophys. J.}\ }\textbf
  {\bibinfo {volume} {792}},\ \bibinfo {pages} {135} (\bibinfo {year}
  {2014})},\ \Eprint {http://arxiv.org/abs/1407.0999} {arXiv:1407.0999
  [astro-ph.SR]} \BibitemShut {NoStop}%
\bibitem [{\citenamefont {Strolger}\ \emph {et~al.}(2015)\citenamefont
  {Strolger} \emph {et~al.}}]{Strolger:2015kra}%
  \BibitemOpen
  \bibfield  {author} {\bibinfo {author} {\bibfnamefont {L.-G.}\ \bibnamefont
  {Strolger}} \emph {et~al.},\ }\href {\doibase 10.1088/0004-637X/813/2/93}
  {\bibfield  {journal} {\bibinfo  {journal} {Astrophys. J.}\ }\textbf
  {\bibinfo {volume} {813}},\ \bibinfo {pages} {93} (\bibinfo {year} {2015})},\
  \Eprint {http://arxiv.org/abs/1509.06574} {arXiv:1509.06574 [astro-ph.GA]}
  \BibitemShut {NoStop}%
\bibitem [{\citenamefont {Ivezi\'c}\ \emph {et~al.}(2019)\citenamefont
  {Ivezi\'c} \emph {et~al.}}]{LSST:2008ijt}%
  \BibitemOpen
  \bibfield  {author} {\bibinfo {author} {\bibfnamefont {v.}~\bibnamefont
  {Ivezi\'c}} \emph {et~al.} (\bibinfo {collaboration} {LSST}),\ }\href
  {\doibase 10.3847/1538-4357/ab042c} {\bibfield  {journal} {\bibinfo
  {journal} {Astrophys. J.}\ }\textbf {\bibinfo {volume} {873}},\ \bibinfo
  {pages} {111} (\bibinfo {year} {2019})},\ \Eprint
  {http://arxiv.org/abs/0805.2366} {arXiv:0805.2366 [astro-ph]} \BibitemShut
  {NoStop}%
\bibitem [{\citenamefont {{Lien}}\ and\ \citenamefont
  {{Fields}}(2009)}]{2009JCAP...01..047L}%
  \BibitemOpen
  \bibfield  {author} {\bibinfo {author} {\bibfnamefont {A.}~\bibnamefont
  {{Lien}}}\ and\ \bibinfo {author} {\bibfnamefont {B.~D.}\ \bibnamefont
  {{Fields}}},\ }\href {\doibase 10.1088/1475-7516/2009/01/047} {\bibfield
  {journal} {\bibinfo  {journal} {JCAP}\ }\textbf {\bibinfo {volume} {2009}},\
  \bibinfo {eid} {047} (\bibinfo {year} {2009})},\ \Eprint
  {http://arxiv.org/abs/0902.0979} {arXiv:0902.0979 [astro-ph.CO]} \BibitemShut
  {NoStop}%
\bibitem [{\citenamefont {Lentz}\ \emph {et~al.}(2012)\citenamefont {Lentz},
  \citenamefont {Mezzacappa}, \citenamefont {Bronson~Messer}, \citenamefont
  {Liebendorfer}, \citenamefont {Hix},\ and\ \citenamefont
  {Bruenn}}]{Lentz:2011aa}%
  \BibitemOpen
  \bibfield  {author} {\bibinfo {author} {\bibfnamefont {E.~J.}\ \bibnamefont
  {Lentz}}, \bibinfo {author} {\bibfnamefont {A.}~\bibnamefont {Mezzacappa}},
  \bibinfo {author} {\bibfnamefont {O.~E.}\ \bibnamefont {Bronson~Messer}},
  \bibinfo {author} {\bibfnamefont {M.}~\bibnamefont {Liebendorfer}}, \bibinfo
  {author} {\bibfnamefont {W.~R.}\ \bibnamefont {Hix}}, \ and\ \bibinfo
  {author} {\bibfnamefont {S.~W.}\ \bibnamefont {Bruenn}},\ }\href {\doibase
  10.1088/0004-637X/747/1/73} {\bibfield  {journal} {\bibinfo  {journal}
  {Astrophys. J.}\ }\textbf {\bibinfo {volume} {747}},\ \bibinfo {pages} {73}
  (\bibinfo {year} {2012})},\ \Eprint {http://arxiv.org/abs/1112.3595}
  {arXiv:1112.3595 [astro-ph.SR]} \BibitemShut {NoStop}%
\bibitem [{\citenamefont {Sukhbold}\ \emph {et~al.}(2016)\citenamefont
  {Sukhbold}, \citenamefont {Ertl}, \citenamefont {Woosley}, \citenamefont
  {Brown},\ and\ \citenamefont {Janka}}]{Sukhbold:2015wba}%
  \BibitemOpen
  \bibfield  {author} {\bibinfo {author} {\bibfnamefont {T.}~\bibnamefont
  {Sukhbold}}, \bibinfo {author} {\bibfnamefont {T.}~\bibnamefont {Ertl}},
  \bibinfo {author} {\bibfnamefont {S.~E.}\ \bibnamefont {Woosley}}, \bibinfo
  {author} {\bibfnamefont {J.~M.}\ \bibnamefont {Brown}}, \ and\ \bibinfo
  {author} {\bibfnamefont {H.~T.}\ \bibnamefont {Janka}},\ }\href {\doibase
  10.3847/0004-637X/821/1/38} {\bibfield  {journal} {\bibinfo  {journal}
  {Astrophys. J.}\ }\textbf {\bibinfo {volume} {821}},\ \bibinfo {pages} {38}
  (\bibinfo {year} {2016})},\ \Eprint {http://arxiv.org/abs/1510.04643}
  {arXiv:1510.04643 [astro-ph.HE]} \BibitemShut {NoStop}%
\bibitem [{\citenamefont {Takiwaki}\ \emph {et~al.}(2014)\citenamefont
  {Takiwaki}, \citenamefont {Kotake},\ and\ \citenamefont
  {Suwa}}]{Takiwaki:2013cqa}%
  \BibitemOpen
  \bibfield  {author} {\bibinfo {author} {\bibfnamefont {T.}~\bibnamefont
  {Takiwaki}}, \bibinfo {author} {\bibfnamefont {K.}~\bibnamefont {Kotake}}, \
  and\ \bibinfo {author} {\bibfnamefont {Y.}~\bibnamefont {Suwa}},\ }\href
  {\doibase 10.1088/0004-637X/786/2/83} {\bibfield  {journal} {\bibinfo
  {journal} {Astrophys. J.}\ }\textbf {\bibinfo {volume} {786}},\ \bibinfo
  {pages} {83} (\bibinfo {year} {2014})},\ \Eprint
  {http://arxiv.org/abs/1308.5755} {arXiv:1308.5755 [astro-ph.SR]} \BibitemShut
  {NoStop}%
\bibitem [{\citenamefont {O'Connor}\ \emph {et~al.}(2018)\citenamefont
  {O'Connor} \emph {et~al.}}]{OConnor:2018sti}%
  \BibitemOpen
  \bibfield  {author} {\bibinfo {author} {\bibfnamefont {E.}~\bibnamefont
  {O'Connor}} \emph {et~al.},\ }\href {\doibase 10.1088/1361-6471/aadeae}
  {\bibfield  {journal} {\bibinfo  {journal} {J. Phys. G}\ }\textbf {\bibinfo
  {volume} {45}},\ \bibinfo {pages} {104001} (\bibinfo {year} {2018})},\
  \Eprint {http://arxiv.org/abs/1806.04175} {arXiv:1806.04175 [astro-ph.HE]}
  \BibitemShut {NoStop}%
\bibitem [{\citenamefont {Burrows}\ \emph {et~al.}(2020)\citenamefont
  {Burrows}, \citenamefont {Radice}, \citenamefont {Vartanyan}, \citenamefont
  {Nagakura}, \citenamefont {Skinner},\ and\ \citenamefont
  {Dolence}}]{Burrows:2019zce}%
  \BibitemOpen
  \bibfield  {author} {\bibinfo {author} {\bibfnamefont {A.}~\bibnamefont
  {Burrows}}, \bibinfo {author} {\bibfnamefont {D.}~\bibnamefont {Radice}},
  \bibinfo {author} {\bibfnamefont {D.}~\bibnamefont {Vartanyan}}, \bibinfo
  {author} {\bibfnamefont {H.}~\bibnamefont {Nagakura}}, \bibinfo {author}
  {\bibfnamefont {M.~A.}\ \bibnamefont {Skinner}}, \ and\ \bibinfo {author}
  {\bibfnamefont {J.}~\bibnamefont {Dolence}},\ }\href {\doibase
  10.1093/mnras/stz3223} {\bibfield  {journal} {\bibinfo  {journal} {Mon. Not.
  Roy. Astron. Soc.}\ }\textbf {\bibinfo {volume} {491}},\ \bibinfo {pages}
  {2715} (\bibinfo {year} {2020})},\ \Eprint {http://arxiv.org/abs/1909.04152}
  {arXiv:1909.04152 [astro-ph.HE]} \BibitemShut {NoStop}%
\bibitem [{\citenamefont {Duan}\ \emph {et~al.}(2010)\citenamefont {Duan},
  \citenamefont {Fuller},\ and\ \citenamefont {Qian}}]{Duan:2010bg}%
  \BibitemOpen
  \bibfield  {author} {\bibinfo {author} {\bibfnamefont {H.}~\bibnamefont
  {Duan}}, \bibinfo {author} {\bibfnamefont {G.~M.}\ \bibnamefont {Fuller}}, \
  and\ \bibinfo {author} {\bibfnamefont {Y.-Z.}\ \bibnamefont {Qian}},\ }\href
  {\doibase 10.1146/annurev.nucl.012809.104524} {\bibfield  {journal} {\bibinfo
   {journal} {Ann. Rev. Nucl. Part. Sci.}\ }\textbf {\bibinfo {volume} {60}},\
  \bibinfo {pages} {569} (\bibinfo {year} {2010})},\ \Eprint
  {http://arxiv.org/abs/1001.2799} {arXiv:1001.2799 [hep-ph]} \BibitemShut
  {NoStop}%
\bibitem [{\citenamefont {Chakraborty}\ \emph {et~al.}(2016)\citenamefont
  {Chakraborty}, \citenamefont {Hansen}, \citenamefont {Izaguirre},\ and\
  \citenamefont {Raffelt}}]{Chakraborty:2016yeg}%
  \BibitemOpen
  \bibfield  {author} {\bibinfo {author} {\bibfnamefont {S.}~\bibnamefont
  {Chakraborty}}, \bibinfo {author} {\bibfnamefont {R.}~\bibnamefont {Hansen}},
  \bibinfo {author} {\bibfnamefont {I.}~\bibnamefont {Izaguirre}}, \ and\
  \bibinfo {author} {\bibfnamefont {G.}~\bibnamefont {Raffelt}},\ }\href
  {\doibase 10.1016/j.nuclphysb.2016.02.012} {\bibfield  {journal} {\bibinfo
  {journal} {Nucl. Phys. B}\ }\textbf {\bibinfo {volume} {908}},\ \bibinfo
  {pages} {366} (\bibinfo {year} {2016})},\ \Eprint
  {http://arxiv.org/abs/1602.02766} {arXiv:1602.02766 [hep-ph]} \BibitemShut
  {NoStop}%
\bibitem [{\citenamefont {Tamborra}\ and\ \citenamefont
  {Shalgar}(2021)}]{Tamborra:2020cul}%
  \BibitemOpen
  \bibfield  {author} {\bibinfo {author} {\bibfnamefont {I.}~\bibnamefont
  {Tamborra}}\ and\ \bibinfo {author} {\bibfnamefont {S.}~\bibnamefont
  {Shalgar}},\ }\href {\doibase 10.1146/annurev-nucl-102920-050505} {\bibfield
  {journal} {\bibinfo  {journal} {Ann. Rev. Nucl. Part. Sci.}\ }\textbf
  {\bibinfo {volume} {71}},\ \bibinfo {pages} {165} (\bibinfo {year} {2021})},\
  \Eprint {http://arxiv.org/abs/2011.01948} {arXiv:2011.01948 [astro-ph.HE]}
  \BibitemShut {NoStop}%
\bibitem [{\citenamefont {Patwardhan}\ \emph {et~al.}(2023)\citenamefont
  {Patwardhan}, \citenamefont {Cervia}, \citenamefont {Rrapaj}, \citenamefont
  {Siwach},\ and\ \citenamefont {Balantekin}}]{Patwardhan:2022mxg}%
  \BibitemOpen
  \bibfield  {author} {\bibinfo {author} {\bibfnamefont {A.~V.}\ \bibnamefont
  {Patwardhan}}, \bibinfo {author} {\bibfnamefont {M.~J.}\ \bibnamefont
  {Cervia}}, \bibinfo {author} {\bibfnamefont {E.}~\bibnamefont {Rrapaj}},
  \bibinfo {author} {\bibfnamefont {P.}~\bibnamefont {Siwach}}, \ and\ \bibinfo
  {author} {\bibfnamefont {A.~B.}\ \bibnamefont {Balantekin}},\ }\enquote
  {\bibinfo {title} {{Many-Body Collective Neutrino Oscillations: Recent
  Developments}},}\ in\ \href {\doibase 10.1007/978-981-15-8818-1_126-1} {\emph
  {\bibinfo {booktitle} {{Handbook of Nuclear Physics}}}},\ \bibinfo {editor}
  {edited by\ \bibinfo {editor} {\bibfnamefont {I.}~\bibnamefont {Tanihata}},
  \bibinfo {editor} {\bibfnamefont {H.}~\bibnamefont {Toki}}, \ and\ \bibinfo
  {editor} {\bibfnamefont {T.}~\bibnamefont {Kajino}}}\ (\bibinfo {year}
  {2023})\ pp.\ \bibinfo {pages} {1--16},\ \Eprint
  {http://arxiv.org/abs/2301.00342} {arXiv:2301.00342 [hep-ph]} \BibitemShut
  {NoStop}%
\bibitem [{\citenamefont {Volpe}(2023)}]{Volpe:2023met}%
  \BibitemOpen
  \bibfield  {author} {\bibinfo {author} {\bibfnamefont {M.~C.}\ \bibnamefont
  {Volpe}},\ }\href@noop {} {\  (\bibinfo {year} {2023})},\ \Eprint
  {http://arxiv.org/abs/2301.11814} {arXiv:2301.11814 [hep-ph]} \BibitemShut
  {NoStop}%
\bibitem [{\citenamefont {Abe}\ \emph {et~al.}(2021)\citenamefont {Abe} \emph
  {et~al.}}]{Super-Kamiokande:2021jaq}%
  \BibitemOpen
  \bibfield  {author} {\bibinfo {author} {\bibfnamefont {K.}~\bibnamefont
  {Abe}} \emph {et~al.} (\bibinfo {collaboration} {Super-Kamiokande}),\ }\href
  {\doibase 10.1103/PhysRevD.104.122002} {\bibfield  {journal} {\bibinfo
  {journal} {Phys. Rev. D}\ }\textbf {\bibinfo {volume} {104}},\ \bibinfo
  {pages} {122002} (\bibinfo {year} {2021})},\ \Eprint
  {http://arxiv.org/abs/2109.11174} {arXiv:2109.11174 [astro-ph.HE]}
  \BibitemShut {NoStop}%
\bibitem [{\citenamefont {Harada}\ \emph {et~al.}(2023)\citenamefont {Harada}
  \emph {et~al.}}]{Super-Kamiokande:2023xup}%
  \BibitemOpen
  \bibfield  {author} {\bibinfo {author} {\bibfnamefont {M.}~\bibnamefont
  {Harada}} \emph {et~al.} (\bibinfo {collaboration} {Super-Kamiokande}),\
  }\href {\doibase 10.3847/2041-8213/acdc9e} {\bibfield  {journal} {\bibinfo
  {journal} {Astrophys. J. Lett.}\ }\textbf {\bibinfo {volume} {951}},\
  \bibinfo {pages} {L27} (\bibinfo {year} {2023})},\ \Eprint
  {http://arxiv.org/abs/2305.05135} {arXiv:2305.05135 [astro-ph.HE]}
  \BibitemShut {NoStop}%
\bibitem [{\citenamefont {Aharmim}\ \emph {et~al.}(2020)\citenamefont {Aharmim}
  \emph {et~al.}}]{SNO:2020gqd}%
  \BibitemOpen
  \bibfield  {author} {\bibinfo {author} {\bibfnamefont {B.}~\bibnamefont
  {Aharmim}} \emph {et~al.} (\bibinfo {collaboration} {SNO}),\ }\href {\doibase
  10.1103/PhysRevD.102.062006} {\bibfield  {journal} {\bibinfo  {journal}
  {Phys. Rev. D}\ }\textbf {\bibinfo {volume} {102}},\ \bibinfo {pages}
  {062006} (\bibinfo {year} {2020})},\ \Eprint
  {http://arxiv.org/abs/2007.08018} {arXiv:2007.08018 [hep-ex]} \BibitemShut
  {NoStop}%
\bibitem [{\citenamefont {Lunardini}\ and\ \citenamefont
  {Peres}(2008)}]{Lunardini:2008xd}%
  \BibitemOpen
  \bibfield  {author} {\bibinfo {author} {\bibfnamefont {C.}~\bibnamefont
  {Lunardini}}\ and\ \bibinfo {author} {\bibfnamefont {O.~L.~G.}\ \bibnamefont
  {Peres}},\ }\href {\doibase 10.1088/1475-7516/2008/08/033} {\bibfield
  {journal} {\bibinfo  {journal} {JCAP}\ }\textbf {\bibinfo {volume} {08}},\
  \bibinfo {pages} {033} (\bibinfo {year} {2008})},\ \Eprint
  {http://arxiv.org/abs/0805.4225} {arXiv:0805.4225 [astro-ph]} \BibitemShut
  {NoStop}%
\bibitem [{\citenamefont {Suliga}\ \emph {et~al.}(2022)\citenamefont {Suliga},
  \citenamefont {Beacom},\ and\ \citenamefont {Tamborra}}]{Suliga:2021hek}%
  \BibitemOpen
  \bibfield  {author} {\bibinfo {author} {\bibfnamefont {A.~M.}\ \bibnamefont
  {Suliga}}, \bibinfo {author} {\bibfnamefont {J.~F.}\ \bibnamefont {Beacom}},
  \ and\ \bibinfo {author} {\bibfnamefont {I.}~\bibnamefont {Tamborra}},\
  }\href {\doibase 10.1103/PhysRevD.105.043008} {\bibfield  {journal} {\bibinfo
   {journal} {Phys. Rev. D}\ }\textbf {\bibinfo {volume} {105}},\ \bibinfo
  {pages} {043008} (\bibinfo {year} {2022})},\ \Eprint
  {http://arxiv.org/abs/2112.09168} {arXiv:2112.09168 [astro-ph.HE]}
  \BibitemShut {NoStop}%
\bibitem [{\citenamefont {Beacom}\ and\ \citenamefont
  {Vagins}(2004)}]{Beacom:2003nk}%
  \BibitemOpen
  \bibfield  {author} {\bibinfo {author} {\bibfnamefont {J.~F.}\ \bibnamefont
  {Beacom}}\ and\ \bibinfo {author} {\bibfnamefont {M.~R.}\ \bibnamefont
  {Vagins}},\ }\href {\doibase 10.1103/PhysRevLett.93.171101} {\bibfield
  {journal} {\bibinfo  {journal} {Phys. Rev. Lett.}\ }\textbf {\bibinfo
  {volume} {93}},\ \bibinfo {pages} {171101} (\bibinfo {year} {2004})},\
  \Eprint {http://arxiv.org/abs/hep-ph/0309300} {arXiv:hep-ph/0309300}
  \BibitemShut {NoStop}%
\bibitem [{\citenamefont {Abusleme}\ \emph {et~al.}(2022)\citenamefont
  {Abusleme} \emph {et~al.}}]{JUNO:2022lpc}%
  \BibitemOpen
  \bibfield  {author} {\bibinfo {author} {\bibfnamefont {A.}~\bibnamefont
  {Abusleme}} \emph {et~al.} (\bibinfo {collaboration} {JUNO}),\ }\href
  {\doibase 10.1088/1475-7516/2022/10/033} {\bibfield  {journal} {\bibinfo
  {journal} {JCAP}\ }\textbf {\bibinfo {volume} {10}},\ \bibinfo {pages} {033}
  (\bibinfo {year} {2022})},\ \Eprint {http://arxiv.org/abs/2205.08830}
  {arXiv:2205.08830 [hep-ex]} \BibitemShut {NoStop}%
\bibitem [{\citenamefont {Zhu}\ \emph {et~al.}(2019)\citenamefont {Zhu},
  \citenamefont {Li},\ and\ \citenamefont {Beacom}}]{Zhu:2018rwc}%
  \BibitemOpen
  \bibfield  {author} {\bibinfo {author} {\bibfnamefont {G.}~\bibnamefont
  {Zhu}}, \bibinfo {author} {\bibfnamefont {S.~W.}\ \bibnamefont {Li}}, \ and\
  \bibinfo {author} {\bibfnamefont {J.~F.}\ \bibnamefont {Beacom}},\ }\href
  {\doibase 10.1103/PhysRevC.99.055810} {\bibfield  {journal} {\bibinfo
  {journal} {Phys. Rev. C}\ }\textbf {\bibinfo {volume} {99}},\ \bibinfo
  {pages} {055810} (\bibinfo {year} {2019})},\ \Eprint
  {http://arxiv.org/abs/1811.07912} {arXiv:1811.07912 [hep-ph]} \BibitemShut
  {NoStop}%
\bibitem [{\citenamefont {M\o{}ller}\ \emph {et~al.}(2018)\citenamefont
  {M\o{}ller}, \citenamefont {Suliga}, \citenamefont {Tamborra},\ and\
  \citenamefont {Denton}}]{Moller:2018kpn}%
  \BibitemOpen
  \bibfield  {author} {\bibinfo {author} {\bibfnamefont {K.}~\bibnamefont
  {M\o{}ller}}, \bibinfo {author} {\bibfnamefont {A.~M.}\ \bibnamefont
  {Suliga}}, \bibinfo {author} {\bibfnamefont {I.}~\bibnamefont {Tamborra}}, \
  and\ \bibinfo {author} {\bibfnamefont {P.~B.}\ \bibnamefont {Denton}},\
  }\href {\doibase 10.1088/1475-7516/2018/05/066} {\bibfield  {journal}
  {\bibinfo  {journal} {JCAP}\ }\textbf {\bibinfo {volume} {05}},\ \bibinfo
  {pages} {066} (\bibinfo {year} {2018})},\ \Eprint
  {http://arxiv.org/abs/1804.03157} {arXiv:1804.03157 [astro-ph.HE]}
  \BibitemShut {NoStop}%
\bibitem [{\citenamefont {Zhuang}\ \emph {et~al.}(2023)\citenamefont {Zhuang},
  \citenamefont {Strigari}, \citenamefont {Jin},\ and\ \citenamefont
  {Sinha}}]{Zhuang:2023dzd}%
  \BibitemOpen
  \bibfield  {author} {\bibinfo {author} {\bibfnamefont {Y.}~\bibnamefont
  {Zhuang}}, \bibinfo {author} {\bibfnamefont {L.~E.}\ \bibnamefont
  {Strigari}}, \bibinfo {author} {\bibfnamefont {L.}~\bibnamefont {Jin}}, \
  and\ \bibinfo {author} {\bibfnamefont {S.}~\bibnamefont {Sinha}},\
  }\href@noop {} {\  (\bibinfo {year} {2023})},\ \Eprint
  {http://arxiv.org/abs/2307.13792} {arXiv:2307.13792 [hep-ph]} \BibitemShut
  {NoStop}%
\bibitem [{\citenamefont {Strumia}\ and\ \citenamefont
  {Vissani}(2003)}]{Strumia:2003zx}%
  \BibitemOpen
  \bibfield  {author} {\bibinfo {author} {\bibfnamefont {A.}~\bibnamefont
  {Strumia}}\ and\ \bibinfo {author} {\bibfnamefont {F.}~\bibnamefont
  {Vissani}},\ }\href {\doibase 10.1016/S0370-2693(03)00616-6} {\bibfield
  {journal} {\bibinfo  {journal} {Phys. Lett. B}\ }\textbf {\bibinfo {volume}
  {564}},\ \bibinfo {pages} {42} (\bibinfo {year} {2003})},\ \Eprint
  {http://arxiv.org/abs/astro-ph/0302055} {arXiv:astro-ph/0302055} \BibitemShut
  {NoStop}%
\bibitem [{\citenamefont {Ricciardi}\ \emph {et~al.}(2022)\citenamefont
  {Ricciardi}, \citenamefont {Vignaroli},\ and\ \citenamefont
  {Vissani}}]{Ricciardi:2022pru}%
  \BibitemOpen
  \bibfield  {author} {\bibinfo {author} {\bibfnamefont {G.}~\bibnamefont
  {Ricciardi}}, \bibinfo {author} {\bibfnamefont {N.}~\bibnamefont
  {Vignaroli}}, \ and\ \bibinfo {author} {\bibfnamefont {F.}~\bibnamefont
  {Vissani}},\ }\href {\doibase 10.1007/JHEP08(2022)212} {\bibfield  {journal}
  {\bibinfo  {journal} {JHEP}\ }\textbf {\bibinfo {volume} {08}},\ \bibinfo
  {pages} {212} (\bibinfo {year} {2022})},\ \Eprint
  {http://arxiv.org/abs/2206.05567} {arXiv:2206.05567 [hep-ph]} \BibitemShut
  {NoStop}%
\bibitem [{\citenamefont {Bulbul}\ \emph {et~al.}(2014)\citenamefont {Bulbul},
  \citenamefont {Markevitch}, \citenamefont {Foster}, \citenamefont {Smith},
  \citenamefont {Loewenstein},\ and\ \citenamefont {Randall}}]{Bulbul:2014sua}%
  \BibitemOpen
  \bibfield  {author} {\bibinfo {author} {\bibfnamefont {E.}~\bibnamefont
  {Bulbul}}, \bibinfo {author} {\bibfnamefont {M.}~\bibnamefont {Markevitch}},
  \bibinfo {author} {\bibfnamefont {A.}~\bibnamefont {Foster}}, \bibinfo
  {author} {\bibfnamefont {R.~K.}\ \bibnamefont {Smith}}, \bibinfo {author}
  {\bibfnamefont {M.}~\bibnamefont {Loewenstein}}, \ and\ \bibinfo {author}
  {\bibfnamefont {S.~W.}\ \bibnamefont {Randall}},\ }\href {\doibase
  10.1088/0004-637X/789/1/13} {\bibfield  {journal} {\bibinfo  {journal}
  {Astrophys. J.}\ }\textbf {\bibinfo {volume} {789}},\ \bibinfo {pages} {13}
  (\bibinfo {year} {2014})},\ \Eprint {http://arxiv.org/abs/1402.2301}
  {arXiv:1402.2301 [astro-ph.CO]} \BibitemShut {NoStop}%
\bibitem [{\citenamefont {Boyarsky}\ \emph {et~al.}(2014)\citenamefont
  {Boyarsky}, \citenamefont {Ruchayskiy}, \citenamefont {Iakubovskyi},\ and\
  \citenamefont {Franse}}]{Boyarsky:2014jta}%
  \BibitemOpen
  \bibfield  {author} {\bibinfo {author} {\bibfnamefont {A.}~\bibnamefont
  {Boyarsky}}, \bibinfo {author} {\bibfnamefont {O.}~\bibnamefont
  {Ruchayskiy}}, \bibinfo {author} {\bibfnamefont {D.}~\bibnamefont
  {Iakubovskyi}}, \ and\ \bibinfo {author} {\bibfnamefont {J.}~\bibnamefont
  {Franse}},\ }\href {\doibase 10.1103/PhysRevLett.113.251301} {\bibfield
  {journal} {\bibinfo  {journal} {Phys. Rev. Lett.}\ }\textbf {\bibinfo
  {volume} {113}},\ \bibinfo {pages} {251301} (\bibinfo {year} {2014})},\
  \Eprint {http://arxiv.org/abs/1402.4119} {arXiv:1402.4119 [astro-ph.CO]}
  \BibitemShut {NoStop}%
\bibitem [{\citenamefont {Spergel}\ and\ \citenamefont
  {Steinhardt}(2000)}]{Spergel:1999mh}%
  \BibitemOpen
  \bibfield  {author} {\bibinfo {author} {\bibfnamefont {D.~N.}\ \bibnamefont
  {Spergel}}\ and\ \bibinfo {author} {\bibfnamefont {P.~J.}\ \bibnamefont
  {Steinhardt}},\ }\href {\doibase 10.1103/PhysRevLett.84.3760} {\bibfield
  {journal} {\bibinfo  {journal} {Phys. Rev. Lett.}\ }\textbf {\bibinfo
  {volume} {84}},\ \bibinfo {pages} {3760} (\bibinfo {year} {2000})},\ \Eprint
  {http://arxiv.org/abs/astro-ph/9909386} {arXiv:astro-ph/9909386} \BibitemShut
  {NoStop}%
\bibitem [{\citenamefont {Aker}\ \emph
  {et~al.}(2021{\natexlab{a}})\citenamefont {Aker} \emph
  {et~al.}}]{KATRIN:2021dfa}%
  \BibitemOpen
  \bibfield  {author} {\bibinfo {author} {\bibfnamefont {M.}~\bibnamefont
  {Aker}} \emph {et~al.} (\bibinfo {collaboration} {KATRIN}),\ }\href {\doibase
  10.1088/1748-0221/16/08/T08015} {\bibfield  {journal} {\bibinfo  {journal}
  {JINST}\ }\textbf {\bibinfo {volume} {16}},\ \bibinfo {pages} {T08015}
  (\bibinfo {year} {2021}{\natexlab{a}})},\ \Eprint
  {http://arxiv.org/abs/2103.04755} {arXiv:2103.04755 [physics.ins-det]}
  \BibitemShut {NoStop}%
\bibitem [{\citenamefont {Aker}\ \emph {et~al.}(2019)\citenamefont {Aker} \emph
  {et~al.}}]{KATRIN:2019yun}%
  \BibitemOpen
  \bibfield  {author} {\bibinfo {author} {\bibfnamefont {M.}~\bibnamefont
  {Aker}} \emph {et~al.} (\bibinfo {collaboration} {KATRIN}),\ }\href {\doibase
  10.1103/PhysRevLett.123.221802} {\bibfield  {journal} {\bibinfo  {journal}
  {Phys. Rev. Lett.}\ }\textbf {\bibinfo {volume} {123}},\ \bibinfo {pages}
  {221802} (\bibinfo {year} {2019})},\ \Eprint
  {http://arxiv.org/abs/1909.06048} {arXiv:1909.06048 [hep-ex]} \BibitemShut
  {NoStop}%
\bibitem [{\citenamefont {Aker}\ \emph
  {et~al.}(2021{\natexlab{b}})\citenamefont {Aker} \emph
  {et~al.}}]{KATRIN:2021fgc}%
  \BibitemOpen
  \bibfield  {author} {\bibinfo {author} {\bibfnamefont {M.}~\bibnamefont
  {Aker}} \emph {et~al.} (\bibinfo {collaboration} {KATRIN}),\ }\href {\doibase
  10.1103/PhysRevD.104.012005} {\bibfield  {journal} {\bibinfo  {journal}
  {Phys. Rev. D}\ }\textbf {\bibinfo {volume} {104}},\ \bibinfo {pages}
  {012005} (\bibinfo {year} {2021}{\natexlab{b}})},\ \Eprint
  {http://arxiv.org/abs/2101.05253} {arXiv:2101.05253 [hep-ex]} \BibitemShut
  {NoStop}%
\bibitem [{\citenamefont {Aker}\ \emph {et~al.}(2022)\citenamefont {Aker} \emph
  {et~al.}}]{KATRIN:2021uub}%
  \BibitemOpen
  \bibfield  {author} {\bibinfo {author} {\bibfnamefont {M.}~\bibnamefont
  {Aker}} \emph {et~al.} (\bibinfo {collaboration} {KATRIN}),\ }\href {\doibase
  10.1038/s41567-021-01463-1} {\bibfield  {journal} {\bibinfo  {journal}
  {Nature Phys.}\ }\textbf {\bibinfo {volume} {18}},\ \bibinfo {pages} {160}
  (\bibinfo {year} {2022})},\ \Eprint {http://arxiv.org/abs/2105.08533}
  {arXiv:2105.08533 [hep-ex]} \BibitemShut {NoStop}%
\bibitem [{\citenamefont {Asaka}\ \emph {et~al.}(2006)\citenamefont {Asaka},
  \citenamefont {Shaposhnikov},\ and\ \citenamefont {Kusenko}}]{Asaka:2006ek}%
  \BibitemOpen
  \bibfield  {author} {\bibinfo {author} {\bibfnamefont {T.}~\bibnamefont
  {Asaka}}, \bibinfo {author} {\bibfnamefont {M.}~\bibnamefont {Shaposhnikov}},
  \ and\ \bibinfo {author} {\bibfnamefont {A.}~\bibnamefont {Kusenko}},\ }\href
  {\doibase 10.1016/j.physletb.2006.05.067} {\bibfield  {journal} {\bibinfo
  {journal} {Phys. Lett. B}\ }\textbf {\bibinfo {volume} {638}},\ \bibinfo
  {pages} {401} (\bibinfo {year} {2006})},\ \Eprint
  {http://arxiv.org/abs/hep-ph/0602150} {arXiv:hep-ph/0602150} \BibitemShut
  {NoStop}%
\bibitem [{\citenamefont {Patwardhan}\ \emph {et~al.}(2015)\citenamefont
  {Patwardhan}, \citenamefont {Fuller}, \citenamefont {Kishimoto},\ and\
  \citenamefont {Kusenko}}]{Patwardhan:2015kga}%
  \BibitemOpen
  \bibfield  {author} {\bibinfo {author} {\bibfnamefont {A.~V.}\ \bibnamefont
  {Patwardhan}}, \bibinfo {author} {\bibfnamefont {G.~M.}\ \bibnamefont
  {Fuller}}, \bibinfo {author} {\bibfnamefont {C.~T.}\ \bibnamefont
  {Kishimoto}}, \ and\ \bibinfo {author} {\bibfnamefont {A.}~\bibnamefont
  {Kusenko}},\ }\href {\doibase 10.1103/PhysRevD.92.103509} {\bibfield
  {journal} {\bibinfo  {journal} {Phys. Rev. D}\ }\textbf {\bibinfo {volume}
  {92}},\ \bibinfo {pages} {103509} (\bibinfo {year} {2015})},\ \Eprint
  {http://arxiv.org/abs/1507.01977} {arXiv:1507.01977 [astro-ph.CO]}
  \BibitemShut {NoStop}%
\bibitem [{\citenamefont {Fuller}\ and\ \citenamefont
  {Schramm}(1992)}]{Fuller:1990mq}%
  \BibitemOpen
  \bibfield  {author} {\bibinfo {author} {\bibfnamefont {G.~M.}\ \bibnamefont
  {Fuller}}\ and\ \bibinfo {author} {\bibfnamefont {D.~N.}\ \bibnamefont
  {Schramm}},\ }\href {\doibase 10.1103/PhysRevD.45.2595} {\bibfield  {journal}
  {\bibinfo  {journal} {Phys. Rev. D}\ }\textbf {\bibinfo {volume} {45}},\
  \bibinfo {pages} {2595} (\bibinfo {year} {1992})}\BibitemShut {NoStop}%
\bibitem [{\citenamefont {Patwardhan}\ and\ \citenamefont
  {Fuller}(2014)}]{Patwardhan:2014iha}%
  \BibitemOpen
  \bibfield  {author} {\bibinfo {author} {\bibfnamefont {A.~V.}\ \bibnamefont
  {Patwardhan}}\ and\ \bibinfo {author} {\bibfnamefont {G.~M.}\ \bibnamefont
  {Fuller}},\ }\href {\doibase 10.1103/PhysRevD.90.063009} {\bibfield
  {journal} {\bibinfo  {journal} {Phys. Rev. D}\ }\textbf {\bibinfo {volume}
  {90}},\ \bibinfo {pages} {063009} (\bibinfo {year} {2014})},\ \Eprint
  {http://arxiv.org/abs/1401.1923} {arXiv:1401.1923 [astro-ph.CO]} \BibitemShut
  {NoStop}%
\bibitem [{\citenamefont {Tremaine}\ and\ \citenamefont
  {Gunn}(1979)}]{Tremaine:1979we}%
  \BibitemOpen
  \bibfield  {author} {\bibinfo {author} {\bibfnamefont {S.}~\bibnamefont
  {Tremaine}}\ and\ \bibinfo {author} {\bibfnamefont {J.~E.}\ \bibnamefont
  {Gunn}},\ }\href {\doibase 10.1103/PhysRevLett.42.407} {\bibfield  {journal}
  {\bibinfo  {journal} {Phys. Rev. Lett.}\ }\textbf {\bibinfo {volume} {42}},\
  \bibinfo {pages} {407} (\bibinfo {year} {1979})}\BibitemShut {NoStop}%
\bibitem [{\citenamefont {Boyarsky}\ \emph {et~al.}(2009)\citenamefont
  {Boyarsky}, \citenamefont {Ruchayskiy},\ and\ \citenamefont
  {Iakubovskyi}}]{Boyarsky:2008ju}%
  \BibitemOpen
  \bibfield  {author} {\bibinfo {author} {\bibfnamefont {A.}~\bibnamefont
  {Boyarsky}}, \bibinfo {author} {\bibfnamefont {O.}~\bibnamefont
  {Ruchayskiy}}, \ and\ \bibinfo {author} {\bibfnamefont {D.}~\bibnamefont
  {Iakubovskyi}},\ }\href {\doibase 10.1088/1475-7516/2009/03/005} {\bibfield
  {journal} {\bibinfo  {journal} {JCAP}\ }\textbf {\bibinfo {volume} {03}},\
  \bibinfo {pages} {005} (\bibinfo {year} {2009})},\ \Eprint
  {http://arxiv.org/abs/0808.3902} {arXiv:0808.3902 [hep-ph]} \BibitemShut
  {NoStop}%
\bibitem [{\citenamefont {Di~Paolo}\ \emph {et~al.}(2018)\citenamefont
  {Di~Paolo}, \citenamefont {Nesti},\ and\ \citenamefont
  {Villante}}]{DiPaolo:2017geq}%
  \BibitemOpen
  \bibfield  {author} {\bibinfo {author} {\bibfnamefont {C.}~\bibnamefont
  {Di~Paolo}}, \bibinfo {author} {\bibfnamefont {F.}~\bibnamefont {Nesti}}, \
  and\ \bibinfo {author} {\bibfnamefont {F.~L.}\ \bibnamefont {Villante}},\
  }\href {\doibase 10.1093/mnras/sty091} {\bibfield  {journal} {\bibinfo
  {journal} {Mon. Not. Roy. Astron. Soc.}\ }\textbf {\bibinfo {volume} {475}},\
  \bibinfo {pages} {5385} (\bibinfo {year} {2018})},\ \Eprint
  {http://arxiv.org/abs/1704.06644} {arXiv:1704.06644 [astro-ph.GA]}
  \BibitemShut {NoStop}%
\bibitem [{\citenamefont {Savchenko}\ and\ \citenamefont
  {Rudakovskyi}(2019)}]{Savchenko:2019qnn}%
  \BibitemOpen
  \bibfield  {author} {\bibinfo {author} {\bibfnamefont {D.}~\bibnamefont
  {Savchenko}}\ and\ \bibinfo {author} {\bibfnamefont {A.}~\bibnamefont
  {Rudakovskyi}},\ }\href {\doibase 10.1093/mnras/stz1573} {\bibfield
  {journal} {\bibinfo  {journal} {Mon. Not. Roy. Astron. Soc.}\ }\textbf
  {\bibinfo {volume} {487}},\ \bibinfo {pages} {5711} (\bibinfo {year}
  {2019})},\ \Eprint {http://arxiv.org/abs/1903.01862} {arXiv:1903.01862
  [astro-ph.CO]} \BibitemShut {NoStop}%
\bibitem [{\citenamefont {Das}\ and\ \citenamefont
  {Sigurdson}(2012)}]{Das:2010ts}%
  \BibitemOpen
  \bibfield  {author} {\bibinfo {author} {\bibfnamefont {S.}~\bibnamefont
  {Das}}\ and\ \bibinfo {author} {\bibfnamefont {K.}~\bibnamefont
  {Sigurdson}},\ }\href {\doibase 10.1103/PhysRevD.85.063510} {\bibfield
  {journal} {\bibinfo  {journal} {Phys. Rev. D}\ }\textbf {\bibinfo {volume}
  {85}},\ \bibinfo {pages} {063510} (\bibinfo {year} {2012})},\ \Eprint
  {http://arxiv.org/abs/1012.4458} {arXiv:1012.4458 [astro-ph.CO]} \BibitemShut
  {NoStop}%
\bibitem [{\citenamefont {Meiksin}(2009)}]{Meiksin:2007rz}%
  \BibitemOpen
  \bibfield  {author} {\bibinfo {author} {\bibfnamefont {A.~A.}\ \bibnamefont
  {Meiksin}},\ }\href {\doibase 10.1103/RevModPhys.81.1405} {\bibfield
  {journal} {\bibinfo  {journal} {Rev. Mod. Phys.}\ }\textbf {\bibinfo {volume}
  {81}},\ \bibinfo {pages} {1405} (\bibinfo {year} {2009})},\ \Eprint
  {http://arxiv.org/abs/0711.3358} {arXiv:0711.3358 [astro-ph]} \BibitemShut
  {NoStop}%
\bibitem [{\citenamefont {Zelko}\ \emph {et~al.}(2022)\citenamefont {Zelko},
  \citenamefont {Treu}, \citenamefont {Abazajian}, \citenamefont {Gilman},
  \citenamefont {Benson}, \citenamefont {Birrer}, \citenamefont {Nierenberg},\
  and\ \citenamefont {Kusenko}}]{Zelko:2022tgf}%
  \BibitemOpen
  \bibfield  {author} {\bibinfo {author} {\bibfnamefont {I.~A.}\ \bibnamefont
  {Zelko}}, \bibinfo {author} {\bibfnamefont {T.}~\bibnamefont {Treu}},
  \bibinfo {author} {\bibfnamefont {K.~N.}\ \bibnamefont {Abazajian}}, \bibinfo
  {author} {\bibfnamefont {D.}~\bibnamefont {Gilman}}, \bibinfo {author}
  {\bibfnamefont {A.~J.}\ \bibnamefont {Benson}}, \bibinfo {author}
  {\bibfnamefont {S.}~\bibnamefont {Birrer}}, \bibinfo {author} {\bibfnamefont
  {A.~M.}\ \bibnamefont {Nierenberg}}, \ and\ \bibinfo {author} {\bibfnamefont
  {A.}~\bibnamefont {Kusenko}},\ }\href {\doibase
  10.1103/PhysRevLett.129.191301} {\bibfield  {journal} {\bibinfo  {journal}
  {Phys. Rev. Lett.}\ }\textbf {\bibinfo {volume} {129}},\ \bibinfo {pages}
  {191301} (\bibinfo {year} {2022})},\ \Eprint
  {http://arxiv.org/abs/2205.09777} {arXiv:2205.09777 [hep-ph]} \BibitemShut
  {NoStop}%
\bibitem [{\citenamefont {Dienes}\ \emph {et~al.}(2022)\citenamefont {Dienes},
  \citenamefont {Huang}, \citenamefont {Kost}, \citenamefont {Thomas},\ and\
  \citenamefont {Yu}}]{Dienes:2021cxp}%
  \BibitemOpen
  \bibfield  {author} {\bibinfo {author} {\bibfnamefont {K.~R.}\ \bibnamefont
  {Dienes}}, \bibinfo {author} {\bibfnamefont {F.}~\bibnamefont {Huang}},
  \bibinfo {author} {\bibfnamefont {J.}~\bibnamefont {Kost}}, \bibinfo {author}
  {\bibfnamefont {B.}~\bibnamefont {Thomas}}, \ and\ \bibinfo {author}
  {\bibfnamefont {H.-B.}\ \bibnamefont {Yu}},\ }\href {\doibase
  10.1103/PhysRevD.106.123521} {\bibfield  {journal} {\bibinfo  {journal}
  {Phys. Rev. D}\ }\textbf {\bibinfo {volume} {106}},\ \bibinfo {pages}
  {123521} (\bibinfo {year} {2022})},\ \Eprint
  {http://arxiv.org/abs/2112.09105} {arXiv:2112.09105 [astro-ph.CO]}
  \BibitemShut {NoStop}%
\end{thebibliography}%
\end{document}